\documentclass{article}
\usepackage{graphicx}  
\usepackage{amsmath}   
\usepackage[compress]{cite}
\usepackage{amssymb}   
\usepackage{bm} 
\usepackage{dcolumn}
\usepackage{color}
\usepackage{mathrsfs}
\usepackage{amsfonts}
\usepackage{varioref}
\usepackage{textcomp}
\usepackage{empheq}
\RequirePackage[colorlinks,citecolor=blue,urlcolor=magenta,linkcolor=blue]{hyperref}
\allowdisplaybreaks
\usepackage[page,toc,titletoc,title]{appendix} 
\newcounter{mnotecount}[section]

\renewcommand{\themnotecount}{\thesection.\arabic{mnotecount}}

\newcommand{\mnotex}[1]
{\protect{\stepcounter{mnotecount}}$^{\mbox{\footnotesize
$
\bullet$\themnotecount}}$ \marginpar{
\raggedright\small\em
$\!\!\!\!\!\!\,\bullet$\themnotecount: #1} }

\labelformat{section}{Section #1} 
\labelformat{subsection}{Section #1} 
\labelformat{subsubsection}{Section #1}
\labelformat{subsubsubsection}{Section #1}
\labelformat{equation}{Eq.~(#1)} 
\labelformat{figure}{Fig.~#1} 
\labelformat{subfigure}{Fig.~\thefigure#1} 
\labelformat{table}{Tab.~#1} 
\labelformat{appendix}{Appendix #1}
\title{\bf Tidal heating of black holes and exotic compact objects on the brane}
\author{Sumanta Chakraborty\footnote{sumantac.physics@gmail.com}$~^{1}$, Sayak Datta\footnote{skdatta@iucaa.in}$~^{2}$, and Subhadip Sau\footnote{subhadipsau2@gmail.com}$~^{1}$
\\
$^{1}${\small{School of Physical Sciences}}\\
{\small{Indian Association for the Cultivation of Science, Kolkata-700032, India}}
\\
$^{2}${\small{IUCAA, Post Bag 4, Ganeshkhind, Pune 411007, India}}}
\begin{document}
  
\maketitle
\begin{abstract}

During the early phase of in-spiral of a binary system, the tidal heating of a compact object due to its companion plays a significant role in the determination of the orbital evolution of the binary. The phenomenon depends crucially on the `hairs', as well as on the nature of the compact object. It turns out that the presence of extra dimension affects both these properties, by incorporating an extra tidal charge for braneworld black holes and also by introducing quantum effects, leading to possible existence of exotic compact objects. It turns out that the phasing information from tidal heating in the gravitational wave waveform can constrain the tidal charge inherited from extra dimension to a value $\sim 10^{-6}$, the most stringent constraint, to date. Moreover, second order effects in tidal heating for exotic compact objects, also reveals an oscillatory behaviour with respect to spin, which has unique signatures.

\end{abstract}
\newpage
\tableofcontents
\section{Introduction}

Understanding the nature of the gravitational interaction at very many different length scales has been one of the prime research area in the last few decades. Most of the tests involving the gravitational interaction were in the weak field regime and they all confirmed the correctness of the Einstein's theory of general relativity \cite{Will:1993ns,Will:2005va,Will:2005yc,Berti:2015itd}. However, there are scenarios, which seems to be in tension with the predictions from General relativity. For example, the accelerated expansion of the universe cannot be described within the realm of general relativity, it either requires exotic matter or, modifications to general relativity \cite{Arun:2017uaw,Caldwell:2003vq,Kunz:2006ca,Peebles:2002gy}. Similarly, the black holes, which are solutions of the Einstein's field equations, harbour singularities, where the description of the spacetime structure itself breaks down, suggesting issues with general relativity \cite{Will:2005va,Will:2005yc,Will:2014bqa}. These, along with several other factors, have contributed in the search for gravitational theories beyond general relativity. However, with only weak field tests it was not possible to constrain, or, better, to even rule out some of these theories. All these have changed after the recent detection of gravitational waves \cite{Abbott:2016blz,TheLIGOScientific:2016pea,Abbott:2016nmj,TheLIGOScientific:2016src,Abbott:2017vtc} from mergers of binary black holes and neutron stars, as well as, an explicit observation of black hole shadow \cite{Akiyama:2019cqa,Akiyama:2019eap}. The gravitational wave observation involves several observables, associated with three distinct parts of the gravitational wave waveform --- (a) the in-spiral phase, (b) the merger phase and (c) the ringdown phase. In this paper we will solely concentrate on a certain aspect of the in-spiral phase, namely the phenomenon of \emph{tidal heating}, in theories beyond general relativity. 

One possible avenue, which requires modifications of general relativity from the perspective of a four dimensional observer, is the presence of an extra spatial dimension. Incorporation of an extra spatial dimension has benefits from several different arenas, most notably in the possible resolution of the gauge hierarchy problem \cite{Randall:1999ee,ArkaniHamed:1998rs,Csaki:2004ay}. The presence of a compact extra dimension can bring down the energy scale of the gravitational interaction from the Planck scale in five dimensions to the weak energy scale in four dimensions, thereby resolving the hierarchy of $\sim \mathcal{O}(10^{15})$ between the weak energy scale and the Planck scale \cite{Randall:1999ee,ArkaniHamed:1998rs,Randall:1999vf,Maartens:2010ar}. As a consequence, even if the gravitational interaction in the five dimensional spacetime is given by general relativity, the effective four dimensional description will involve further corrections over and above the general relativistic prediction due to the presence of the extra spatial dimension. The corrections to the effective gravitational field equations on the brane (equivalently, the four-dimensional hypersurface we live in) are dependent on two factors --- (a) the projection of the bulk Weyl tensor and (b) a particular quadratic combination involving energy-momentum tensor on the brane \cite{Shiromizu:1999wj,Dadhich:2000am,Harko:2004ui,Aliev:2005bi,Chakraborty:2015taq,Chakraborty:2014xla}. In vacuum spacetime, the sole modification to the Einstein's equations on the brane is from the bulk Weyl tensor and it behaves as a Maxwell field \emph{but} with its energy-momentum tensor having an overall \emph{negative} sign \cite{Dadhich:2000am,Harko:2004ui}. This provides a tantamount evidence to look for the presence of extra dimensions. Following which, we have studied the effect of the \emph{negative} tidal charge on the phenomenon of tidal heating, as two black holes on the four dimensional brane are moving around each other in a binary system. For other observational avenues, within or outside the domain of gravitational wave, to test the existence of a negative tidal charge, see \cite{Chakraborty_2018,Chakravarti:2018vlt,Chakravarti:2019aup,Banerjee:2019nnj,Banerjee:2019sae,Rahman:2018oso, Hou:2021okc}.

The presence of extra dimension can also affect the Physics of the binary black hole system in a non-trivial manner. In particular, it can modify the nature of the compact object, i.e., it may be more appropriate to consider these objects as exotic compact objects, rather than black holes. This is primarily because of two reasons, first of all, the AdS/CFT correspondence suggests that the gravitational field equations on the brane are modified by the presence of the quantum stress tensor of the CFT and hence the horizon will involve quantum corrections over and above the classical value. Secondly, the complete extension of the black hole solution to the bulk spacetime is not known and hence the horizon of the black hole on the brane is better regarded as an apparent horizon, than an event horizon. Following which, we may consider the solutions to the effective gravitational field equations on the brane to be an exotic compact object, with its surface differing from the location of the black hole horizon by some quantum corrections related to the bulk properties, which could be due to the back-reaction from the CFT on the brane \cite{Dey:2020lhq,Dey:2020pth}. In this work, we will treat the solutions of the effective gravitational field equations on the brane as black holes and also as exotic compact objects to understand the effect of the \emph{negative} tidal charge and the back-reaction due to quantum effects from the bulk on the in-spiral regime of two compact objects on the brane, through tidal heating, in a separate manner.    

The regime of the in-spiral phase, that we are interested in, corresponds to the situation when the separation between the orbiting compact objects in a binary are much larger than the characteristic size of these compact objects. For example, if these compact objects are black holes, then the above statement demands that the separation between them must be much larger than the horizon radius of the individual black holes. In this case, post-Newtonian approximations can be used to a very good accuracy. However, most of the studies regarding the early in-spiral phase, devotes their attention to the energy and angular momentum carried out to infinity through gravitational waves. On the contrary, in this work we are interested in knowing the energy and angular momentum flow down the horizon of a black hole or, the surface of an exotic compact object. Though the flow down the surface of the compact objects are smaller compared to the gravitational wave emission to infinity, these effects are important for accurate orbital evolution of a binary system. In the binary system, the tidal field of a compact object on its companion will deform its companion, whose ``hairs", e.g., mass and angular momentum will evolve with time. This phenomenon, known as \emph{tidal heating} \cite{Hartle:1973zz,Hughes:2001jr,PoissonWill}, is related to the absorption/emission from the compact objects and has been studied in several works \cite{Hughes:2001jr, Alvi:2001mx, Chatziioannou:2012gq, Isoyama:2017tbp, Datta:2019epe}. This phenomenon of tidal heating can help to figure out the presence of additional hairs, possibly implying the presence of an extra spatial dimension, and also to distinguish a black hole from an exotic compact object \cite{Datta:2019euh, Maselli:2017cmm, Datta:2020gem,Datta:2020rvo, Datta:2019epe,Datta:2021row}. This is what we will explore in this work. 

The paper is organized as follows: In \ref{tidal_heat_bh} we discuss the phenomenon of tidal heating for braneworld black holes. In particular, we will study the effect of an extra spatial dimension on the tidal heating of a brane-localized black hole in a binary system for both equatorial, as well as non-equatorial motion. Using the effect of tidal heating in the phase of gravitational wave waveform, we have been able to provide a strong constraint on the tidal charge parameter in \ref{tidal_heat_const}. Finally, implications of tidal heating for exotic compact objects in the braneworld, in the case of stationary companion have been detailed in \ref{tidal_heat_eco}, with special emphasis on the second order effects. We conclude with a discussion on our results in \ref{conclusion}. Some relevant computations have been presented in \ref{AppC} to \ref{AppG}.

\emph{Notations and Conventions:} We set the fundamental constants $c$ and $G$ to unity. All the four dimensional quantities are denoted by Greek indices ($\mu,\nu,\alpha,\cdots$) and we adopt mostly positive signature convention, such that the flat spacetime metric reads, $\textrm{diag}(-1,+1,+1,+1)$.

\section{Tidal heating for a braneworld black hole}\label{tidal_heat_bh}

In this section we will derive the phenomenon of tidal heating associated with a rotating black hole in a binary system living on the four-dimensional brane. There are three key steps in the derivation of the tidal heating for a black hole in a binary --- (a) determination of the area of the black hole and the connection between the change of area with the change of black hole parameters, which is akin to the laws of black hole mechanics; (b) the rate of change of area and angular momentum of the black hole, when the binary companion is stationary and (c) its generalization for non-stationary case, when the binary black hole system is in the equatorial as well as non-equatorial orbits. In what follows we will work out each of these steps in an explicit manner for the rotating black hole localized on the brane. 

\subsection{Area of a braneworld black hole}

The computation of tidal heating for a black hole crucially depends on its area and how the area depends on various hairs of the spacetime. For example, in the case of Kerr spacetime, this will be the mass $M$ and rotation parameter $a\equiv (J/M)$ of the black hole, where $J$ is the angular momentum. In the case of a rotating braneworld black hole, in addition to the mass $M$ and angular momentum $J$, we have an additional parameter, referred to as the `tidal charge' $Q$, inheriting the bulk degrees of freedom. Thus the braneworld black hole has three hairs and hence the area of the horizon of the black hole will also depend on all these three parameters. 

The computation of the area of a rotating braneworld black hole is most straightforward in the null coordinate system $(u,r,\theta,\phi)$, where $u$ is the null coordinate. In this coordinate system, the line element for the rotating braneworld black hole takes the following form \cite{Aliev:2005bi,Aliev_2009},
\begin{align}\label{brane_sol}
ds^{2}=-\left(1-\dfrac{2Mr+Q}{\rho^{2}}\right)du^{2} &-2du\,dr -\frac{2a\sin^{2}\theta\left(2Mr+Q\right)}{\rho^{2}}~du\,d\phi +2a\sin^{2}\theta d\phi \,dr 
\nonumber
\\
&+\rho^{2}\,d\theta^{2}+\dfrac{\sin^{2}\theta}{\rho^{2}}\left[(r^{2}+a^{2})^{2}-a^{2}\Delta \sin^{2}\theta \right]d\phi^{2}~,
\end{align}
where, $\rho^{2}\equiv r^{2}+a^{2}\cos^{2}\theta$ and $\Delta \equiv r^{2}-2Mr+a^{2}-Q$. The location of the black hole horizon can be obtained from the solution of the following algebraic equation, namely $\Delta=0$. Given the above expression for $\Delta$, the condition $\Delta=0$ yields two independent solutions: $r=r_{\pm}$, where, $r_{\pm}=M\pm \sqrt{M^{2}-a^{2}+Q}$. Interestingly note that for positive values of $Q$, the extremal limit (where the two roots of the equation $\Delta=0$ coincide) will correspond to $(a/M)>1$, which is a distinct signature of the braneworld scenario. We will explore possible consequences of this result later in this work. 

From the above discussion it is clear that the horizon corresponds to a $r=\textrm{constant}$ surface. Thus in order to determine the area of the horizon, we will consider its two-dimensional cross-section, formed at the junction of the $r=\textrm{constant}$ and the $u=\textrm{constant}$ hypersurfaces. Therefore, from \ref{brane_sol} we obtain the following expression for the determinant of the induced metric on this two-dimensional cross-section,
\begin{align}\label{tran_det}
h^{(2)}=(r^{2}+a^{2})^{2}\sin^{2}\theta - a^{2}\Delta \sin^{4}\theta~.
\end{align} 
Given the determinant of the induced metric, the area of this two-dimensional cross-section at the junction of $u=\textrm{constant}$ and $r=\textrm{constant}$ surface is given by,
\begin{align}\label{Area_Gen}
A=\int_{0}^{\pi}\int_{0}^{2\pi} \sqrt{h^{(2)}} \,d\theta d\phi = 2\pi \int_{0}^{\pi} \sqrt{(r^{2}+a^{2})^{2}-a^{2}\Delta+a^{2}\Delta \cos^{2}\theta}\, \, \sin \theta d\theta~,
\end{align}
where, we have performed the integral over $\phi$. The above expression for area, derived so far, is completely general, holding true for any $r=\textrm{constant}$ surface, with no reference to the horizon. In order to determine the area of the two-dimensional cross-section of the horizon, we must substitute $r=r_{+}$ in \ref{Area_Gen}. Then $\Delta$ will vanish identically and hence the expression for the area of the two-dimensional cross-section of the event horizon becomes,
\begin{flalign}\label{area_bh}
A_{\rm BH}=4\pi \left(r_{+}^{2}+a^{2}\right)~.
\end{flalign}
Using the expression for $r_{+}$ for the braneworld black hole, as described above, the area of the event horizon in terms of the `hairs' of the spacetime is given by,
\begin{align}
A_{\rm BH}=4\pi \left[2M^{2}+Q+2M\sqrt{M^{2}-a^{2}+Q}\right]=8\pi M^{2}\left[1+\frac{Q}{2M^{2}}+\sqrt{1-\chi^{2}+\frac{Q}{M^{2}}} \right]~,
\end{align}
where, we have defined the following dimensionless quantity: $\chi\equiv (a/M)$. Keeping our later purposes in mind, it will be useful to take the derivative of the area of the event horizon with respect to the mass $M$, angular momentum $J$ and the tidal charge parameter $Q$, which yields,
\begin{flalign}
\partial_{M}A_{\rm BH}&=\dfrac{16\pi M}{\sqrt{1-\chi^{2}+\frac{Q}{M^{2}}}}\left[1+\frac{Q}{2M^{2}}+\sqrt{1-\chi^{2}+\frac{Q}{M^{2}}}\right]~,
\\
\partial_{J}A_{\rm BH}&=-\dfrac{8\pi \chi}{\sqrt{1-\chi^{2}+\frac{Q}{M^{2}}}}~;
\qquad 
\partial_{Q}A_{\rm BH}=4\pi\left[1+\dfrac{1}{\sqrt{1-\chi^{2}+\frac{Q}{M^{2}}}}\right]~.
\end{flalign}
Combining these results, we may write down the following relation,
\begin{align}\label{derivative_relation}
d\left(\frac{A_{\rm BH}}{8\pi}\right)=\dfrac{2MdM}{\sqrt{1-\chi^{2}+\frac{Q}{M^{2}}}}\left[1+\frac{Q}{2M^{2}}+\sqrt{1-\chi^{2}+\frac{Q}{M^{2}}}\right]-\dfrac{\chi dJ}{\sqrt{1-\chi^{2}+\frac{Q}{M^{2}}}}+\frac{dQ}{2}\left[1+\dfrac{1}{\sqrt{1-\chi^{2}+\frac{Q}{M^{2}}}}\right]~.
\end{align}
Note that all the quantities in the square bracket, along with the coefficient of $dJ$ are expressed in terms of the two dimensionless quantities, $\chi$ and $(Q/M^{2})$. This is the practice we will follow throughout this work, by expressing all the relevant expressions in terms of the two dimensionless quantities introduced above. The above expression is merely the law of black hole mechanics taking into account the tidal charge parameter as well. However, in practical situations we may ignore the change in area due to a change in the tidal charge parameter. This is because the tidal charge parameter is originating from the bulk (equivalently, five-dimensional spacetime) and hence any change in the tidal charge will reflect a corresponding change in the bulk spacetime. For example, the tidal charge parameter is related to the length of the extra dimension \cite{Chamblin_2001} and hence any change in the tidal charge parameter will reflect as a change in the length of the extra dimension. However changing any parameters of the extra dimension is a high energy process, which cannot be realized in any of the processes on the brane, e.g., emission of gravitational waves from a binary black hole system, as considered here. Thus for all intents and purposes of the present work, we may ignore any change in the tidal charge parameter. This consideration as well as the above expression for the change in area will be extensively used in the later discussions.  

\subsection{Tidal heating with a stationary companion}

In this section we will derive the tidal heating or, the mass and the angular moment influx into a black hole (denoted as BBH1, or braneworld black hole 1) as the second black hole (denoted as BBH2) is held stationary, i.e., for a stationary companion in a binary system. Effectively, the stationary companion, namely BBH2 will tidally deform the BBH1 and we would like to find out the tidal field of BBH2 acting on BBH1. 
   
For this purpose we will work in the Local Asymptotic Rest Frame (denoted by LARF) of the BBH1, in which the three-space coordinates of the stationary BBH2 are taken to be $(b,\theta_{0},\phi_{0})$. In such a scenario, the expansion of the Newtonian gravitational potential (effectively the $g_{00}$ component in the LARF coordinate system) due to the stationary BBH2, yields \cite{Alvi:2001mx},
\begin{align}
\Phi(r,\theta,\phi)=-4\pi \left(\frac{M_{2}}{b}\right)\sum_{\ell=0}^{\infty}\sum_{m=-l}^{l}\left(\frac{1}{2l+1}\right)\left(\frac{r}{b}\right)^{l}Y_{\ell m}^{*}(\theta_{0},\phi_{0})Y_{\ell m}(\theta,\phi)~,
\end{align}
where, $M_{2}$ is the mass of the stationary companion BBH2. The above expansion of the gravitational potential is true in the region $r<b$, which is where the BBH1 is located. Due to this external tidal field, the BBH1 will experience a tidal force, which is given by, $\mathcal{E}_{ij}=\partial_{i}\partial_{j}\Phi$ \cite{MembraneParadigm,Alvi:2001mx}. The most relevant angular mode for our analysis corresponds to the $l=2$ mode and the following combination of the tidal field $\left(\mathcal{E}_{\hat{\phi}\hat{\phi}}-\mathcal{E}_{\hat{\theta}\hat{\theta}}-2i\mathcal{E}_{\hat{\theta}\hat{\phi}}\right)$ is the one responsible for tidal deformation. In the present context, the above combination reads \cite{Alvi:2001mx}
\begin{align}\label{tidal_field}
\mathcal{E}_{\hat{\phi}\hat{\phi}}-\mathcal{E}_{\hat{\theta}\hat{\theta}}-2i\mathcal{E}_{\hat{\theta}\hat{\phi}}=8\pi \frac{\sqrt{6}M_{2}}{5b^{3}}\sum_{m=-2}^{2}~_{2}Y_{2m}(\theta,\phi)Y^{*}_{2m}(\theta_{0},\phi_{0})~,
\end{align}
where, $\,_{2}Y_{2m}(\theta,\phi)$ is the spin-weighted spherical harmonics \cite{Goldberg:1966uu}. Note that due to the tidal field $\Phi$, arising from the stationary companion BBH2, the BBH1 will be perturbed. If we are interested in a regime which is far away from the horizons of both the black holes, however at a much smaller distance compared to the separation between the two black holes (i.e., $\{M_{1},M_{2}\}\ll r\ll b$), we can consider the perturbation of the BBH1 to be linear. Thus, in this intermediate spacetime region, the perturbations will satisfy the Teukolsky-like equation (see \cite{Teukolsky:1973ha}), which must be solved with an appropriate boundary condition in order to determine the rate of change of area, leading to tidal heating. The Teukolsky equation is in terms of the Weyl scalars $\Psi_{0}$ and $\Psi_{4}$. Since we are interested in matter influx through the horizon, we took $\Psi_{0}$ as the primary field. Furthermore, we impose the boundary condition on $\Psi_{0}$, such that, asymptotically it should have the form given by the tidal field combination $\left(\mathcal{E}_{\hat{\phi}\hat{\phi}}-\mathcal{E}_{\hat{\theta}\hat{\theta}}-2i\mathcal{E}_{\hat{\theta}\hat{\phi}}\right)$.  

In other words, the Teukolsky-like equation governing the linear gravitational perturbation of the rotating branewold black hole BBH1, due to the binary companion BBH2, is a second order differential equation and hence it requires two boundary conditions in order to determine the solution. One of the boundary condition is imposed in the asymptotic regime, which in the present context corresponds to $r\gg M_{1}$, but $r\ll b$. Such that \ref{tidal_field} provides the necessary boundary condition on the gravitational perturbation and hence we obtain,    
\begin{align}\label{far_bound_cond}
\Psi_{0}\rightarrow 8\pi \frac{\sqrt{6}M_{2}}{5b^{3}}\sum_{m=-2}^{2}~_{2}Y_{2m}(\theta,\phi)Y^{*}_{2m}(\theta_{0},\phi_{0})~.
\end{align}
Since the spacetime described by \ref{brane_sol} depicts a black hole, the other boundary condition is imposed on the black hole horizon located at $r=r_{+}$ and it corresponds to purely ingoing gravitational perturbation. As we will see later, for exotic compact objects (henceforth as ECOs), it is the boundary condition at the surface of the ECO, which will be different, yielding a modified solution to the gravitational perturbation, satisfying the Teukolsky-like equation. 

At this outset, let us briefly comment on the applicability of the Teukolsky-like equation in the context of gravitational perturbation of rotating braneworld black hole. First of all, the effective gravitational field equations on the four-dimensional brane takes the form, $^{(4)}G_{\mu \nu}+E_{\mu \nu}=0$, where $E_{\mu \nu}$ is an appropriate projection of the bulk Weyl tensor, signalling the non-vacuum nature of the field equations. Due to the symmetries of the Weyl tensor, it follows that, $E^{\mu}_{\mu}=0$, and also from the Bianchi identity, we obtain $\,^{(4)}\nabla_{\mu}E^{\mu}_{\nu}=0$, where $\,^{(4)}\nabla_{\mu}$ is the four-dimensional covariant derivative operator. Thus, $E_{\mu \nu}$ behaves in an identical manner to that of a Maxwell stress-tensor, with an overall negative sign, explaining the origin of the tidal charge term, with the crucial negative sign, in \ref{brane_sol}. The perturbation of the above equation will involve perturbation of the four-dimensional Einstein tensor $^{(4)}G_{\mu \nu}$, as well as the perturbation of the extra-dimensional contribution from $E_{\mu \nu}$. This will in general lead to non-separable equations for the electromagnetic and gravitational perturbations \cite{Berti:2005eb,dudley1979covariant}. However, under reasonable assumptions, as described in \cite{dudley1979covariant,PhysRevLett.39.367}, the perturbation equations for electromagnetic and gravitational perturbations become separable. In the present scenario as well, we are interested in the gravitational perturbation of the four-dimensional brane, whose energy scale is much smaller than the energy scale of the bulk spacetime, we can perturb $^{(4)}G_{\mu \nu}$, keeping $E_{\mu \nu}$, contribution from the extra-dimension, unchanged. In this case, following \cite{Berti:2005eb,dudley1979covariant,PhysRevLett.39.367}, it is evident that the gravitational perturbations will satisfy a Teukolsky-like equation, as considered here.

For the purpose of solving the Teukolsky-like equation, we note that the existence of a timelike Killing vector field $(\partial/\partial t)^{\mu}$ and rotational Killing vector field $(\partial/\partial \phi)^{\mu}$ ensures that the Weyl scalar can be decomposed as,
\begin{align}
\Psi_{0}=\sum_{l=0}^{\infty}\sum_{m=-l}^{l}~_{2}S_{lm}(\theta)R_{lm}(r)e^{-i\omega t}e^{im\phi}~,
\end{align}
where the angular part $\,_{2}S_{lm}(\theta)$ satisfies the following differential equation \cite{Teukolsky:1972my,MTB},
\begin{align}
\frac{1}{\sin \theta}\dfrac{d}{d\theta}\left[\sin \theta\dfrac{d  {}_{2}S_{lm}}{d\theta}\right]+ \left[(a\omega \cos\theta)^{2}- 4a\omega \cos\theta+2+ {}_{2}A_{lm} -\dfrac{(m+2\cos \theta)^{2}}{1-\cos^{2}\theta}\right]  {}_{2}S_{lm}=0~.
\end{align}
The above equation coincides with that of the spin-weighted spherical harmonics, with $~_{2}A_{lm}$ acting as the separation constant between the radial and the angular part of the gravitational perturbation. Finally the radial function $R_{lm}(r)$ satisfies the following differential equation \cite{Teukolsky:1972my,MTB}, 
\begin{align}\label{Eq_5.5}
\frac{1}{\Delta^{2}}\dfrac{d}{dr}\left[\Delta^{3} \dfrac{dR_{lm}}{dr}\right]+ \left[ \dfrac{K^{2}-4i(r-M)K}{\Delta}+ 4i\dfrac{dK}{dr}-\lambda \right]R_{lm } =0~,
\end{align}
where, $K= (r^{2}+a^{2})\omega -am$ and $\lambda= {}_{2}A_{lm}+(a\omega)^{2}-2am\omega$, is related to the separation constant $~_{2}A_{lm}$ appearing in the angular part of the perturbation equation. As we have emphasized earlier, the scenario relevant for our purpose of the computation of the tidal heating in a binary black hole system corresponds to $l=2$. Additionally, in the near horizon and in the small frequency limit, the radial perturbation equation can be solved in an analytic manner. In order to obtain the solution of the gravitational perturbation it is advantageous to introduce the following geometrical quantities,
\begin{align}\label{redefinition}
x\equiv \left(\dfrac{r-r_{+}}{r_{+}-r_{-}}\right)~;
\qquad 
\gamma_{m}\equiv i\dfrac{(r_{+}^{2}+a^{2})(m\Omega_H - \omega)}{\left(r_{+}-r_{-}\right)}
\equiv -i\dfrac{(r_{+}^{2}+a^{2})\bar{\omega}}{\left(r_{+}-r_{-}\right)}~,
\end{align}
where, $\bar{\omega}\equiv \omega-m\Omega_H$ and $\Omega_H=a/(r_{+}^{2}+a^{2})$ is the angular velocity of the horizon. Under this change of variable from the radial coordinate $r$ to the newly defined coordinate $x$, the radial differential equation governing the gravitational perturbation takes the following form,
\begin{align}
\frac{1}{x(1+x)}\partial_{x}\Big\{[x(1+x)]^{3}\partial_{x}R_{2m}\Big\}+\Big[-\gamma_{m}^{2} +2\gamma_{m}(1+2x)\Big]R_{2m} =0~.
\end{align}
The solution to the above differential equation can be obtained in terms of the hypergeometric functions and in the present context the solution to the radial part of the gravitational perturbation equation takes the following form,
\begin{align}
R_{2m} = (1+x)^{\gamma_{m}}\Big[A_{m} x^{-\gamma_{m}} {}_{2}F_{1} (0,5;3-2\gamma_{m};-x)+ C_{m} x^{\gamma_{m}-2} {}_{2}F_{1}(-2+2\gamma_{m},3+2\gamma_{m};2\gamma_{m}-1;-x)\Big]~,
\end{align}
where $A_{m}$ and $C_{m}$ are arbitrary constants, whose values we need to determine from the boundary conditions discussed above. The above solution to the radial part of the gravitational perturbation can also be expressed in a more useful form. This is achieved by using the following identity for the hypergeometric function: ${}_{2}F_{1} (a,b;c;z) = (1-z)^{c-a-b} {}_{2}F_{1}(c-a,c-b;c;z)$, which transforms the radial part of the gravitational perturbation $R_{2m}$ to the following form,
\begin{align}
R_{2m}&=(1+x)^{\gamma_{m}}\Big[A_{m} x^{-\gamma_{m}} {}_{2}F_{1} (0,5;3-2\gamma_{m};-x)+ C_{m} x^{\gamma_{m}-2} (1+x)^{-2\gamma_{m}-2} {}_{2}F_{1}(1,-4;2\gamma_{m}-1;-x)\Big]
\nonumber
\\
&=(1+x)^{\gamma_{m}}\Big[A_{m} x^{-\gamma_{m}}+ C_{m} x^{\gamma_{m}-2} (1+x)^{-2\gamma_{m}-2} {}_{2}F_{1}(1,-4;2\gamma_{m}-1;-x)\Big]~.
\end{align}
Here we have used the result, ${}_{2}F_{1} (0,b;c;z)=1$ and also note that the other hypergeometric function ${}_{2}F_{1}(1,-4;2\gamma_{m}-1;-x)$ can be expressed as a fourth order polynomial function in $x$. In this form it is easier to determine the unknown coefficients $A_{m}$ and $C_{m}$ appearing in $R_{2m}$. For this purpose, we need to find out the near horizon as well as the asymptotic behaviour, i.e., the $x \to \infty$ and $x\rightarrow 0$ limit of the radial perturbation $R_{2m}$, respectively. In the asymptotic limit, using the identity
\begin{align}
\displaystyle{\lim_{x\to \infty}} {}_{2}F_{1}(a,b;c;x)  = \dfrac{\Gamma(b-a)\Gamma(c)}{\Gamma(b)\Gamma(c-a)}(-x)^{-a} + \dfrac{\Gamma(a-b)\Gamma(c)}{\Gamma(a)\Gamma(c-b)} (-x)^{-b}~,
\end{align}
we obtain,
\begin{align}\label{far_zone}
\displaystyle{\lim_{x \to \infty}} R_{2m} (x)=A_{m}+C_{m}\dfrac{6}{(1+\gamma_{m})(1+2\gamma_{m})\gamma_{m}}~.
\end{align}
On the other hand, in the near horizon limit, we can use the following identity for the hypergeometric function, namely $\lim _{x\rightarrow 0}{}_{2}F_{1}(a,b;c;x)=1$, to obtain,
\begin{align}\label{near_horizon}
\displaystyle{\lim_{x \to 0}} R_{2m}(x)&=A_{m}\exp\left[i\bar{\omega}r_{*}\right]+\frac{C_{m}}{\Delta^{2}}\exp\left[-i\bar{\omega}r_{*}\right]~,
\end{align}
where, we have absorbed a few constants in the coefficients $A_{m}$ and $C_{m}$, respectively. The quantity $\bar{\omega}$ has already been defined earlier and has the expression, $\omega-m\Omega$, while the tortoise coordinate $r_{*}$ can be defined as,
\begin{align}
r_{*}=\int dr~\frac{r^{2}+a^{2}}{(r-r_{+})(r-r_{-})}=r-r_{+}+\left(\frac{r_{+}^{2}+a^{2}}{r_{+}-r_{-}}\right)\log\left(\frac{r-r_{+}}{r_{+}} \right)-\left(\frac{r_{-}^{2}+a^{2}}{r_{+}-r_{-}}\right)\log\left(\frac{r-r_{-}}{r_{-}} \right)~.
\end{align}
Such that near $r=r_{+}$, we obtain the following relation between the tortoise coordinate $r_{*}$ and the redefined radial coordinate $x$ , 
\begin{align}
x\simeq\frac{r_{+}}{(r_{+}-r_{-})}\exp\left[\left(\frac{r_{+}-r_{-}}{r_{+}^{2}+a^{2}}\right)r_{*}\right]~,
\end{align}
which have been used to derive the final form of the near horizon solution in \ref{near_horizon}. As emphasized earlier, in the case of black hole spacetime, in the near horizon regime, the perturbation modes must be purely ingoing, which demands $R_{2m}\sim e^{-i\bar{\omega}r_{*}}$. Therefore, compatibility of the radial perturbation variable $R_{2m}(r)$ in \ref{near_horizon}, with the above boundary condition, demands $A_{m}=0$. The other arbitrary constant $C_{m}$ can be determined by taking the asymptotic limit $x\rightarrow \infty$. Thus matching the asymptotic solution for the radial perturbation variable derived in \ref{far_zone} with the boundary condition given by \ref{far_bound_cond}, we immediately obtain the following expression for the unknown coefficient $C_{m}$, as,
\begin{align}\label{arb_constant}
C_{m}=\dfrac{8\pi M_{2}}{5\sqrt{6}b^{3}} \gamma_{m}({\gamma_{m}+1})(2\gamma_{m}+1)Y^{*}_{2m}(\theta_{0},\phi_{0})~.
\end{align} 
Therefore, the solution of the Teukolsky-like equation for the linear gravitational perturbation of the rotating braneworld black hole, satisfying proper asymptotic and near horizon boundary conditions, takes the following form,
\begin{align}
\Psi_{0}=e^{-i\omega t}\sum_{m=-2}^{2} C_{m}~x^{\gamma_{m}-2} (1+x)^{-2-\gamma_{m}}~ {}_{2}F_{1} (1,-4,-1+2\gamma_{m};-x)~ {}_{2}Y_{2m}(\theta,\phi)  
\end{align}
where $x$ is related to the radial coordinate $r$ through \ref{redefinition} and the constant $C_{m}$ is given by \ref{arb_constant}. However, note that, in the coordinate system considered here, the perturbation at the event horizon is not regular, as it scales as $\Delta^{-2}$. In order to obtain a regular perturbation at the event horizon we have to transform to a new null basis, introduced in \cite{Hawking:1972hy} and known as the Hartle-Hawking tetrad. The explicit transformation to the Hartle-Hawking tetrad can be found in \cite{Teukolsky_thesis,Teukolsky:1973ha,Teukolsky:1974yv}, following which we quote the final expression for the transformed Weyl scalar component in the Hartle-Hawking frame, 
\begin{align}\label{HH_Weyl_BH}
\Psi_{0}^{\rm HH}=\dfrac{\Delta^{2}}{4(r^{2}+a^{2})^{2}}e^{-i\omega t}\sum_{m=-2}^{2} C_{m}~x^{\gamma_{m}-2} (1+x)^{-2-\gamma_{m}}~ {}_{2}F_{1} (1,-4,-1+2\gamma_{m};-x)~ {}_{2}Y_{2m}(\theta,\phi)~.
\end{align}
Here, the superscript `HH' stands for the Weyl scalar in the null tetrad basis, introduced by Hartle and Hawking in \cite{Hawking:1972hy}. In the near horizon regime, i.e., as $x\rightarrow 0$, a factor of $x^{2}$ originates from the $\Delta^{2}$ term, which combines with the $x^{\gamma_{m}-2}$, thereby yielding $x^{\gamma_{m}}$. Similarly, in this limit, any term involving a factor of $(1+x)$ can be approximated as unity, such that the Weyl scalar in the Hartle-Hawking null tetrad becomes, 
\begin{align}\label{Weyl_HH}
\Psi_{0}^{\rm HH}(x\rightarrow 0)=\dfrac{\left(1-\chi_{1}^{2}+\frac{Q}{M_{1}^{2}}\right)^{2}}{\left[\left(1+\sqrt{1-\chi_{1}^{2}+\frac{Q}{M_{1}^{2}}}\right)+\frac{Q}{2M_{1}^{2}}\right]^{2}}\sum_{m=-2}^{2} C_{m}~x^{\gamma_{m}}~ {}_{2}Y_{2m}(\theta,\phi)~.
\end{align}
Note that, the above expression is for the linear gravitational perturbation of the BBH1 due to the tidal field of the BBH2 in the binary system. Thus, having derived the relevant gravitational perturbation, let us now compute the rate of change of area and angular momentum due to tidal heating. This can be achieved following \cite{Hawking:1972hy}, which we will briefly review here for completeness. The first order gravitational perturbation will cause the horizon to be perturbed, in particular the convergence and shear of the null congruence on the horizon will be modified. Additionally, the rotation of the black hole demands a certain periodicity in the perturbations of convergence and shear, which imposes the conditions that the first order perturbation of the convergence should vanish, while the second order perturbation will depend on the square of the first order perturbation of the shear. Since the convergence of the null congruence on the horizon is related to the rate of change of area of the horizon, it follows that \cite{Hawking:1972hy}  
\begin{align}\label{stat_ini_area}
\frac{dA_{\rm BH1}}{dt}=\frac{2}{\kappa_{+1}}\int |\sigma_{\rm HH}|^{2}\sqrt{h^{(2)}}d^{2}x=\frac{2(r_{+1}^{2}+a_{1}^{2})}{\kappa_{+1}}\int |\sigma_{\rm HH}|^{2}d\Omega~,
\end{align}
where, in deriving the second equality we have used the expression for the determinant of the induced metric on the black hole horizon from \ref{tran_det}. Moreover, the first order perturbation in the shear is related to the Weyl scalar $\Psi_{0}^{\rm HH}$ and hence one can express the rate of change of area of the BBH1, due to the binary companion BBH2, in terms of the angular average of $|\Psi_{0}^{\rm HH}|^{2}$. Given the expression for the Weyl scalar $\Psi_{0}^{\rm HH}$ in the near horizon regime in \ref{Weyl_HH}, the angular average of the magnitude of the Weyl scalar $\Psi_{0}^{\rm HH}$ yields,
\begin{align}\label{ang_avg}
\int d\Omega \vert \Psi_{0}^{\rm HH}\vert^{2}&=\dfrac{\left(1-\chi_{1}^{2}+\frac{Q}{M_{1}^{2}}\right)^{4}}{\left[\left(1+\sqrt{1-\chi_{1}^{2}+\frac{Q}{M_{1}^{2}}}\right)+\frac{Q}{2M_{1}^{2}}\right]^{4}}\sum_{m=-2}^{2}\sum_{m'=-2}^{2} C_{m}C_{m'}^{*}~x^{\gamma_{m}+\gamma_{m'}^{*}}\int d\Omega ~{}_{2}Y_{2m}(\theta,\phi)~{}_{2}Y_{2m'}^{*}(\theta,\phi)
\nonumber
\\
&=\dfrac{\left(1-\chi_{1}^{2}+\frac{Q}{M_{1}^{2}}\right)^{4}}{\left[\left(1+\sqrt{1-\chi_{1}^{2}+\frac{Q}{M_{1}^{2}}}\right)+\frac{Q}{2M_{1}^{2}}\right]^{4}}\sum_{m=-2}^{2}\vert C_{m}\vert^{2}~.
\end{align}
In order to arrive at the last expression, we have used the orthonormality property of the spin-weighted spherical harmonics $_{2}Y_{2m}(\theta,\phi)$. The only remaining bit corresponds to the computation related to $|C_{m}|^{2}$, where the arbitrary constant $C_{m}$ has the expression given by \ref{arb_constant}. It turns out that the $m=0$ mode appearing in the summation of \ref{ang_avg} does not contribute \cite{Alvi:2001mx, Datta:2020rvo} and hence by taking into account the expression for $\gamma_{m}$ from \ref{redefinition} and summing $|C_{m}|^{2}$ appropriately, we obtain (for a derivation, see \ref{AppC}),
\begin{align}\label{mag_arb_cons}
\sum_{m\neq 0}\vert C_{m}\vert^{2}&=\sum_{m\neq 0}\dfrac{64\pi^{2}M_{2}^{2}}{25\times 6b^{6}}\vert\gamma_{m}\vert^{2}({\gamma_{m}+1})({\gamma_{m}^{*}+1})(2\gamma_{m}+1)(2\gamma_{m}^{*}+1)Y^{*}_{2m}(\theta_{0},\phi_{0})Y_{2m}(\theta_{0},\phi_{0})
\nonumber
\\
&=\dfrac{2\pi M_{2}^{2}}{5b^{6}} \dfrac{\chi_{1}^{2}\sin^{2}\theta_{0}}{\left(1-\chi_{1}^{2}+\frac{Q}{M_{1}^{2}}\right)^{3}} 
\bigg[1- \dfrac{3}{4}\chi_{1}^{2}+\frac{Q}{M_{1}^{2}}\left\{2-\dfrac{3}{4}\chi_{1}^{2}+\frac{Q}{M_{1}^{2}}\right\} 
+\dfrac{15}{4}\chi_{1}^{2}\left\{1+\frac{Q}{M_{1}^{2}}\right\}\sin^{2}\theta_{0}\bigg]~.
\end{align}
Finally, the substitution of the summation of $\vert C_{m}\vert^{2}$ over appropriate $m$ values derived in \ref{mag_arb_cons}, in the expression for the angular average of $\vert \Psi_{0}^{\rm HH}\vert^{2}$ given by \ref{ang_avg}, yields the following expression,
\begin{align}\label{ang_avg_final}
\int d\Omega \vert \Psi_{0}^{\rm HH}\vert^{2}&=\dfrac{2\pi M_{2}^{2}}{5b^{6}}\dfrac{\left(1-\chi_{1}^{2}+\frac{Q}{M_{1}^{2}}\right)\chi_{1}^{2}\sin^{2}\theta_{0}}{\left[\left(1+\sqrt{1-\chi_{1}^{2}+\frac{Q}{M_{1}^{2}}}\right)+\frac{Q}{2M_{1}^{2}}\right]^{4}}\Bigg[\mathcal{A}+\mathcal{B}\sin^{2}\theta_{0}\Bigg]
\\
\mathcal{A}&\equiv 1-\dfrac{3}{4}\chi_{1}^{2}+\frac{Q}{M_{1}^{2}} \left\{2-\dfrac{3}{4} \bar{\chi}_{1}^{2}+\frac{Q}{M_{1}^{2}}\right\}~;
\quad 
\mathcal{B}\equiv \dfrac{15}{4}\chi_{1}^{2}\left(1+\frac{Q}{M_{1}^{2}}\right)~.
\end{align}
To reiterate, we have expressed the angular average in terms of the dimensionless parameters, $\chi_{1}=(a_{1}/M_{1})$ and $(Q/M_{1}^{2})$. Note that in the $Q\rightarrow 0$ limit, the above expression reduces to the one derived in \cite{Alvi:2001mx}, depicting internal consistency of this result. Therefore, we obtain the following expression for the rate of change of area of the BBH1, due to the tidal field of its binary companion as (see \ref{AppD} for derivation),
\begin{align}\label{stat_area_change}
\frac{dA_{\rm BH1}}{dt}=\frac{64\pi M_{1}^{5}M_{2}^{2}}{5b^{6}}\frac{\chi_{1}^{2}\sin^{2}\theta_{0}}{\sqrt{1-\chi_{1}^{2}+\frac{Q}{M_{1}^{2}}}}\left[\mathcal{A}+\mathcal{B}\sin^{2}\theta_{0}\right]~.
\end{align}
Note that, the rate of change of area is proportional to the dimensionless rotation parameter $\chi_{1}$ and hence for non-rotating black hole there is no change in area. This is expected, since for non-rotating black hole the area depends on the mass of the black hole alone, which does not change in the stationary approximation. It is often useful to express the above expression for the rate of change of area in dimensionless units. Thus, using the total mass $M\equiv M_{1}+M_{2}$ of the binary system, we obtain,
\begin{align}
\frac{1}{M}\frac{dA_{\rm BH1}}{dt}=\frac{64\pi M_{1}^{3}M^{3}\eta^{2}}{5b^{6}}\frac{\chi_{1}^{2}\sin^{2}\theta_{0}}{\sqrt{1-\chi_{1}^{2}+\frac{Q}{M_{1}^{2}}}}\left[\mathcal{A}+\mathcal{B}\sin^{2}\theta_{0}\right]~,
\end{align}
where, we have introduced the dimensionless mass ratio: $\eta \equiv (M_{1}M_{2}/M^{2})$. As we have seen earlier the area of the braneworld black hole depends on the hairs of the black hole, namely the mass $M$, angular momentum $J$ and tidal charge $Q$ and hence the rate of change of area will induce change in all these three hairs. However, in the stationary approximation the rate of change of mass identically vanishes, i.e., $(dM_{1}/dt)=0$. Additionally, as emphasized before, the tidal charge parameter is inherited from the higher dimensional spacetime and it is not possible to modify the structure of the extra dimension using the low-energy physics on the brane. Thus we can safely assume $(dQ/dt)=0$ as well. Hence the rate of change of the angular momentum $J$ can be obtained from \ref{derivative_relation}, which for $(dM_{1}/dt)=0=(dQ/dt)$, yields,
\begin{align}
\frac{dJ_{1}}{dt}&=-\frac{\sqrt{1-\chi_{1}^{2}+\frac{Q}{M_{1}^{2}}}}{8\pi\chi_{1}}\left(\frac{dA_{\rm BH1}}{dt}\right)
=-\frac{8 M_{1}^{5}M_{2}^{2}}{5b^{6}}\chi_{1}\sin^{2}\theta_{0}\left[\mathcal{A}+\mathcal{B}\sin^{2}\theta_{0}\right]~.
\end{align}
It is instructive to rewrite the above expression for the rate of change of angular momentum in a dimensionless form and that can be achieved through the division of the above expression by its Newtonian counterpart. In particular, the rate of change of angular momentum in the Newtonian limit becomes, $(dJ/dt)_{\rm N}=(32/5)(M_{1}^{2}M_{2}^{2}/b^{3})\sqrt{(M/b)}$, where $M=M_{1}+M_{2}$ is the total mass of the black hole binary and $b$ is the separation between the black holes in the binary system. This yields,
\begin{align}
\frac{dJ_{1}}{dt}=-\left(\frac{dJ}{dt}\right)_{\rm N}\frac{M_{1}^{3}}{4b^{2}\sqrt{bM}}\chi_{1}\sin^{2}\theta_{0}\left[\mathcal{A}+\mathcal{B}\sin^{2}\theta_{0}\right]~.
\end{align}
These provide the respective expressions for the rate of change of area and the rate of change of angular momentum of the BBH1 due to the tidal field of the BBH2 under stationary approximation, i.e., assuming $dM_{1}/dt=0$. Furthermore, these results reduce to the rate of change of area and angular momentum of a Kerr black hole in a binary system in the limit $Q\rightarrow 0$, as one can explicitly verify.

\subsection{Rate of change of geometrical quantities on the equatorial orbit}

In the previous section, we have worked out the rate of change of area and angular momentum of a braneworld black hole in a binary system for the specific case of a stationary companion. Let us now consider a generalized scenario, which is also the most important one from the astrophysical point of view, in which both the black holes are on the equatorial plane and rotating around each other with angular velocity $\Omega$. However for generality, rather than considering the equatorial plane alone, we will derive the rate of change of area, mass and angular momentum as the motion of the binary system is confined to a plane, given by $\theta_{0}=\textrm{constant}$. In this case of motion confined to a particular plane, the rate of change of area will be given by (for a derivation, see \cite{MembraneParadigm}),
\begin{align}\label{area_change}
\frac{\kappa_{+1}}{8\pi}\frac{dA_{\rm BH1}}{dt}=\left(\Omega-\Omega_{\rm H1}\right)^{2}I(\theta_{0})~,
\end{align}
as well as, the rate of change of mass and the rate of change of angular momentum will have the following expressions,
\begin{align}\label{mass_momentum_change}
\frac{dM_{1}}{dt}=\Omega\left(\Omega-\Omega_{\rm H1}\right)I(\theta_{0})~;
\qquad
\frac{dJ_{1}}{dt}=\left(\Omega-\Omega_{\rm H1}\right)I(\theta_{0})~.
\end{align}
Note that in the stationary case, the angular velocity $\Omega=0$ and hence $(dM_{1}/dt)=0$, which is consistent with our result derived in the previous section. In this case as well, since the low-energy processes on the brane cannot possibly affect the physics of the extra-dimension, it follows that we can safely assume that $(dQ/dt)=0$. The above expressions for the rate of change of area, as well as the rate of change of mass and the rate of change of angular momentum must be consistent with \ref{derivative_relation}. This can be immediately checked by substituting the expressions for the rate of change of area from \ref{area_change} and the rate of change of mass and angular momentum from \ref{mass_momentum_change} in \ref{derivative_relation}. 

Having argued about the correctness of the results given in \ref{area_change} and \ref{mass_momentum_change}, let us now compute the integral $I(\theta_{0})$. In general, this integral $I(\theta_{0})$ is a complicated function of the black hole hairs, however for all the intent and purposes of this work we can approximate it to be the corresponding expression in the stationary situation, $I_{0}(\theta_{0})$. In order to obtain the stationary integral $I_{0}(\theta_{0})$, we substitute $\Omega=0$ in \ref{area_change} and use the expression for the rate of change of area in the stationary case from \ref{stat_area_change}, to obtain,
\begin{align}\label{integral_general}
I_{0}(\theta_{0})=\frac{1}{\Omega_{H1}^{2}}\frac{\kappa_{+}}{8\pi}\frac{dA_{\rm BH1}}{dt}=\frac{16M_{1}^{5}M_{2}^{2}}{5b^{6}}\sin^{2}\theta_{0}\left[\mathcal{A}+\mathcal{B}\sin^{2}\theta_{0}\right]\left(r_{+1}+\frac{Q}{2M_{1}}\right)~.
\end{align}
Again, note that in the $Q\rightarrow 0$ limit, we get back the corresponding expression for the Kerr black hole, coinciding with the one given in \cite{Alvi:2001mx}. So far, we have considered the general case of the motion of the binary system on a constant $\theta=\theta_{0}$ plane. However, our main interest is in the motion of the binary black holes confined to the equatorial plane (i.e., $\theta_{0}=(\pi/2)$). In this case, the above expression for the integral $I_{0}(\theta_{0}=\pi/2)$ takes the following form,
\begin{align}\label{integral}
I_{0}(\pi/2)=\frac{16M_{1}^{5}M_{2}^{2}}{5b^{6}}\left(r_{+}+\frac{Q}{2M_{1}}\right)\left[1+3\chi_{1}^{2}+\frac{Q}{M_{1}^{2}}\left(2+3\chi_{1}^{2}+\frac{Q}{M_{1}^{2}}\right) \right]~.
\end{align}
The next step is to compute expression for the quantity $(\Omega-\Omega_{\rm H1})$. First of all, the angular velocity of the BBH1, to the leading Newtonian order will be given by $\Omega=(\hat{\bf{L}}_{\rm orb}\cdot \hat{\bf{J}}_{1})\Omega_{\rm N}$, where $\Omega_{\rm N}=(v^{3}/M)$ is the common angular velocity of both the black holes, with $M=M_{1}+M_{2}$ and $\hat{\bf{L}}_{\rm orb}$ is the unit vector along the direction of the orbital angular momentum. In addition, we have $v=\sqrt{(M/b)}$, where $b$ is the relative separation between the two black holes in the binary. Finally, substituting the expression for $\Omega_{\rm H1}$, we obtain,
\begin{align}\label{ang_vel}
\Omega-\Omega_{\rm H1}=\frac{1}{2\left[r_{+1}+\frac{Q}{2M_{1}}\right]}\left\{-\chi_{1}+2\left(\hat{\bf{L}}_{\rm orb}\cdot \hat{\bf{J}}_{1}\right)\frac{M_{1}}{M} \left[1+\sqrt{1-\chi_{1}^{2}+\frac{Q}{M_{1}^{2}}}+\frac{Q}{2M_{1}^{2}}\right]v^{3}\right\}~.
\end{align}
We now have all the necessary ingredients required to compute the rate of change of angular momentum, mass and black hole area. The expression for the integral $I_{0}(\theta_{0})$ on the equatorial plane is given by \ref{integral} and the expression for $(\Omega-\Omega_{\rm H1})$ has been presented in \ref{ang_vel}. Thus the rate of change of angular momentum, mass and area of the BBH1 take the following form, 
\begin{align}
\frac{dJ_{1}}{dt}&=\left(\frac{dJ}{dt}\right)_{\rm N}\left(\frac{M_{1}}{M}\right)^{3}\frac{v^{5}}{4}\left\{-\chi_{1}+2\left(\hat{\bf{L}}_{\rm orb}\cdot \hat{\bf{J}}_{1}\right)\frac{v^{3}}{M} \left[r_{+1}+\frac{Q}{2M_{1}}\right]\right\}
\nonumber
\\
&\hskip 3 cm \times \left[1+3\chi_{1}^{2}+\frac{Q}{M_{1}^{2}}\left(2+3\chi_{1}^{2}+\frac{Q}{M_{1}^{2}}\right) \right]~;
\quad
\left(\frac{dJ}{dt}\right)_{\rm N}\equiv \frac{32}{5}\eta^{2}Mv^{7}~,
\\
\frac{dM_{1}}{dt}&=\left(\frac{dE}{dt}\right)_{\rm N}\left(\frac{M_{1}}{M}\right)^{3}\frac{v^{5}}{4}\left\{-\chi_{1}\left(\hat{\bf{L}}_{\rm orb}\cdot \hat{\bf{J}}_{1}\right)+2\frac{v^{3}}{M} \left[r_{+1}+\frac{Q}{2M_{1}}\right]\right\}
\nonumber
\\
&\hskip 3 cm \times \left[1+3\chi_{1}^{2}+\frac{Q}{M_{1}^{2}}\left(2+3\chi_{1}^{2}+\frac{Q}{M_{1}^{2}}\right) \right]~;
\quad
\left(\frac{dE}{dt}\right)_{\rm N}\equiv \frac{v^{3}}{M}\left(\frac{dJ}{dt}\right)_{\rm N}~,
\label{mass_rate}
\\
\frac{1}{M}\left(\frac{dA_{\rm BH1}}{dt}\right)&=\frac{64\pi M_{1}^{3}M^{3}\eta^{2}}{5b^{6}}\frac{1}{\sqrt{1-\chi_{1}^{2}+\frac{Q}{M_{1}^{2}}}}\left\{-\chi_{1}+2\left(\hat{\bf{L}}_{\rm orb}\cdot \hat{\bf{J}}_{1}\right)\frac{v^{3}}{M} \left[r_{+1}+\frac{Q}{2M_{1}}\right]\right\}^{2}
\nonumber
\\
&\hskip 2 cm \times \left[1+3\chi_{1}^{2}+\frac{Q}{M_{1}^{2}}\left(2+3\chi_{1}^{2}+\frac{Q}{M_{1}^{2}}\right) \right]\frac{1+\sqrt{1-\chi_{1}^{2}+\frac{Q}{M_{1}^{2}}}}{\left[1+\sqrt{1-\chi_{1}^{2}+\frac{Q}{M_{1}^{2}}}+\frac{Q}{2M_{1}^{2}}\right]}~,
\end{align}
where, $\eta \equiv M_{1}M_{2}/M^{2}$ and we have used the result, $(\hat{\bf{L}}_{\rm orb}\cdot \hat{\bf{J}}_{1})^{2}=1$. It is to be noted that the above set of equations, describing the rate of change of mass, angular momentum and area, has been expressed solely in terms of the dimensionless combinations $\chi\equiv (a/M)$ and $q\equiv (Q/M^{2})$, respectively. This is the trend we will follow in the subsequent sections as well, where all the physical quantities of interest will be expressed in terms of these dimensionless black hole hairs. Furthermore, the rate of change of mass and the rate of change of angular momentum are related to each other, such that, 
\begin{align}
\frac{(dJ_{1}/dt)}{(dJ/dt)_{\rm N}}=\frac{(dM_{1}/dt)}{(dE/dt)_{\rm N}}\left(\hat{\bf{L}}_{\rm orb}\cdot \hat{\bf{J}}_{1}\right)~.
\end{align}
Thus, for aligned braneworld black hole, i.e., with $\hat{\bf{L}}_{\rm orb}\cdot \hat{\bf{J}}_{1}=+1$, the dimensionless rate of change of angular momentum coincides with the dimensionless rate of change of mass. While for the case of anti-aligned braneworld black hole, with $\hat{\bf{L}}_{\rm orb}\cdot \hat{\bf{J}}_{1}=-1$, the dimensionless rate of change of angular momentum coincides with the negative of the dimensionless rate of change of mass. The above expressions are for the BBH1, due to the tidal field of BBH2. We can easily obtain the respective expressions for BBH2, due to the tidal field of BBH1, by simply exchanging $1\leftrightarrow 2$ in the subscripts of the above expressions.  

\begin{figure}
\includegraphics[scale=0.41]{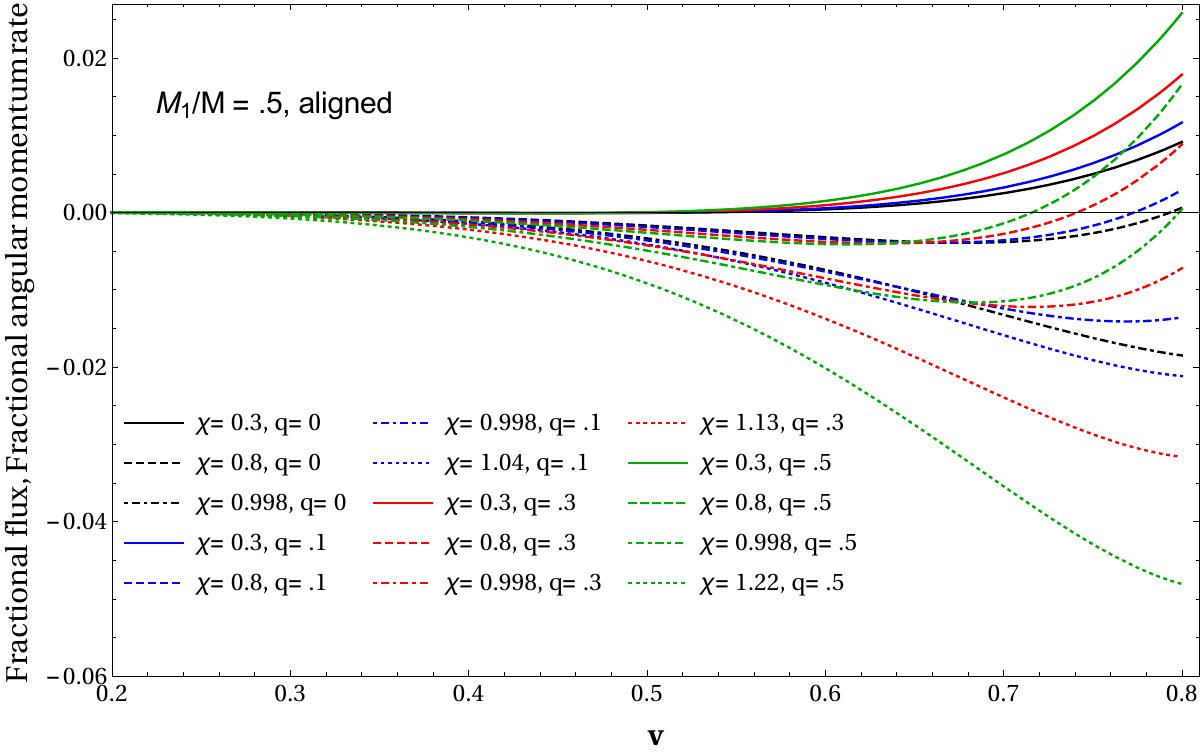}
\hskip .5 cm
\includegraphics[scale=0.41]{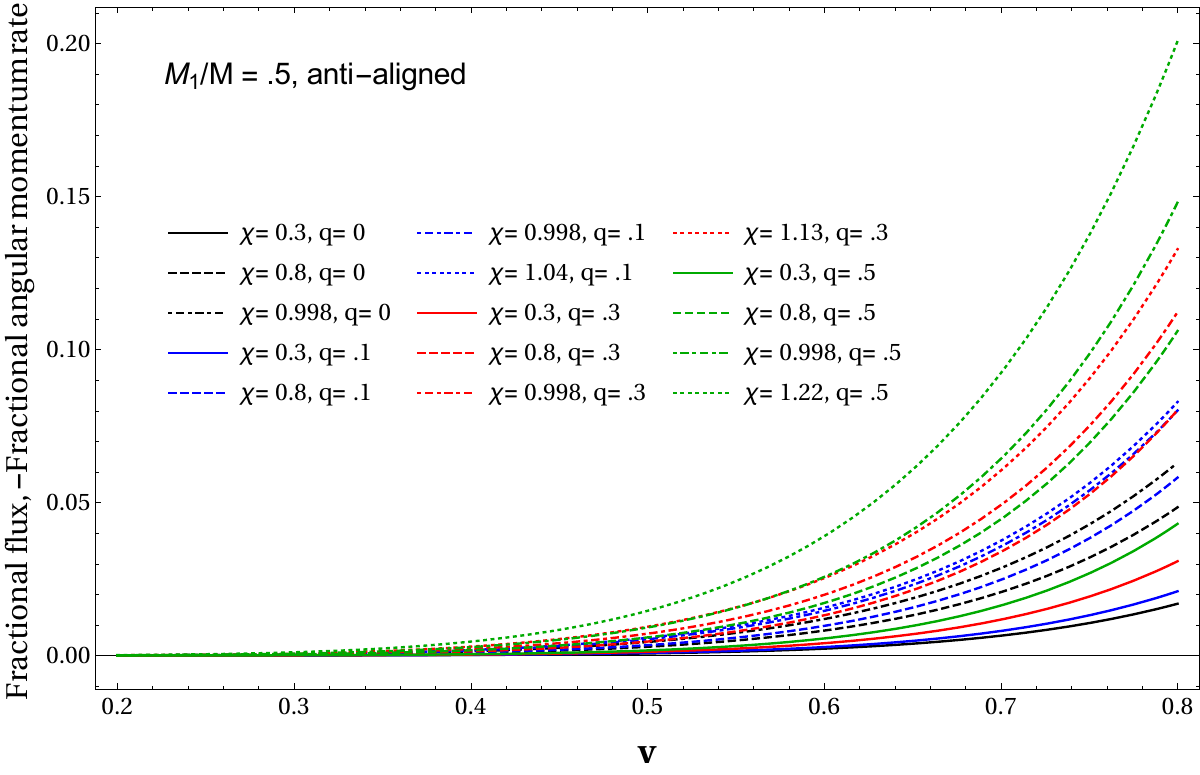}
\\
\includegraphics[scale=0.41]{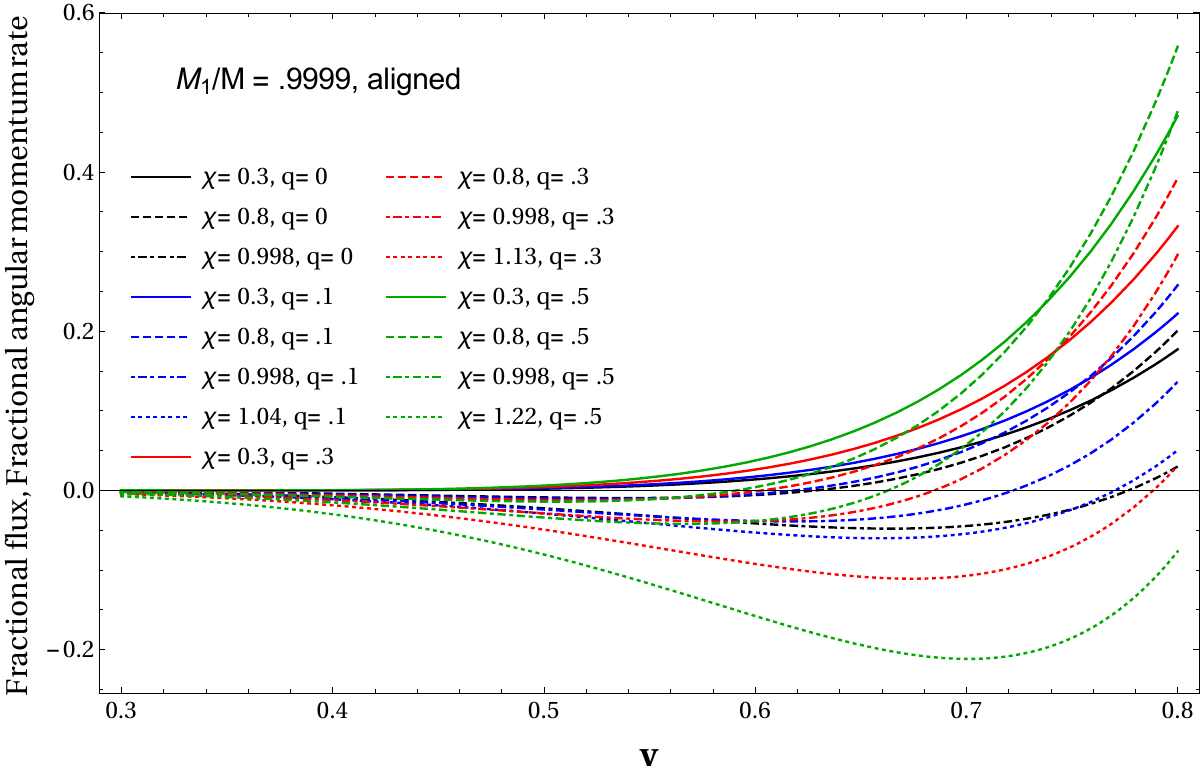}
\hskip .5    cm
\includegraphics[scale=0.41]{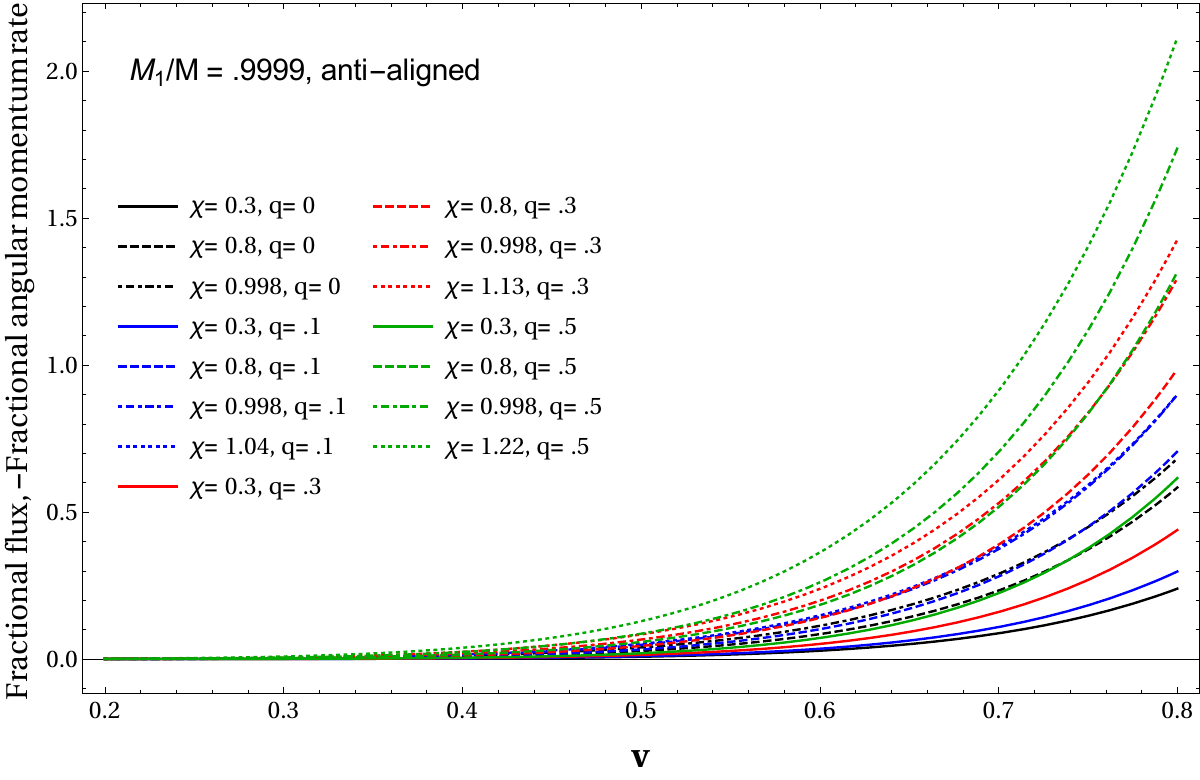}
\caption{The dimensionless rate of change of angular momentum $(dJ_{1}/dt)(dJ/dt)_{\rm N}^{-1}$ of the BBH1 has been plotted against the relative velocity $v$ between the orbitting black holes in the binary for both aligned and anti-aligned cases, i.e., with $\hat{\bf{L}}_{\rm orb}\cdot \hat{\bf{J}}_{1}=\pm1$. The plots are for various choices of the dimensionless spin parameter $\chi=(a/M)$ and the dimensionless tidal charge parameter $q=(Q/M^{2})$.  The plots on the upper row are for the case of equal mass ratio black holes, with $(M_{1}/M)=0.5$, while the plots on the bottom row are for the extreme mass ratio, with $(M_{1}/M)=0.9999$. All the plots are on the equatorial plane. Note that for larger values of $q$, the spin parameter $\chi$ can take values larger than unity and the rate of change of angular momentum, as well as the rate of change of mass differs significantly from the general relativistic expression. The deviation increases as the relative velocity between the two black holes become larger.}
\label{angular_mass_rate_equatorial}
\end{figure}

The above expressions for the rate of change of angular momentum and mass, normalized to their Newtonian values, have been plotted in \ref{angular_mass_rate_equatorial}\footnote{For all the figures except \ref{fig:contour} and \ref{lnepsilon} we have maintained the following convention. The colour of a curve represents the value of $q$ (defined as the dimensionless combination $Q/M^{2}$) and the style of the line plot represents the $\chi$ (defined as $a/M$) value. Black, blue, red, and green colour represent the $q$ value 0, 0.1, 0.3, and 0.5 respectively. Solid, dashed, dot-dashed, and dotted curves represent $\chi$ values equal to 0.3, 0.8, 0.998 and over-spinning, respectively.}. As evident from the previous discussion, the dimensionless rate of change of angular momentum coincides with the dimensionless rate of change of mass, for aligned spin and orbital angular momentum, while these two are negative of one another, in the case of anti-aligned spin and orbital angular momentum. For equal mass ratio binary black hole, the dimensionless rate of change of angular momentum decreases, while the dimensionless rate of change of mass increases, for anti-aligned spin and orbital angular momentum (see the plot on the top right corner of \ref{angular_mass_rate_equatorial}). On the other hand, for aligned spin and orbital angular momentum, the dimensionless rate of change of mass and angular momentum increases with an increase in the relative velocity between the black holes in the binary for small spin parameter. However for larger spin parameter both the dimensionless rate of change of mass and angular momentum decreases with an increase in the relative velocity between the black holes in the binary (see the plot on the top left corner of \ref{angular_mass_rate_equatorial}). 

For extreme mass ratio in-spiral, both for the aligned as well as anti-aligned spin and orbital angular momentum, the dimensionless rate of change of mass and angular momentum ultimately increases with an increase in the relative velocity between the black holes in the binary. Even though for aligned spin and orbital angular momentum, the dimensionless rate of change of mass and angular momentum decreases for smaller relative velocity between the black holes in the binary (see the plots on the lower row of \ref{angular_mass_rate_equatorial}). As we can see from \ref{angular_mass_rate_equatorial}, presence of tidal charge affects the rate of change of mass and angular momentum significantly, since it differs from general relativity values by an appreciable amount.

This finishes our discussion regarding the rate of change of various geometrical quantities, in particular the hairs of a braneworld black hole due to tidal effects of its companion in a binary system on the equatorial plane. We will now integrate these rates and hence obtain the total change of these geometrical quantities in the next section.

\subsection{Total changes of geometrical quantities during in-spiral on equatorial plane}

The above expressions are for the rate of change of the quantities of interest for black holes. Given these, we will try to determine the total change of these quantities during the in-spiral of black holes towards each other in a binary system. To the leading Newtonian order, the orbital separation $b$ among the binary black holes evolve during the in-spiral in the following manner \cite{Alvi:2001mx}: $b(t)=b_{0}[1-(t/\tau_{0})]^{1/4}$, where the constant $\tau_{0}$ takes the following form, $\tau_{0}\equiv (5/256)b_{0}^{4}(\eta M^{3})^{-1}$. Here $b_{0}$ is the initial separation between the black holes. Following which we can parametrize the orbit by the separation $b$, rather than the time coordinate $t$ and hence compute the total change in mass, angular momentum and area as the binary system spirals down from the initial separation $b_{0}$ to a final separation $b$. Integrating the rate of change of the respective quantities during the total phase of in-spiral of the binary system from infinity, i.e., with $b_{0}=\infty$, we obtain the following changes,
\begin{align}
\frac{\Delta J_{1}}{M_{1}^{2}}&=\frac{\eta M_{1}}{4M}\left\{-\frac{\chi_{1}}{4}\left(\frac{M}{b}\right)^{2}+\frac{2}{7}\left(\hat{\bf{L}}_{\rm orb}\cdot \hat{\bf{J}}_{1}\right)\frac{M_{1}}{M} \left[1+\sqrt{1-\chi_{1}^{2}+\frac{Q}{M_{1}^{2}}}+\frac{Q}{2M_{1}^{2}}\right]\left(\frac{M}{b}\right)^{7/2}\right\}
\nonumber
\\
&\hskip 3 cm \times \left[1+3\chi_{1}^{2}+\frac{Q}{M_{1}^{2}}\left(2+3\chi_{1}^{2}+\frac{Q}{M_{1}^{2}}\right) \right]~,
\\
\frac{\Delta M_{1}}{M_{1}}&=\frac{\eta M_{1}^{2}}{4M^{2}}\left\{-\frac{\chi_{1}}{7}\left(\hat{\bf{L}}_{\rm orb}\cdot \hat{\bf{J}}_{1}\right)\left(\frac{M}{b}\right)^{7/2}+\frac{1}{5}\left(\hat{\bf{L}}_{\rm orb}\cdot \hat{\bf{J}}_{1}\right)\frac{M_{1}}{M} \left[1+\sqrt{1-\chi_{1}^{2}+\frac{Q}{M_{1}^{2}}}+\frac{Q}{2M_{1}^{2}}\right]\left(\frac{M}{b}\right)^{5}\right\}
\nonumber
\\
&\hskip 3 cm \times \left[1+3\chi_{1}^{2}+\frac{Q}{M_{1}^{2}}\left(2+3\chi_{1}^{2}+\frac{Q}{M_{1}^{2}}\right) \right]~,
\\
\frac{\Delta A_{\rm BH1}}{A_{\rm BH1}}&=\frac{\eta M_{1}^{2}}{8M}\left(\frac{1}{r_{+1}+\frac{Q}{2M_{1}}}\right)\frac{1}{\sqrt{1-\chi_{1}^{2}+\frac{Q}{M_{1}^{2}}}}\left(\frac{1+\sqrt{1-\chi_{1}^{2}+\frac{Q}{M_{1}^{2}}}}{1+\sqrt{1-\chi_{1}^{2}+\frac{Q}{M_{1}^{2}}}+\frac{Q}{2M_{1}^{2}}}\right)\Bigg\{\frac{\chi_{1}^{2}}{2}\left(\frac{M}{b}\right)^{2}
\nonumber
\\
&\hskip 2 cm -\frac{8\chi_{1}}{7}\left(\frac{M}{b}\right)^{7/2}\left(\hat{\bf{L}}_{\rm orb}\cdot \hat{\bf{J}}_{1}\right)\frac{1}{M}\left[r_{\rm +1}+\frac{Q}{2M_{1}}\right]+\frac{4}{5}\left(\frac{M}{b}\right)^{5}\frac{1}{M^{2}} \left[r_{+1}+\frac{Q}{2M_{1}}\right]^{2}\Bigg\}
\nonumber
\\
&\hskip 3 cm \times \left[1+3\chi_{1}^{2}+\frac{Q}{M_{1}^{2}}\left(2+3\chi_{1}^{2}+\frac{Q}{M_{1}^{2}}\right) \right]~.
\end{align}
In the derivation of the above expressions, we have used the result $(db/dt)=-(b_{0}^{4}/4\tau_{0})b^{-3}$, which follows from the evolution of the separation $b(t)$ between the black holes in the binary, as mentioned above. Note that the above derivation requires the final separation $b_{\rm final}$ between the binary black holes to satisfy the condition, $b_{\rm final}\gg 2M_{1,2}$. This is because the above formalism, based on the linear gravitational perturbation, becomes invalid as the separation between the black holes becomes comparable to their horizon radii. As in the previous section, here also the corresponding change for the BBH2 due to the tidal field of BBH1 can be obtained by simply exchanging $1\leftrightarrow 2$ in the above expressions.    

\begin{figure}
\includegraphics[scale=0.41]{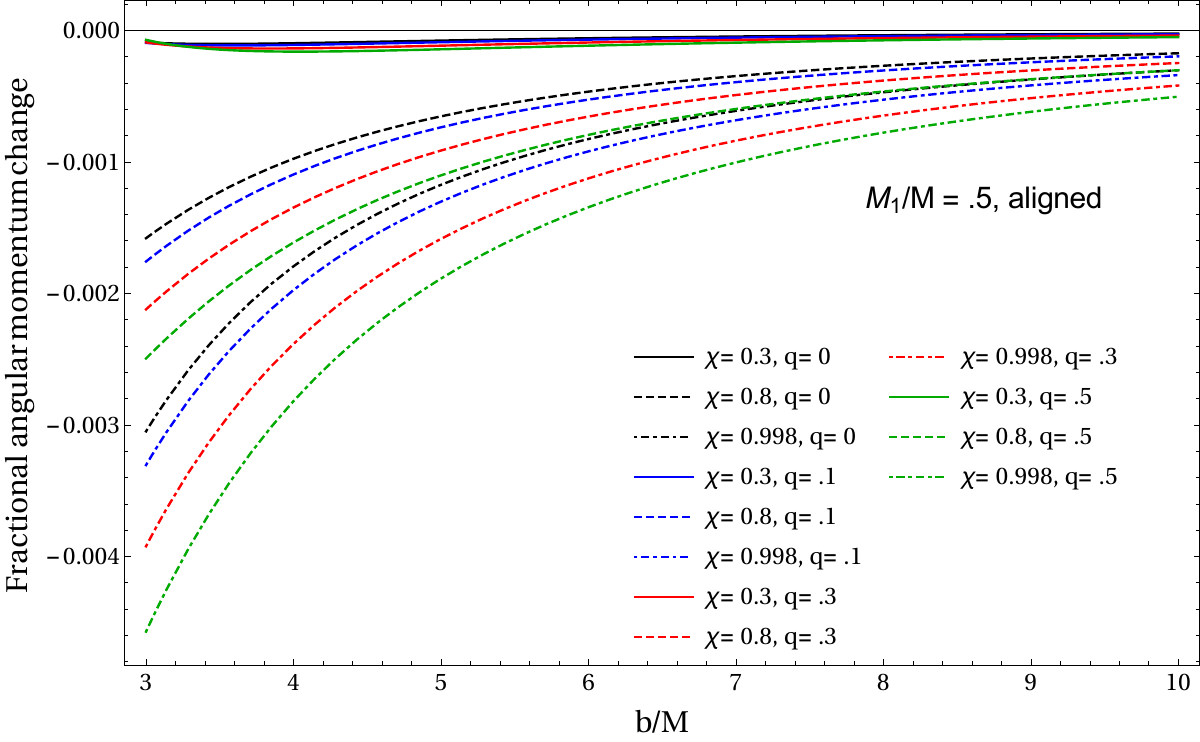}
\hskip .5 cm
\includegraphics[scale=0.41]{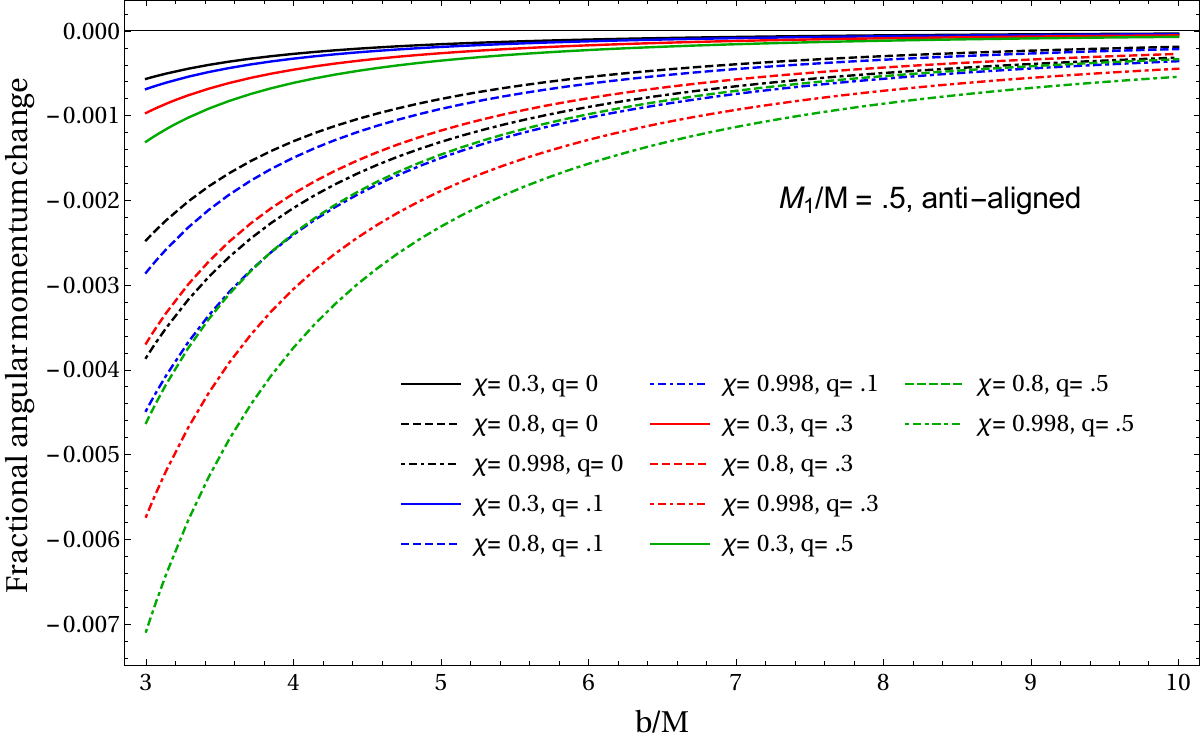}
\\
\includegraphics[scale=0.41]{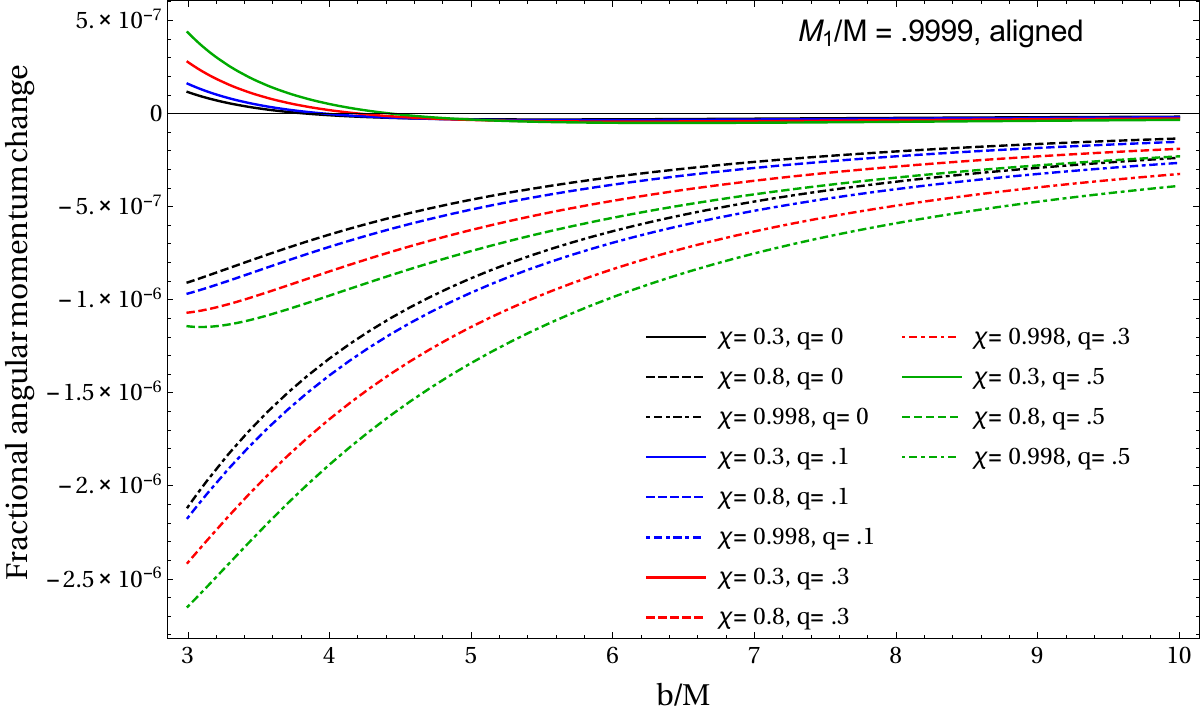}
\hskip .5 cm
\includegraphics[scale=0.41]{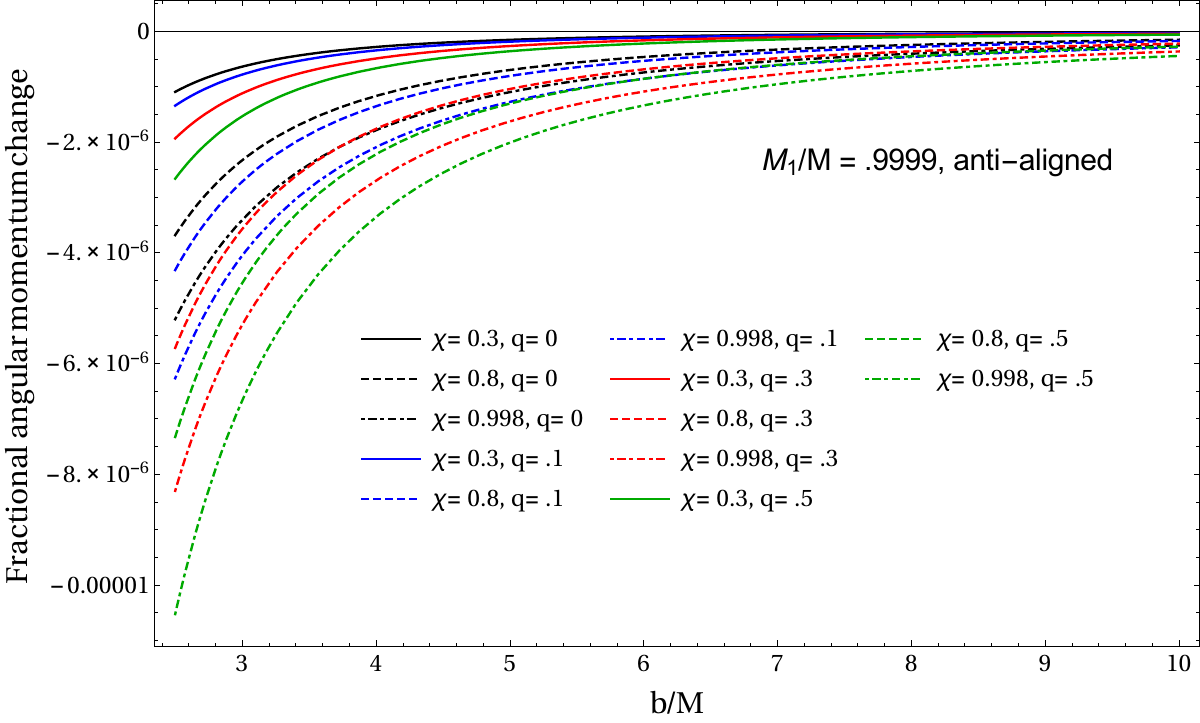}
\caption{The total change in the angular momentum (normalized by the mass squared of the BBH1), during the in-spiral of the black hole from a large distance to a finite separation $b$, has been plotted against the dimensionless ratio $(b/M)$ for different mass ratios as well as for various choices of the spin parameter $\chi$ and tidal charge parameter $q$. For binaries with aligned spin, the angular momentum decreases for large values of $\chi$, but it increases for smaller values (see figures on the left hand side), as the separation $(b/M)$ goes to smaller and smaller values. While for binaries with anti-aligned spin the angular momentum always decreases irrespective of the choices of $\chi$ and $q$. Larger values of $q$ results into larger change in the angular momentum compared to the corresponding situation in general relativity. See text for more discussions.}
\label{angular_total}
\end{figure}

Given the above expression for the total change in the angular momentum, which depends on the separation $b$ between the black holes, the mass ratio $M_{1}/M$, the spin parameter $\chi$ and the tidal charge $q$ (all in dimensionless units), we have plotted the same in \ref{angular_total}. The plots are for $(\Delta J_{1}/M_{1}^{2})$, against the dimensionless separation $(b/M)$ between the black holes in the binary, for various choices of $(M_{1}/M)$, $\chi$ and $q$. For black holes in the binary system with aligned spin, the total change in the angular momentum for $(M_{1}/M)=0.5$ is larger than the total change for extreme mass-ratio by a factor of $10^{3}$. Also in comparison to the general relativistic prediction, presence of the tidal charge modifies the change in the angular momentum by $25\%$, for extreme mass-ratio, while for $(M_{1}/M)=0.5$, the change in angular momentum is modified by $33\%$ (see the left hand plots in \ref{angular_total}). On the other hand, for anti-aligned spin, the angular momentum always decreases, irrespective of the mass ratio. In this case the presence of the tidal charge modifies the general relativistic prediction by $75\%$ for mass ratio $(M_{1}/M)=0.5$, while for extreme mass ratio the general relativistic prediction is modified by $33\%$. Moreover, for anti-aligned spin the angular momentum decreases as the black holes come closer, irrespective of the other parameters in the problem (see the right hand plots in \ref{angular_total}). While for aligned spin, the change in the angular momentum can be positive for smaller rotation parameter, as the left hand plots in \ref{angular_total} demonstrates.    
 
\begin{figure}
\includegraphics[scale=0.41]{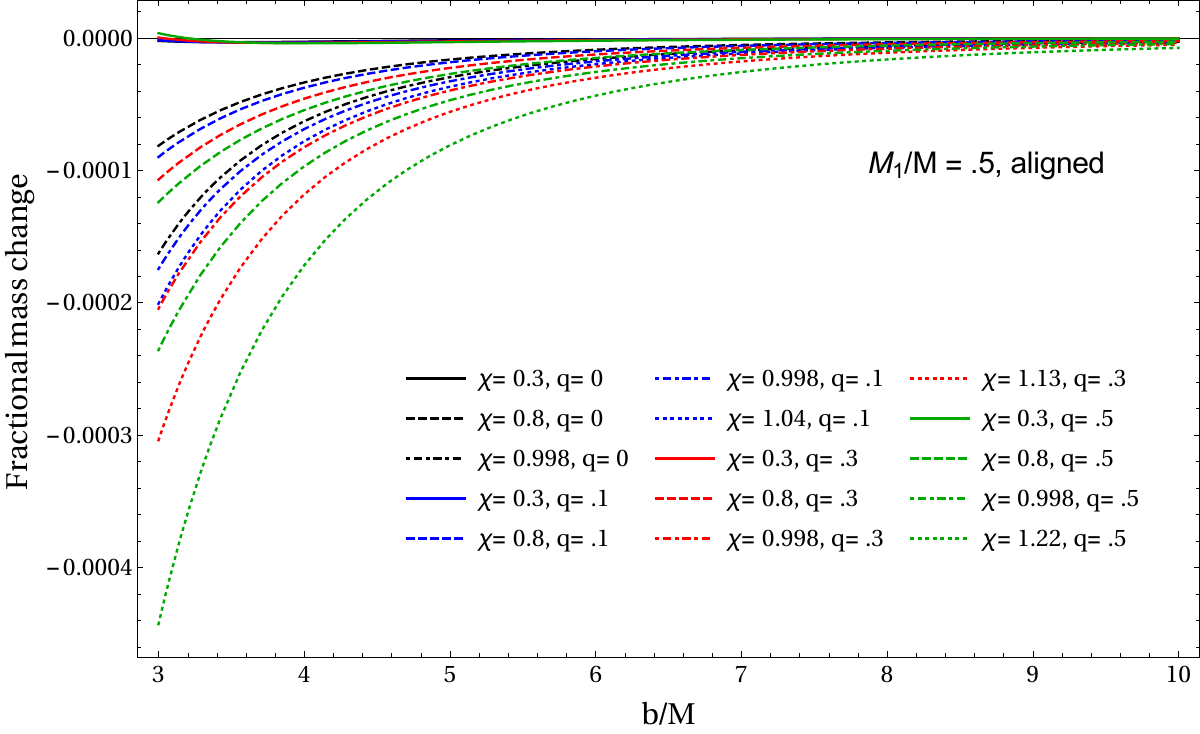}
\hskip .5 cm
\includegraphics[scale=0.41]{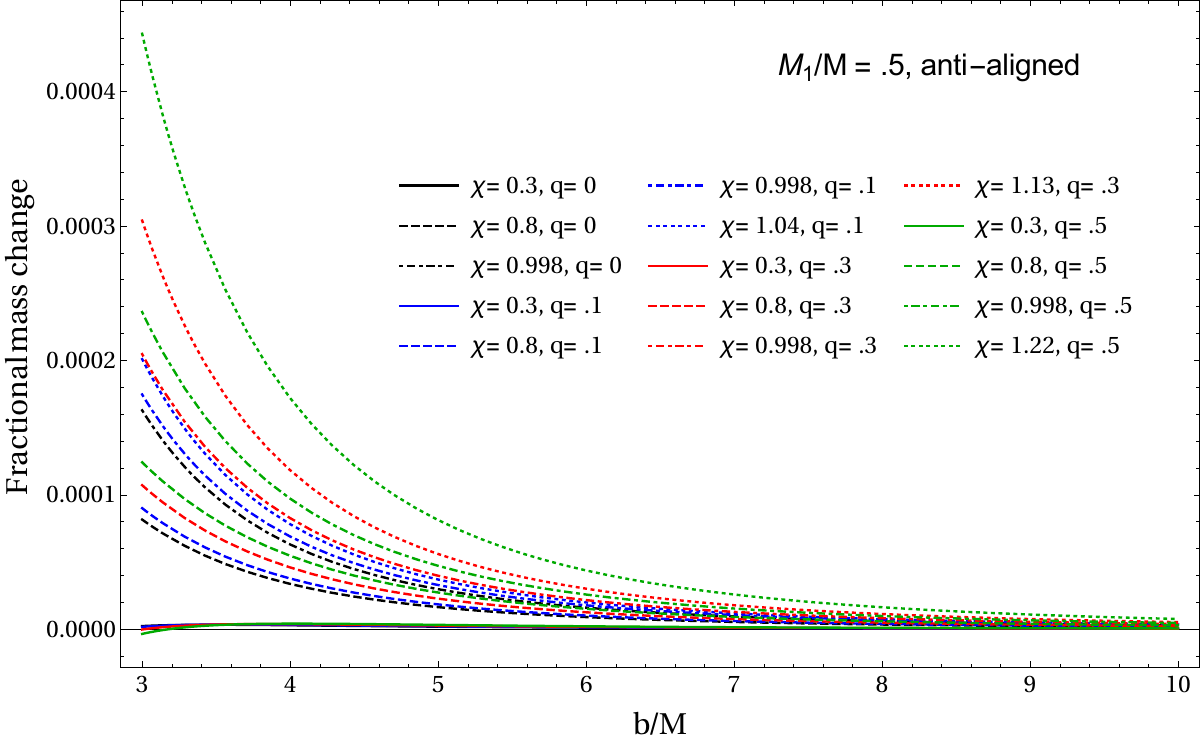}
\\
\includegraphics[scale=0.41]{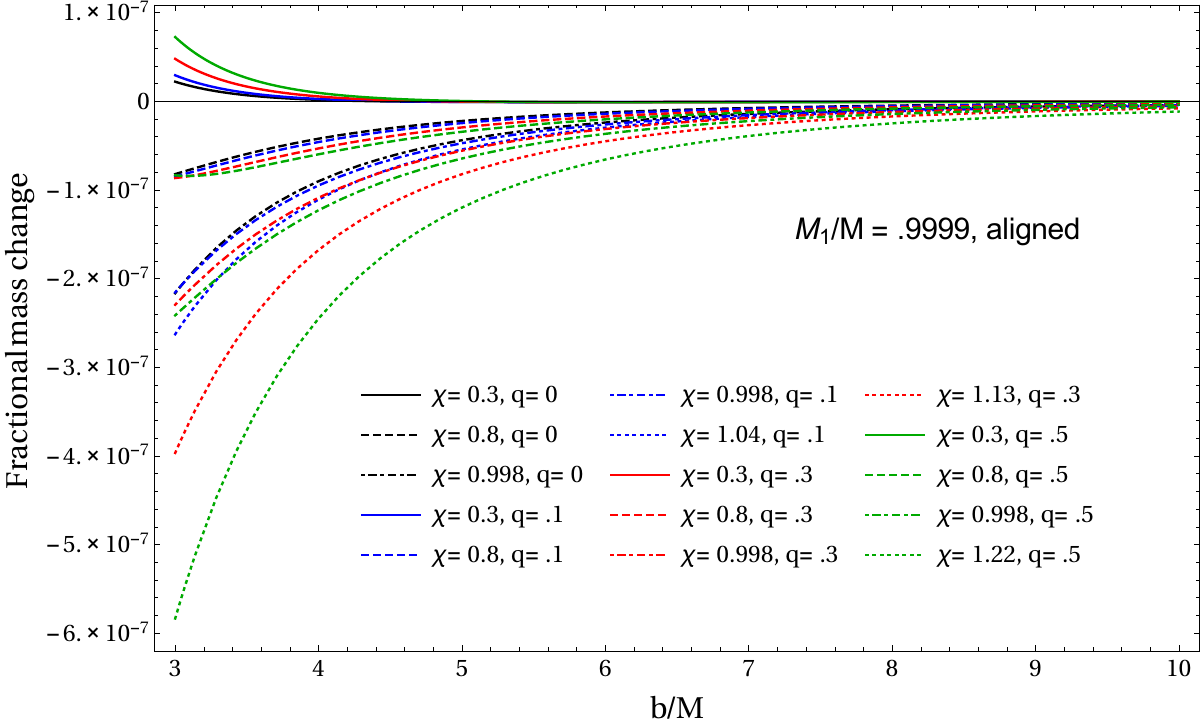}
\hskip .5 cm
\includegraphics[scale=0.41]{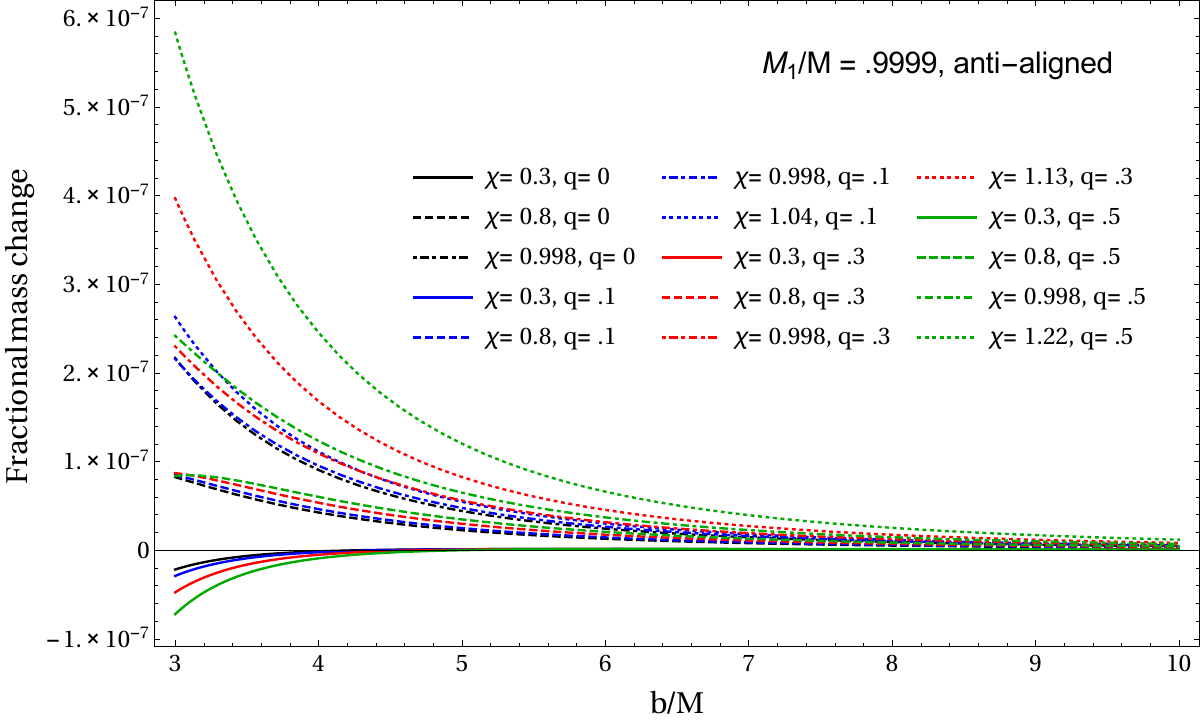}
\caption{The total change in the mass (normalized by the mass of the BBH1), during the in-spiral of the black hole from a large distance to a finite separation $b$, has been plotted for different mass ratios as well as for various choices of the spin parameter $\chi$ and tidal charge parameter $q$ against the dimensionless ratio $(b/M)$. For binaries with aligned spin, the mass decreases for large values of $\chi$, but it increases for smaller values (see figures on the left hand side), as the black holes in the binary approach closer to each other. While for binaries with anti-aligned spin, the change of mass depicts an exactly opposite behaviour. As in the earlier scenarios, larger values of $q$ results into larger change in the mass compared to the corresponding situation in general relativity.}
\label{mass_total}
\end{figure}

The fractional change of mass has been depicted in \ref{mass_total}, for both equal and extreme mass ratio black holes, as well as for black holes with both aligned and anti-aligned spins. In contrast to the change in the angular momentum, the change in mass can be either positive or negative as the black holes in the binary come close to each other, irrespective of their spins being aligned/anti-aligned (see \ref{mass_total}). However, for aligned spin black holes, the change in the mass is mostly negative, as the spin parameter increases. While an exactly opposite scenario is observed for black holes with anti-aligned spin. Again the change in the mass of the braneworld black hole, due to tidal effects from the companion is larger by a factor $10^{3}$ for equal-mass black holes, in comparison to the extreme mass ratio black holes. Additionally, the maximal change in the mass, for equal mass ratio braneworld black hole binaries, between $q=0$ and $q=0.5$ is $125\%$. While for extreme mass ratio in-spiral the maximal change in the mass is given by $170\%$. Similar numbers hold true for braneworld black holes with anti-aligned spin as well.   

\subsection{Non-equatorial orbits and average rates}

We have discussed the rate of change of mass, angular momentum and area in the case of stationary companion as well as when the binary system is moving on the equatorial plane. Using the formula for the rate of change of mass, angular momentum and area, we have derived the total change in the respective quantities as the two black holes in-spiral down from their initial separation to a final separation, which is larger than the horizon radii of the individual black holes. 

However, due to the spin-orbit and spin-spin coupling, generically the companion of a black hole in the binary system is not confined to the equatorial plane and hence it is necessary to consider the motion of the binary system \emph{off} the equatorial plane. The spin and the orbital precession frequencies being much smaller than the Newtonian angular velocity $\Omega_{\rm N}$, we can approximate the orbit to be confined to a single plane for a few orbital periods, with the normal to the plane given by the instantaneous direction of the orbital angular momentum $\hat{\bf{L}}_{\rm orb}(t)$. 

For the binary black hole system under consideration we can use the Boyer-Lindquist coordinates, denoted by $(t,r,\theta,\phi)$, in the LARF associated with the BBH1. For a few orbital periods the radial separation $b$ can be taken to be a constant, while the angular coordinates of the BBH2 will change over time. We denote the initial direction of the BBH2, with the BBH1 at the origin, by $\vec{\bf{X}}_{0}$ and the BBH2 is moving in a plane with unit normal vector $\hat{\bf{L}}_{\rm orb}(t)$ with an orbital angular velocity $\Omega_{\rm N}$. The motion of the BBH2 is in the plane spanned by the vectors $\vec{\bf{X}}_{0}$ and $\hat{\bf{L}}_{\rm orb}(t)\times \vec{\bf{X}}_{0}$, such that the position of the BBH2 at a later instant is given by, $\vec{\bf{X}}=\vec{\bf{X}}_{0} \cos(\Omega_{\rm N}t)+(\hat{\bf{L}}_{\rm orb}(t)\times \vec{\bf{X}}_{0})\sin(\Omega_{\rm N}t)$. Equating the position $\vec{\bf{X}}(t)$ of the BBH2 at a later instant by $b[\sin \theta(t)\cos\phi(t),\sin\theta(t)\sin\phi(t),\cos\theta(t)]$, along with $\vec{\bf{X}}_{0}$ to be a vector on the equatorial plane, we obtain, 
\begin{align}\label{angle_time}
\sin^{2}\theta(t)=1-\left(1-\mathcal{N}_{1}^{2}\right)\sin^{2}(\Omega_{\rm N}t)~; 
\quad
\dot{\phi}(t)=\frac{\Omega_{\rm N}\mathcal{N}_{1}}{\sin^{2}\theta(t)}~;
\quad
\mathcal{N}_{1}\equiv \hat{\bf{L}}_{\rm orb}(t)\cdot \hat{\bf J}_{1}(t)~.
\end{align}
In what follows, we require $\sin^{2}\theta(t)$ and $\dot{\phi}(t)$ to be slowly varying functions of time, such that over the timescale in which the horizon is affected, these quantities remain almost constant. This requires the orbital plane of the binary to be located near the equatorial plane. Furthermore, we also require, $|\dot{\theta}|\ll |\dot{\phi}-\Omega_{\rm H1}|$. With these approximations, we can simply replace the angular velocity $\Omega$ by the instantaneous rate $\dot{\phi}$ and hence the rate of change of angular momentum, mass and area of the BBH1 is given by,
\begin{align}
\frac{dJ_{1}}{dt}&=\left(\dot{\phi}(t)-\Omega_{\rm H1}\right)I_{0}[\theta(t)]~;
\quad
\frac{dM_{1}}{dt}=\dot{\phi}(t)\left(\dot{\phi}(t)-\Omega_{\rm H1}\right)I_{0}[\theta(t)]~,
\label{nonequa1}
\\
\frac{\kappa_{+1}}{8\pi}\frac{dA_{1}}{dt}&=\left(\dot{\phi}(t)-\Omega_{\rm H1}\right)^{2}I_{0}[\theta(t)]~.
\label{nonequa2}
\end{align}
Here, $\theta(t)$ and $\dot{\phi}(t)$ are given by \ref{angle_time} and the integral $I_{0}$ is given by \ref{integral_general}. In the present context, rather than the rate expressions presented above, the orbit-averaged quantities are of primary interest. This is obtained by integrating the rate over a complete orbital period with the properties of the binary held constant. The average of some quantity $\mathcal{Q}$ will be denoted by $\langle \mathcal{Q}\rangle$, which is given by the following integral, 
\begin{align}\label{avg_int}
\langle \mathcal{Q}\rangle=\frac{\Omega_{\rm N}}{2\pi}\int_{0}^{2\pi/\Omega_{\rm N}}dt~\mathcal{Q}~.
\end{align}
Thus the averaged rate of change of mass, angular momentum and area of the BBH1, due to the tidal field of the BBH2, moving in a non-equatorial plane  takes the following dimensionless forms,
\begin{align}
\left\langle\frac{dJ_{1}}{dt}\right\rangle&=\left(\frac{dJ}{dt}\right)_{\rm N}\left(\frac{M_{1}^{3}}{64b^{2}\sqrt{bm}}\right)\left(r_{+1}+\frac{Q}{2M_{1}}\right)
\nonumber
\\
&\hskip 0.5 cm \times \Bigg\{16(\mathcal{A}+\mathcal{B})\left[2\Omega_{\rm N}\mathcal{N}_{1}-\Omega_{\rm H1}(1+\mathcal{N}_{1}^{2})\right]+4\mathcal{B}(\mathcal{N}_{1}^{2}-1)\left[4\mathcal{N}_{1}\Omega_{\rm N}-\Omega_{\rm H1}(3\mathcal{N}_{1}^{2}+1)\right]\Bigg\}~,
\\
\left\langle\frac{dM_{1}}{dt}\right\rangle&=\left(\frac{dE}{dt}\right)_{\rm N}\left(\frac{M_{1}^{3}}{16Mb}\right)\left(r_{+1}+\frac{Q}{2M_{1}}\right)\Omega_{\rm N}\mathcal{N}_{1}
\nonumber
\\
&\hskip 0.5 cm \times \Bigg\{-8\Omega_{\rm H1}(\mathcal{A}+\mathcal{B})+4\mathcal{B}\Omega_{\rm H1}\left(1-\mathcal{N}_{1}^{2}\right)+8\mathcal{B}\Omega_{\rm N}\mathcal{N}_{1}+8\mathcal{A}\Omega_{\rm N}\textrm{sign}(\mathcal{N}_{1})\Bigg\}~,
\\
\frac{1}{M}\left\langle \frac{dA_{1}}{dt}\right\rangle&=\left(\frac{8\pi r_{+1} M_{1}^{3}M^{3}\eta^{2}}{5b^{6}\sqrt{1-\chi_{1}^{2}+\frac{Q}{M_{1}^{2}}}}\right)\left(r_{+1}+\frac{Q}{2M_{1}}\right)
\nonumber
\\
&\hskip 1 cm \times \Bigg\{16\mathcal{A}\left[\Omega_{\rm H1}^{2}\left(1+\mathcal{N}_{1}^{2}\right)-4\Omega_{\rm N}\mathcal{N}_{1}\Omega_{\rm H1}+2\Omega_{\rm N}^{2}\mathcal{N}_{1}\textrm{sign}(\mathcal{N}_{1})\right]
\nonumber
\\
&\hskip 1.5 cm+4\mathcal{B}\left[8\Omega_{\rm N}^{2}\mathcal{N}_{1}^{2}-8\Omega_{\rm N}\mathcal{N}_{1}\Omega_{\rm H1}\left(1+\mathcal{N}_{1}^{2}\right)+\Omega_{\rm H1}^{2}\left(3\mathcal{N}_{1}^{4}+2\mathcal{N}_{1}^{2}+3\right)\right]\Bigg\}~.
\end{align}
The above expressions are for the BBH1, due to the tidal field of the BBH2. While the corresponding expressions for the BBH2 can be obtained by the exchange $1\leftrightarrow 2$. Furthermore, in the derivation of the above results we have used the following useful identitites, which we have listed here for ready reference,
\begin{align}\label{int_relation}
\int_{0}^{2\pi/\Omega_{\rm N}}\frac{dt}{\sin^{2}\theta(t)}&=\frac{2\pi}{\Omega_{\rm N}}\frac{(-1)^{\textrm{Floor}(\frac{1}{2}-\frac{\textrm{Arg}\mathcal{N}_{1}}{\pi})}}{\mathcal{N}_{1}}=\frac{2\pi}{\Omega_{\rm N}}\frac{\textrm{sign}(\mathcal{N}_{1})}{\mathcal{N}_{1}}~,
\nonumber
\\
\int_{0}^{2\pi/\Omega_{\rm N}}dt~\sin^{2}\theta(t)&=\frac{\pi}{\Omega_{\rm N}}\left(1+\mathcal{N}_{1}^{2}\right)~,
\nonumber
\\
\int_{0}^{2\pi/\Omega_{\rm N}}dt~\sin^{4}\theta(t)&=\frac{2\pi}{\Omega_{\rm N}}\left[\mathcal{N}_{1}^{2}+\frac{3}{8}\left(1-\mathcal{N}_{1}^{2}\right)^{2}\right]~.
\end{align}
Thus we have explicitly determined the averaged rates for mass, angular momentum and area of a braneworld black hole due to the tidal field of a companion, moving on a non-equatorial plane. 

\begin{figure}
\includegraphics[scale=0.41]{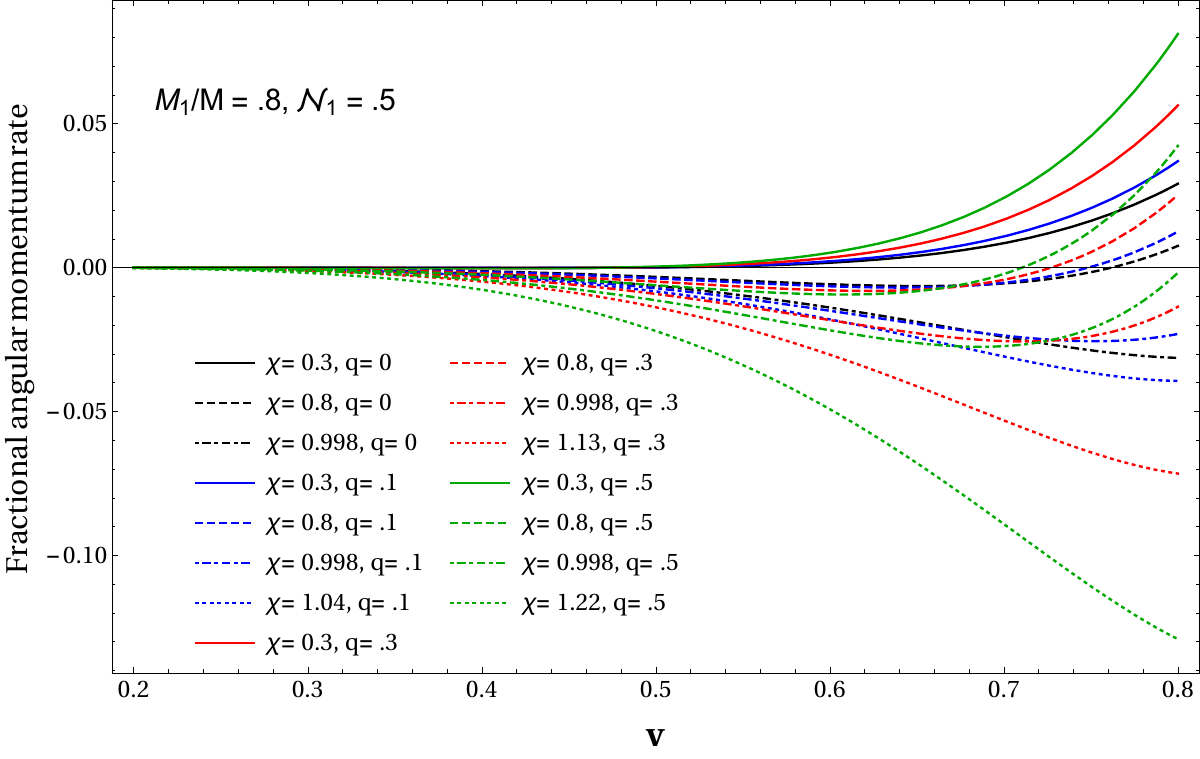}
\hskip .5 cm
\includegraphics[scale=0.41]{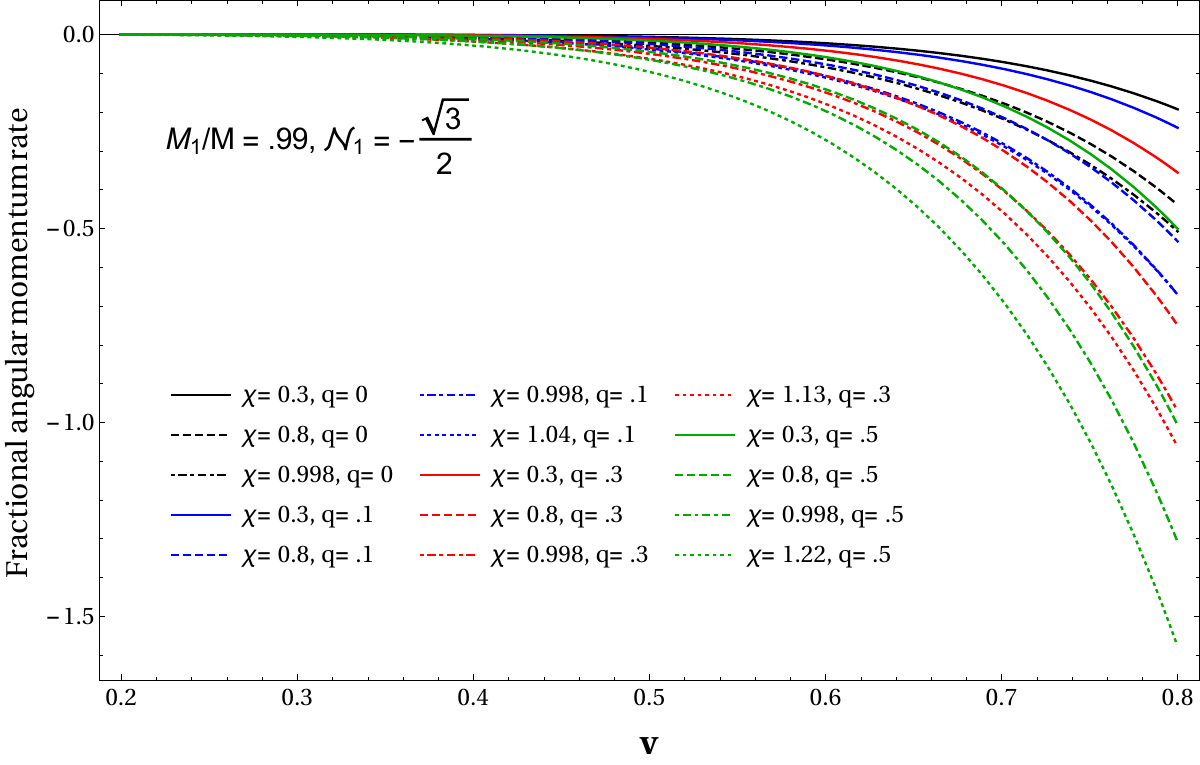}
\caption{The rate of change of angular momentum $(dJ_{1}/dt)$ of the BBH1 has been plotted in dimensionless units, i.e., it is scaled by the Newtonian rate of change of angular momentum $(dJ/dt)_{\rm N}$. The plots are for various choices of the dimensionless spin parameter $\chi=(a/M)$ and the dimensionless tidal charge parameter $q=(Q/M^{2})$, against the relative velocity $v$ between the orbiting black holes in the binary. The plot on the left is for the mass ratio $(M_{1}/M)=0.8$ on a plane with $\theta=(\pi/3)$, while the plot on the right is for the extreme mass ratio $(M_{1}/M)=0.99$ and on a plane with $\theta=(5\pi/6)$. Note that for larger values of $q$, the spin parameter $\chi$ can take values larger than unity and the rate of change of angular momentum differs significantly from the general relativistic expression. The deviation increases as the relative velocity between the two black holes become larger.}
\label{angular_rate}
\end{figure}

The rate of change of angular momentum (normalized by the Newtonian rate $(dJ/dt)_{\rm N}$) has been plotted against the relative velocity $v$ between the orbiting black holes in \ref{angular_rate} for various different choices of the dimensionless spin parameter $\chi$ and tidal charge parameter $q\equiv (Q/M^{2})$. Note that in obtaining these plots we have used the relation between the relative velocity $v$ and the separation $b$ between the two objects, which reads, $v=\sqrt{(M/b)}$. As evident from \ref{angular_rate}, the rate of change of angular momentum changes significantly with the inclusion of the tidal charge $q$ as the relative velocity between the orbiting black holes increases. For example, for a binary black hole system with $(M_{1}/M)=0.8$ on a plane $\theta=(\pi/3)$ with relative velocity $v\simeq 0.8$, the relative change in the angular momentum can be as large as $300\%$ as $q$ changes from the value $0$ to $0.5$. For example, the maximal difference between $q=0$ and $q=0.1$ is $25\%$, while the difference between $q=0$ and $q=0.3$ is $125\%$ and finally the maximal difference between $q=0$ and $q=0.5$ is $300\%$. However, for smaller velocities the change in the rate of angular momentum is smaller. Further, for black hole binary with extreme mass ratio (the right hand figure in \ref{angular_rate}), the rate of change of angular momentum is strictly negative, i.e., the angular momentum should decrease with time. While for black holes with intermediate mass ratio, positive values of $q$ with smaller values of angular momentum provides a positive rate of change of angular momentum.   

\begin{figure}
\includegraphics[scale=0.41]{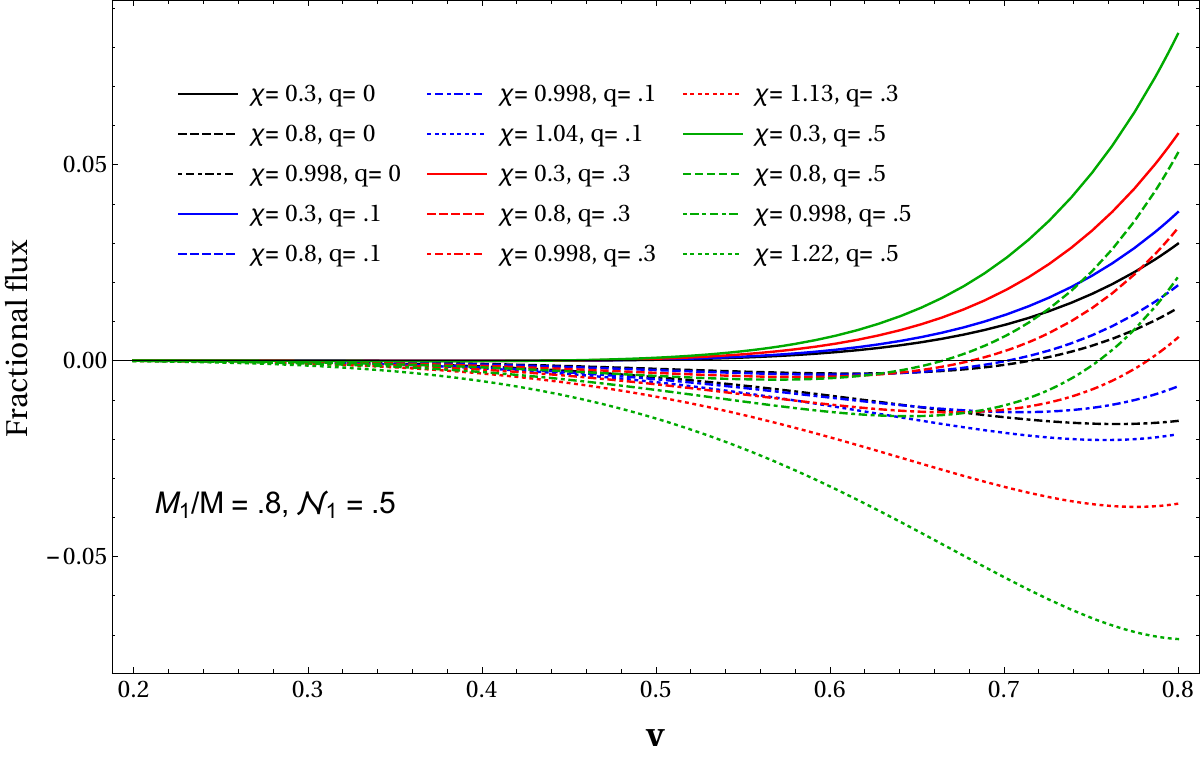}
\hskip .5 cm
\includegraphics[scale=0.41]{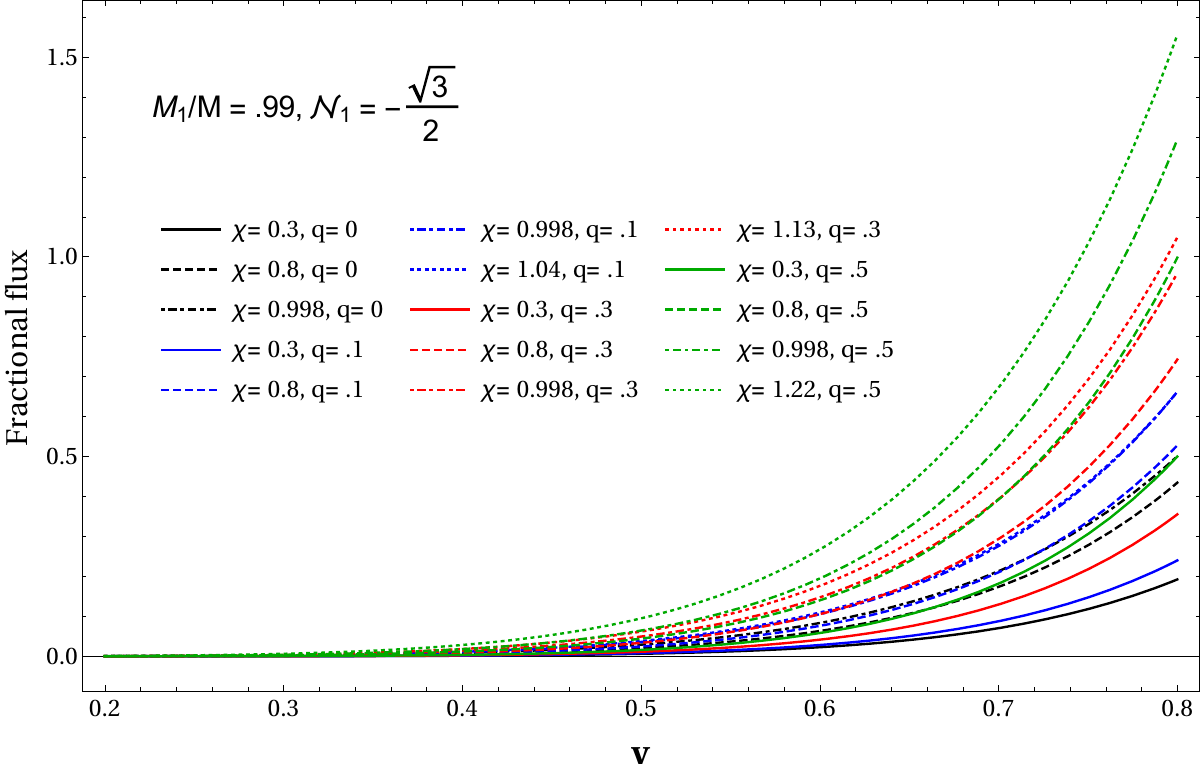}
\caption{The rate of change of mass $(dM_{1}/dt)$ of the BBH1 has been plotted in dimensionless units, i.e., it is scaled by the Newtonian rate of change of energy $(dE/dt)_{\rm N}$. The plots are for various choices of the dimensionless spin parameter $\chi=(a/M)$ and the dimensionless tidal charge parameter $q=(Q/M^{2})$, against the relative velocity $v$ between the orbiting black holes in the binary. The plot on the left is for the mass ratio $(M_{1}/M)=0.8$ on a plane with $\theta=(\pi/3)$, while the plot on the right is for the extreme mass ratio $(M_{1}/M)=0.99$ and on a plane with $\theta=(5\pi/6)$. Note that for larger values of $q$, the spin parameter $\chi$ can take values larger than unity and the rate of change of mass differs from the general relativistic expression, with the deviation increasing as the relative velocity between the two black holes become larger and larger.}
\label{mass_rate_fig}
\end{figure}
 
Alike the rate of change of angular momentum, the rate of change of mass (normalized by the Newtonian rate $(dE/dt)_{\rm N}$) has been plotted against the relative velocity $v$ between the orbiting black holes in \ref{mass_rate_fig} for various different choices of the dimensionless spin parameter $\chi$ and tidal charge parameter $q\equiv (Q/M^{2})$. As evident from \ref{mass_rate}, for a binary black hole system with $(M_{1}/M)=0.8$ on a plane $\theta=(\pi/3)$ with relative velocity $v\simeq 0.8$, the relative change in the mass can be as large as $350\%$ as $q$ changes from the value $0$ to $0.5$. For smaller values of $q$, e.g., for $q=0.1$ the relative change in the mass rate compared to the general relativistic scenario is about $33\%$. Note that for black hole binary with extreme mass ratio (the right hand figure in \ref{mass_rate_fig}), the rate of change of mass is strictly positive. Thus for extreme mass ratio in-spiral, the angular momentum of the massive black hole in the binary always decreases, while the mass of the same increases. Further, note that for intermediate mass ratio black holes in the binary (with $(M_{1}/M)=0.8$), as evident from the left hand figures in \ref{angular_rate} and \ref{mass_rate_fig}, the rate of change of angular momentum and mass is strictly negative for all possible values of relative velocity, provided the spin parameter is close to its extremal value.

\section{Possible constraints on the tidal charge from tidal heating of braneworld black hole}\label{tidal_heat_const}

In this section, we will try to find out possible constraints on the dimensionless tidal charge parameter $q$, that can be achieved from gravitational wave observations. For this purpose, we will primarily focus on the extreme mass ratio in-spirals (EMRI). As long as EMRI is concerned, even though a detailed numerical analysis is necessary to comment on the observability of the tidal charge parameter $q$, it is possible to obtain an order of magnitude estimation for the same. We will do it by using available results in the literature \cite{Datta:2019epe,Flanagan:1997kp,Lindblom:2008cm}. For this purpose, we define a parameter $H$ as follows:
\begin{equation}\label{define_H}
H \equiv \frac{\dot{M}^{\rm BW}_{\rm BH}}{\dot{M}^{\rm GR}_{\rm BH}} \equiv 1+\alpha~.
\end{equation}
Here, $\dot{M}^{\rm BW}_{\rm BH}$ and $\dot{M}^{\rm GR}_{\rm BH}$, refer to the rate of change of mass for the braneworld black hole and for a black hole in general relativity, respectively. Note that the function $H$ is defined in a similar fashion as in Ref. \cite{Datta:2019euh}, which captures the presence and the modification of tidal heating due to the presence of extra dimension. As evident from \ref{define_H}, in the absence of any extra dimension, the parameter $\alpha$ identically  vanishes. Thus any non-zero and \emph{positive} value of $\alpha$ will denote the existence of extra dimension. Following which, it is instructive to express the above definition of $\alpha$ as, $\dot{M}^{\rm BW}_{\rm BH} = \dot{M}^{\rm GR}_{\rm BH} + \alpha \dot{M}^{\rm GR}_{\rm BH}$. 

\begin{figure}
\centering
\includegraphics[scale=0.8]{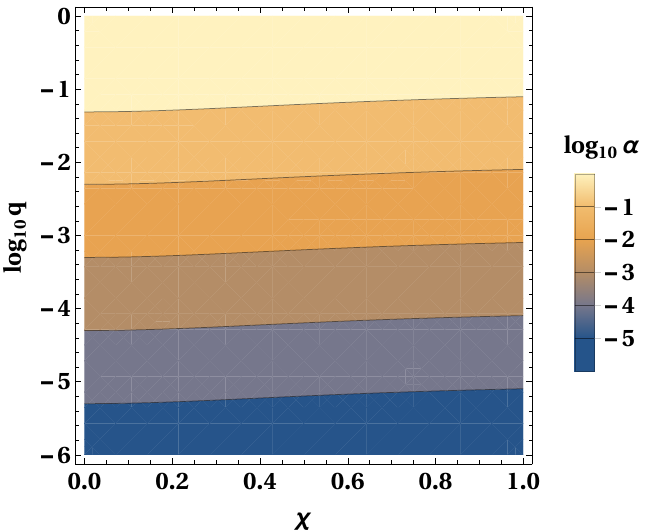}
\caption{In this figure we have plotted the contours of constant $\alpha$, whose expression in terms of the tidal charge $q$ and dimensionless spin parameter $\chi$ is given by \ref{Eq:alpha}. The contours and the respective colour bar represent different values of $\log_{10}\alpha$. In the $x$ axis we present the dimensionless spin parameter $\chi$ and in the $y$ axis we present the values of $\log_{10}q$.}
\label{fig:contour}
\end{figure}

In general, the quantity $\alpha$ is a frequency-dependent function, which is obvious from its definition given by \ref{define_H} and the expression for $\dot{M}^{\rm BW}_{\rm BH}$ in \ref{mass_rate}. Since our objective is to find an order of magnitude estimate for the tidal charge $q$, we will primarily focus on the leading order contribution to the frequency (i.e., the 2.5 PN term only). Note that, in such a case the frequency dependence of $\dot{M}^{\rm BW}_{\rm BH}$ will coincide with the frequency dependence of $\dot{M}^{\rm GR}_{\rm BH}$, and hence $\alpha$ becomes frequency independent. Substituting the relevant expressions for the rate of change of mass of braneworld black hole and black hole in general relativity, we obtain,
\begin{equation}\label{Eq:alpha}
1+\alpha=\frac{1 + 3 \chi^2 + q (2 + 3 \chi^2 + q)}{1 + 3 \chi^2}~.
\end{equation}
Given the above expression for $\alpha$, we have plotted the contours of constant values of $\alpha$ in \ref{fig:contour}, against the tidal charge $q$ and the dimensionless spin parameter $\chi$. We primarily focus on small values of $\alpha$, since we are interested in finding out the smallest possible value of $\alpha$, allowed observationally.
Since $\dot{M}$ is nothing but the energy exchange with the orbit, the $\alpha$ dependent term will modify the in-spiral rate, which in exchange, will affect the emitted gravitational waves by introducing de-phasing. 

In \cite{Datta:2019epe}, the effect of such de-phasing in the gravitational wave waveform due to tidal heating of black holes and ECOs in general relativity has been studied for EMRI. Even though these studies were done keeping ECOs in mind, the similarity of the mathematical expression for rate of change of mass with our current work makes it easy to use their estimates for our purpose. In \cite{Datta:2019epe}, the rate of change of mass of an ECO is related to its reflectivity $\mathcal{R}$ through the following relation,
\begin{equation}
\dot{M}_{\rm ECO} = \dot{M}_{\rm BH} - |\mathcal{R}|^2 \dot{M}_{\rm BH}~.
\end{equation}
From which, it follows that $|\mathcal{R}|^2$ can be constrained down to the value $\leq 5 \times 10^{-5}$ for signal-to-noise ratio (defined as $\rho$ and denoted by SNR) $\sim 20$ with a dimensionless spin parameter $\chi=0.8$ for the supermassive compact object in the EMRI.

This bound on the reflectivity can be obtained along the following lines: two waveforms can be considered indistinguishable in the context of parameter estimation purposes if the mismatch $\mathfrak{M}$ between them and the SNR $\rho$ satisfies the following relation: $\mathfrak{M} \lesssim 1/(2\rho^2)$ \cite{Flanagan:1997kp,Lindblom:2008cm} (see \ref{mismatch} for detailed definition). For an EMRI it follows that with $\rho \approx 20$ (or, with $\rho \approx 100)$, $\mathfrak{M} \lesssim 10^{-3}$ (or, $\mathfrak{M} \lesssim 5 \times 10^{-5})$. Using which, in Ref. \cite{Datta:2019epe} the above bound on the reflectivity $|\mathcal{R}|^2$, namely $|\mathcal{R}|^2\leq 5 \times 10^{-5}$ has been obtained, with direct observable consequences for LISA. For a similar system, requiring the de-phasing to be smaller than one Radian, a weaker constraint $|\mathcal{R}|^2 \lesssim 10^{-4}$ can also be derived \cite{Datta:2019epe}.

The conclusion of Ref. \cite{Datta:2019epe} regarding the constraints on the reflectivity $|\mathcal{R}|^2$ can be easily translated to constraints on the tidal charge $q$. The key difference between the two scenarios being, the reflectivity \emph{decreases} the rate of change of mass, while the presence of tidal charge \emph{increases} the same. This implies that presence of extra dimension will affect the de-phasing, which will be of opposite sign to that in Ref. \cite{Datta:2019epe}. But magnitude wise they will be similar as long as $\alpha \sim |\mathcal{R}|^2$. Therefore, a similar level of mismatch arising from this extra contribution is sufficient enough for the order of magnitude estimation for the constraint on the tidal charge. Hence, the conclusions regarding $|\mathcal{R}|^2$ can be translated to $\alpha$, and hence to the tidal charge $q$, in a straightforward manner. In particular, for a supermassive compact object with dimensionless spin parameter $\chi \sim 0.8$ and SNR $\rho \sim 20$, it follows that even a small value of $\alpha$ ($\lesssim 5 \times 10^{-5}$) can have mismatch $\mathfrak{M}\gtrsim 10^{-3}$. While for an increased SNR $(\rho \sim 100)$, even smaller values of $\alpha$ can result into a mismatch $\mathfrak{M} \sim 10^{-5}$ and hence will lead to stronger constraint on the tidal charge $q$ \footnote{This comment is not entirely accurate, since 4 PN contribution has been ignored. Nevertheless as an order of magnitude estimation this result is indeed important.}. From \ref{fig:contour} and the above constraint on $\alpha$, it follows that the tidal charge parameter satisfies the following constraint: $q\lesssim \mathcal{O}(10^{-6})$. Larger values of the tidal charge parameter will have significant de-phasing effects in EMRI, which may have observable consequences in the upcoming LISA experiment.

\section{Tidal heating for exotic compact objects on the brane}\label{tidal_heat_eco}

So far we have discussed the phenomenon of tidal heating for black holes on the brane. However, in the braneworld scenario it is also possible to consider the solution on the brane to be an ECO, rather than a black hole. The above can be motivated from several points of view, which have been detailed in \cite{Dey:2020lhq,Dey:2020pth} and we briefly review the key points here. First of all, from the AdS/CFT correspondence it follows that the gravitational field equations on the brane differ from the Einstein's equations by the semiclassical correction term, arising out of the presence of the conformal fields on the brane. As a consequence, the vacuum solutions of the Einstein's equations will be modified by the quantum corrections, which will shift the horizon location and thus the solution will behave more like an ECO, rather than a black hole. Moreover, the horizon of the solution on the brane is not necessarily an event horizon. This is because, the event horizon is related to the global properties of the spacetime, which requires knowledge of the bulk spacetime as well. However, given the solution on the brane, it is difficult to generate the full bulk geometry by `evolving' the brane solution along the direction of the extra dimension. As a consequence, the horizon of the brane solution is more akin to an apparent horizon, rather than an event horizon. Thus we may have both ingoing as well as outgoing waves on the horizon of the brane solution and hence it is more appropriate, in this sense, to describe it as an ECO, rather than a black hole. Intriguingly, the tidal charge parameter $Q$ (or, equivalently the dimensionless tidal charge parameter $q$) appearing in the solution on the brane, can be related to the size of the extra dimension. This is because, with an increase in the tidal charge parameter $Q$, the horizon penetrates less and less into the bulk spacetime, i.e., the black hole becomes more and more localized on the brane. This provides the connection between the physics of the extra dimension and the tidal charge parameter $Q$, as well as the reason behind interpreting the solution on the brane as an ECO.    

Following which, in this section we will work out the relevant results for tidal heating with the interpretation that the braneworld solution, presented in \ref{brane_sol} is an ECO. In particular, we would like to explore what new effects the ECOs bring in to the phenomenon of tidal heating. The first step of such an analysis is the computation of the area of the surface of the ECO, which we perform in the next section. 

\subsection{Area of an exotic compact object in the braneworld}

In this section we will compute the area of the surface of an ECO in the braneworld scenario. First of all, we take the surface of the ECO to be located at, $r=r_{\rm s}\equiv r_{+}(1+\epsilon)$, where $r_{+}$ is the location of the event horizon of the black hole solution, presented in \ref{brane_sol} and $\epsilon$ depicts the departure of the surface of the ECO from that of the would-be braneworld black hole. Even though the parameter $\epsilon$ need not be a constant for an actual ECO, but for the purpose of this paper, we will assume that the dimensionless parameter $\epsilon$ is a constant. It is evident that on $r=r_{\rm s}$, $\Delta=(r-r_{+})(r-r_{-})$ does not vanish and hence is not the horizon, or, for that matter it is not even a null surface. Thus the area of the surface of the ECO can be obtained by taking a cue from \ref{Area_Gen}, which takes the following form,
\begin{align}\label{area_eco}
A_{\rm ECO}=2\pi \int_{-1}^{1}\sqrt{\bar{\alpha} +\bar{\alpha} \beta z^{2}} \, dz 
\end{align}
where, we have defined, $\bar{\alpha} \equiv (r^{2}+a^{2})^{2} -a^{2} \Delta$ and $\bar{\alpha} \beta \equiv a^{2}\Delta$. To emphasize, the above area is the cross-section of $r=r_{\rm s}$ and any $u=\textrm{constant}$ hypersurface. Thus one may substitute $r=r_{+}(1+\epsilon)$ in the above expression and hence compute the expansion of the area in powers of $\epsilon$. We will directly present the results for the braneworld geometry considered above, from which the case for an Kerr-like ECO can be straightforwardly obtained and verified with the results presented in \cite{Datta:2020rvo}. For this purpose, expanding out the coefficients $\bar{\alpha}$ and $\bar{\alpha} \beta$ given above, upto quadratic powers of $\epsilon$, the area $A_{\rm ECO}$, given in \ref{area_eco}, takes the following form,
\begin{align}\label{area_eco_N}
A_{\rm ECO}&=2\pi \int_{-1}^{+1} dz ~\sqrt{(r_{+}^{2}+a^{2})^{2}+ c_{1}(z)\epsilon+ c_{2}(z)\epsilon^{2}} 
\nonumber
\\
&\equiv A_{\rm BH} + \epsilon A^{(1)}_{\rm ECO} + \epsilon^{2} A^{(2)}_{\rm ECO}+\mathcal{O}(\epsilon^{3})~,
\end{align}
where, for notational convenience we have introduced the coefficients $c_{1}(z)$ and $c_{2}(z)$, whose explicit expressions are,
\begin{align}
c_{1}(z)&\equiv 4r_{+}^{2}(r_{+}^{2}+ a^{2})-a^{2} r_{+} (r_{+} - r_{-})(1-z^{2})~,
\nonumber
\\
c_{2}(z)&\equiv 2r_{+}^{2} (3r_{+}^{2}+a^{2}) - r_{+}^{2} a^{2} (1-z^{2})~.
\end{align}
Here, $A_{\rm BH}$ is the area of the black hole spacetime, given by \ref{area_bh}, while $A^{(1)}_{\rm ECO}$ and $A^{(2)}_{\rm ECO}$ are the first and second order corrections over and above the horizon area of the would-be braneworld black hole spacetime, respectively. As expected, in the limit $\epsilon \rightarrow 0$, the above expression for area reduces to that of the braneworld black hole. Evaluating the coefficients $A^{(1)}_{\rm ECO}$ and $A^{(2)}_{\rm ECO}$ of each of the powers of $\epsilon$ individually, we obtain (for a derivation, see \ref{AppE}),
\begin{align}\label{first_order_area_eco}
A^{(1)}_{\rm ECO}&=\dfrac{4\pi}{3M}\dfrac{1}{\left(1+\dfrac{Q}{2Mr_{+}}\right)}\left[r_{+}^{3}+2M^{2}r_{+}+3Mr_{+}^{2}+Q(2r_{+}+M)\right]~,
\end{align}
where, we have used the following two relations, $r_{+}^{2}+a^{2}=2M r_{+}+Q$ and $a^{2}= r_{+}(2M- r_{+})+Q$, pertaining to the exterior geometry of the ECO in the braneworld scenario. Note that in the limit of $Q\rightarrow 0$, we get back the first order change in area for a Kerr-like ECO, which takes the form, 
\begin{align}
A^{(1)}_{\rm ECO}\Big\vert_{Q=0}&=\dfrac{4\pi r_{+}}{3M}\left[r_{+}^{2} + 3Mr_{+} +2M^{2}  \right]~,
\end{align}
which matches with the result presented in \cite{Datta:2020rvo}. In a similar manner, the second order contribution to the area for the ECO on the brane takes the following form,
\begin{align}\label{second_order_area_eco}
A^{(2)}_{\rm ECO}&=\dfrac{2\pi r_{+}}{15M^{3}\left(1-\dfrac{Q^{2}}{2Mr_{+}} \right)^{3}}
\Bigg[-r_{+}^{4}-4Mr_{+}^{3}+22M^{2}r_{+}^{2}+12 M^{3}r_{+}-4M^{4}
\nonumber
\\
&\hskip 1 cm +Q^{2}\bigg\{3r_{+}^{2}-22Mr_{+}-20M^{2}+\dfrac{4M^{3}}{r_{+}}\bigg\}
+Q^{4} \bigg\{\dfrac{21}{4}+\dfrac{12 M}{r_{+}}-\dfrac{M^{2}}{r_{+}^{2}} \bigg\} 
+Q^{6} \left\{ -\dfrac{5}{2r_{+}^{2}}\right\} \Bigg]
\end{align} 
Again, one can explicitly verify that in the limit $Q\rightarrow 0$, for Kerr-like ECO, the second order correction to the area coincides with the expression presented in \cite{Datta:2020rvo}. Thus we have explicitly derived how the area of an ECO in the braneworld scenario depends on the parameters of the solution presented in \ref{brane_sol} and on the location of the surface $r_{\rm s}$, which reduces to the respective expression for Kerr-like ECO under appropriate limit.

Having derived the expression of area for the surface of the ECO, it is now time to compute the tidal heating associated with the ECO in the context of the braneworld. The computation of tidal heating will require knowledge about the rate of change of area of the surface of the ECO. This in turn needs to be related to the rate of change in mass and angular momentum of the ECO. As in the case of black hole spacetime, here also in the stationary case, the rate of change of the angular momentum can be related to the rate of change of area. For this purpose, we want to relate the rate of change of area to the rate of change of angular momentum and mass. Again, the tidal charge parameter can be taken to be constant, since low-energy physics on the brane is unlikely to affect parameters, like the tidal charge, inherited from the bulk spacetime. Thus we obtain, 
\begin{align}
\frac{dA_{\rm ECO}}{dt}=\left(\frac{\partial A_{\rm ECO}}{\partial M}\right)\frac{dM}{dt}+\left(\frac{\partial A_{\rm ECO}}{\partial a}\right)\frac{da}{dt}~.
\end{align}
Since the area of the surface of the ECO has been expanded upto $\mathcal{O}(\epsilon^{2})$, we expand both sides of the above equation as well upto $\mathcal{O}(\epsilon^{2})$, yielding,
\begin{align}
\left(\frac{dA_{\rm ECO}}{dt}\right)^{(0)}&+\epsilon \left(\frac{dA_{\rm ECO}}{dt}\right)^{(1)}+\epsilon^{2} \left(\frac{dA_{\rm ECO}}{dt}\right)^{(2)}
\nonumber
\\
&=\left(\frac{\partial A_{\rm ECO}^{(0)}}{\partial M}+\epsilon \frac{\partial A_{\rm ECO}^{(1)}}{\partial M}+\epsilon^{2}\frac{\partial A_{\rm ECO}^{(2)}}{\partial M} \right)\frac{dM}{dt}
+\left(\frac{\partial A_{\rm ECO}^{(0)}}{\partial a}+\epsilon \frac{\partial A_{\rm ECO}^{(1)}}{\partial a}+\epsilon^{2}\frac{\partial A_{\rm ECO}^{(2)}}{\partial a} \right)\frac{da}{dt}~.
\end{align}
Note that we can use the following relation, $a=(J/M)$ in order to relate the rate of change of area with the rate of change of mass and the rate of change of angular momentum $J$, respectively. From the expression for the area of the surface of the ECO as presented in \ref{area_eco_N} and its first order correction over and above the would-be braneworld black hole spacetime, whose expression is given by \ref{first_order_area_eco}, we obtain,
\begin{align}
\frac{\partial A^{(1)}_{\rm ECO}}{\partial M}&=\frac{4\pi r_{+}^{2}}{3M^{2}}\frac{1}{\left(1+\dfrac{Q}{2Mr_{+}}\right)}\frac{1}{\sqrt{M^{2}-a^{2}+Q}}\left[a^{2}+8M^{2}+2Mr_{+}+\frac{4MQ}{r_{+}}-3Q\right]
\nonumber
\\
&+\frac{4\pi}{3M^{2}}\frac{1}{\left(1+\dfrac{Q}{2Mr_{+}}\right)^{2}}\left(\frac{Q}{2Mr_{+}}+\frac{Q}{2r_{+}\sqrt{M^{2}-a^{2}+Q}}\right)\left[r_{+}^{3}+3Mr_{+}^{2}+2M^{2}r_{+}+Q\left(2r_{+}+M\right)\right]~,
\end{align}
as well as,
\begin{align}
\frac{\partial A^{(1)}_{\rm ECO}}{\partial a}&=-\frac{4\pi a}{3M}\frac{1}{\sqrt{M^{2}-a^{2}+Q}}\frac{1}{\left(1+\dfrac{Q}{2Mr_{+}}\right)}\left[3r_{+}^{2}+6Mr_{+}+2M^{2}+2Q\right]
\nonumber
\\
&-\frac{4\pi a}{3M}\frac{1}{\sqrt{M^{2}-a^{2}+Q}}\frac{1}{\left(1+\dfrac{Q}{2Mr_{+}}\right)^{2}}\left(\frac{Q}{2Mr_{+}^{2}}\right)\left[r_{+}^{3}+3Mr_{+}^{2}+2M^{2}r_{+}+Q\left(2r_{+}+M\right)\right]~.
\end{align}
The situation of particular interest is when the companion ECO in the binary system is stationary. In which case, the rate of change of mass identically vanishes and hence the rate of change of angular momentum can be related to the rate of change of area and its expansion in various powers of $\epsilon$ as,
\begin{align}\label{rate_ang_mom}
\frac{1}{M}\frac{dJ}{dt}=\frac{\left(\frac{dA_{\rm ECO}}{dt}\right)^{(0)}+\epsilon \left(\frac{dA_{\rm ECO}}{dt}\right)^{(1)}+\epsilon^{2} \left(\frac{dA_{\rm ECO}}{dt}\right)^{(2)}}{\left(\frac{\partial A_{\rm ECO}^{(0)}}{\partial a}+\epsilon \frac{\partial A_{\rm ECO}^{(1)}}{\partial a}+\epsilon^{2}\frac{\partial A_{\rm ECO}^{(2)}}{\partial a} \right)}~.
\end{align}
The terms in the denominator can be determined by taking derivatives of the coefficients of various powers of $\epsilon$ in the expansion of the area of the surface of ECO. In what follows we will determine the terms in the numerator, i.e., the rate of change of area of an ECO in a binary system to quadratic order in $\epsilon$. Then using \ref{rate_ang_mom}, the rate of change of area can be related to the rate of change of angular momentum and hence one can derive the tidal heating of the binary system, in the context of ECOs on the brane.

\subsection{Tidal heating for stationary exotic compact objects on the brane}

In this section, we will describe the behaviour of the gravitational perturbation with appropriate boundary conditions and hence the tidal heating for ECOs living on the brane, whose exterior geometry is given by \ref{brane_sol}. Here also we concentrate on the perturbed Weyl scalar, as in the case of black hole spacetime, which can be determined by solving the Teukolsky equation. If the companion ECO in the binary system living on the brane, denoted by BECO2 (here, BECO is the short form of braneworld exotic compact object), is located at $(b,\theta_{0},\phi_{0})$ in the LARF coordinate system of BECO1, then the tidal field will be given by, \ref{tidal_field}. Thus it will be considered as giving rise to an appropriate boundary condition at infinity, identical to the corresponding situation in the context of braneworld black holes. Following our previous computation, the solution for the Weyl scalar in the near-horizon regime (for, $l=2=s$), takes the following form, 
\begin{align}
R_{2m}(r)=(1+x)^{\gamma_{m}} \left[ C_{1} x^{-\gamma_{m}} {}_{2}F_{1} (0,5,3-2\gamma_{m};-x) + C_{2} x^{\gamma_{m}-2} (1+x)^{-2-2\gamma_{m}} {}_{2}F_{1}(1,-4,-1+2\gamma_{m}; -x)\right]~,
\end{align}
where $x$ and $\gamma_{m}$ have been defined in \ref{redefinition} and $C_{1}$ and $C_{2}$ are two arbitrary constants. In order to obtain these arbitrary constants, namely $C_{1}$ and $C_{2}$, it is instructive to determine the asymptotic behaviours, i.e, $x \to \infty$ and  $x\to 0$ limits of the radial perturbation $R_{2m}(r)$.  These have already been computed in \ref{far_zone} and \ref{near_horizon} respectively. This suggests to rewrite the above solution for the radial part of the Weyl scalar, using the result ${}_{2}F_{1} (0,b;c;z)=1$ as,
\begin{align}
R^{\rm ECO}_{2m}(r)=(1+x)^{\gamma_{m}}C_{m}^{\rm BH}\left[\mathcal{R}x^{-\gamma_{m}}+\mathcal{T}x^{\gamma_{m}-2} (1+x)^{-2-2\gamma_{m}} {}_{2}F_{1}(1,-4,-1+2\gamma_{m}; -x)\right]~,
\end{align}
where $C_{m}^{\rm BH}$ is given by \ref{arb_constant} and again ${}_{2}F_{1}(1,-4,-1+2\gamma_{m}; -x)$ is a fourth order polynomial in $x$. Here $\mathcal{R}$ is the reflectivity and $\mathcal{T}$ is the transitivity of the surface of the ECO, living on the brane. Therefore, the Weyl scalar in the Hartle-Hawking frame takes the following form,
\begin{align}
\Psi_{0}^{\rm HH}&=\dfrac{\Delta^{2}}{4(r^{2}+a^{2})^{2}}\sum_{m=-2}^{2} C_{m}^{\rm BH}~{}_{2}Y_{2m}(\theta,\phi)~\Big[\mathcal{R}(1+x)^{\gamma_{m}}x^{-\gamma_{m}}
\nonumber
\\
&\hskip 5 cm +\mathcal{T}x^{\gamma_{m}-2} (1+x)^{-2-\gamma_{m}} {}_{2}F_{1}(1,-4,-1+2\gamma_{m}; -x)\Big]~.
\end{align}
Note that in the case of black hole, the reflectivity $\mathcal{R}$ will be identically zero and the transitivity $\mathcal{T}$ will be identity, thereby obtaining \ref{HH_Weyl_BH}.  For tidal heating of ECOs, we need to derive the expression for the above Weyl scalar on the surface $r=r_{\rm s}$ of the ECO. For that purpose, it will be useful to determine a series expansion of $\Psi_{0}^{\rm HH}$ to quadratic powers of $\epsilon$, to be consistent with the expansion of area of the surface of the ECO, presented in the previous section. This yields,
\begin{align}
\Psi_{0}^{\rm HH}=\Psi_{0}^{\rm HH~(0)}+\epsilon \Psi_{0}^{\rm HH~(1)}+\epsilon^{2}\Psi_{0}^{\rm HH~(2)}+\mathcal{O}(\epsilon^{3})~,
\end{align}
where, the coefficients appearing with various powers of $\epsilon$, in the expansion of the relevant projection of Weyl scalar are given by,  
\begin{align}
\Psi_{0}^{\rm HH~(0)}&=\frac{(r_{+}-r_{-})^{4}}{4(r_{+}^{2}+a^{2})^{2}}\sum_{m}\mathcal{T}C_{m}^{\rm BH}x^{\gamma_{m}}~{}_{2}Y_{2m}~,
\\
\Psi_{0}^{\rm HH~(1)}&=\frac{r_{+}(r_{+}-r_{-})^{3}}{4(r_{+}^{2}+a^{2})^{2}}\sum_{m}\mathcal{T}C_{m}^{\rm BH}x^{\gamma_{m}}\left(\frac{4}{2\gamma_{m}-1}-2-\gamma_{m}\right)~{}_{2}Y_{2m}~,
\\
\Psi_{0}^{\rm HH~(2)}&=\frac{r_{+}^{2}(r_{+}-r_{-})^{2}}{4(r_{+}^{2}+a^{2})^{2}}\sum_{m}C_{m}^{\rm BH}~{}_{2}Y_{2m}
\Big[\mathcal{T}x^{\gamma_{m}}\left(\frac{12}{\gamma_{m}(2\gamma_{m}-1)}+\frac{(2+\gamma_{m})(3+\gamma_{m})}{2}-\frac{4(\gamma_{m}+2)}{2\gamma_{m}-1}\right)+\mathcal{R}x^{-\gamma_{m}}\Big]~.
\end{align}
The above provides the expansion of the Weyl scalar, characterizing the gravitational perturbation, in powers of $\epsilon$. As in the case of black holes,   for ECOs as well, we will be looking for the computation of the angular average of $|\Psi_{0}^{\rm HH}|^{2}$, since it will determine the rate of change of area of the BECO1 due to the tidal field of BECO2. Keeping terms upto $\mathcal{O}(\epsilon^{2})$, we obtain, 
\begin{align}\label{expansion_eco_weyl}
\int d\Omega \big|\Psi_{0}^{\rm HH}\big|^{2}&=\dfrac{\left(1-\chi^{2}+\frac{Q}{M^{2}}\right)^{4}}{\left[\left(1+\sqrt{1-\chi^{2}+\frac{Q}{M^{2}}}\right)+\frac{Q}{2M^{2}}\right]^{4}}\big|\mathcal{T}\big|^{2}\sum_{m=-2}^{2}\vert C_{m}^{\rm BH}\vert^{2}
\nonumber
\\
&+\epsilon \left(\frac{r_{+}(r_{+}-r_{-})^{7}}{16\left(r_{+}^{2}+a^{2}\right)^{4}}\right)\sum_{m=-2}^{2}\big|\mathcal{T}\big|^{2}\big|C_{m}^{\rm BH}\big|^{2}\frac{4\left(3-4\gamma_{m}^{2}\right)}{4\gamma_{m}^{2}-1}
\nonumber
\\
&+\epsilon^{2}\left(\frac{r_{+}^{2}(r_{+}-r_{-})^{6}}{16\left(r_{+}^{2}+a^{2}\right)^{4}}\right)\sum_{m=-2}^{2}\big|C_{m}^{\rm BH}\big|^{2}\Bigg[10\big|\mathcal{T}\big|^{2}+\mathcal{R}\mathcal{T}^{*}x^{-2\gamma_{m}}+\mathcal{R}^{*}\mathcal{T}x^{2\gamma_{m}}\Bigg]~.
\end{align}
Given the above expression for the angular average of the $|\Psi_{0}^{\rm HH}|^{2}$, we can derive the rate of change of area of the BECO1, due to the tidal field of BECO2. Following \ref{AppG} we obtain,
\begin{align}
\frac{dA_{\textrm{ECO}1}}{dt}=\left(\frac{dA_{\rm ECO1}}{dt}\right)^{(0)}+\epsilon \left(\frac{dA_{\rm ECO1}}{dt}\right)^{(1)}+\epsilon^{2} \left(\frac{dA_{\rm ECO1}}{dt}\right)^{(2)}~.
\end{align}
Here, the coefficients of various powers of $\epsilon$, in the above expansion for the rate of change of area of the surface of the BECO1 are given by,
\begin{align}
\left(\frac{dA_{\rm ECO1}}{dt}\right)^{(0)}&\equiv \frac{2\left(r_{+1}^{2}+a_{1}^{2}\right)}{\kappa_{+1}^{3}}\dfrac{\left(1-\chi_{1}^{2}+\frac{Q}{M_{1}^{2}}\right)^{4}}{\left[\left(1+\sqrt{1-\chi_{1}^{2}+\frac{Q}{M_{1}^{2}}}\right)+\frac{Q}{2M_{1}^{2}}\right]^{4}}\big|\mathcal{T}_{1}\big|^{2}\sum_{m=-2}^{2}\vert C_{m}^{\rm BH}\vert^{2}
\nonumber
\\
&=32M_{1}^{5}\left(1-\chi_{1}^{2}+\frac{Q}{M_{1}^{2}}\right)^{-1/2}\frac{2\pi M_{2}^{2}}{5b^{6}}\chi_{1}^{2}\sin^{2}\theta_{0}\big|\mathcal{T}_{1}\big|^{2}\left(\mathcal{A}+\mathcal{B}\sin^{2}\theta_{0}\right)=\big|\mathcal{T}_{1}\big|^{2}\left(\frac{dA_{\rm BH1}}{dt}\right)~,
\label{area_change_eco0}
\\
\left(\frac{dA_{\rm ECO1}}{dt}\right)^{(1)}&\equiv \frac{2\left(r_{+1}^{2}+a_{1}^{2}\right)}{\kappa_{+1}^{3}}\Bigg\{\Big[-3\frac{\kappa^{(1)}_{1}}{\kappa_{+1}}+\frac{q^{(1)}_{(\pi/2),1}}{2}\Big]
\nonumber
\\
&\hskip 2 cm \times\Big[\dfrac{\left(1-\chi_{1}^{2}+\frac{Q}{M_{1}^{2}}\right)^{4}}{\left[\left(1+\sqrt{1-\chi_{1}^{2}+\frac{Q}{M_{1}^{2}}}\right)+\frac{Q}{2M_{1}^{2}}\right]^{4}}\big|\mathcal{T}_{1}\big|^{2}\sum_{m=-2}^{2}\vert C_{m}^{\rm BH}\vert^{2}\Big]
\nonumber
\\
&+\left(\frac{r_{+1}(r_{+1}-r_{-1})^{7}}{16\left(r_{+1}^{2}+a_{1}^{2}\right)^{4}}\right)\sum_{m=-2}^{2}\big|\mathcal{T}_{1}\big|^{2}\big|C_{m}^{\rm BH1}\big|^{2}\frac{4\left(3-4\gamma_{m,1}^{2}\right)}{4\gamma_{m,1}^{2}-1}\Bigg\}~,
\label{area_change_eco1}
\\
\left(\frac{dA_{\rm ECO1}}{dt}\right)^{(2)}&\equiv \frac{2\left(r_{+1}^{2}+a_{1}^{2}\right)}{\kappa_{+1}^{3}}\Bigg\{\Big[-3\frac{\kappa^{(2)}_{1}}{\kappa_{+1}}+6\left(\frac{\kappa^{(1)}_{1}}{\kappa_{+1}}\right)^{2}+\left(\frac{q^{(2)}_{(\pi/2),1}}{2}-\frac{\left(q^{(1)}_{(\pi/2),1}\right)^{2}}{8} \right)+\frac{q^{(1)}_{(\pi/2),1}}{2}\left(\frac{\kappa^{(1)}_{1}}{\kappa_{+1}}\right)\Big]
\nonumber
\\
&\hskip 2 cm \times\Big[\dfrac{\left(1-\chi_{1}^{2}+\frac{Q}{M_{1}^{2}}\right)^{4}}{\left[\left(1+\sqrt{1-\chi_{1}^{2}+\frac{Q}{M_{1}^{2}}}\right)+\frac{Q}{2M_{1}^{2}}\right]^{4}}\big|\mathcal{T}_{1}\big|^{2}\sum_{m=-2}^{2}\vert C_{m}^{\rm BH}\vert^{2}\Big]
\nonumber
\\
&+\left(\frac{r_{+1}^{2}(r_{+1}-r_{-1})^{6}}{16\left(r_{+1}^{2}+a_{1}^{2}\right)^{4}}\right)\sum_{m=-2}^{2}\big|C_{m}^{\rm BH1}\big|^{2}\Bigg[10\big|\mathcal{T}_{1}\big|^{2}+\mathcal{R}_{1}\mathcal{T}_{1}^{*}x^{-2\gamma_{m,1}}+\mathcal{R}_{1}^{*}\mathcal{T}_{1}x^{2\gamma_{m,1}}\Bigg]
\nonumber
\\
&+\left[\left(\frac{r_{+1}(r_{+1}-r_{-1})^{7}}{16\left(r_{+1}^{2}+a_{1}^{2}\right)^{4}}\right)\sum_{m=-2}^{2}\big|\mathcal{T}_{1}\big|^{2}\big|C_{m}^{\rm BH1}\big|^{2}\frac{4\left(3-4\gamma_{m,1}^{2}\right)}{4\gamma_{m,1}^{2}-1}\right]\left(\frac{\kappa^{(1)}_{1}}{\kappa_{+1}}+\frac{q^{(1)}_{(\pi/2),1}}{2}\right)\Bigg\}~.
\label{area_change_eco2}
\end{align}
In the above expressions, $\mathcal{R}_{1}$ is the reflectivity and $\mathcal{T}_{1}$ is the transitivity of the surface of the BECO1, while $\gamma_{m,1}$ and $C_{m}^{\rm BH1}$ have been defined earlier in \ref{redefinition} and \ref{arb_constant}, respectively, with $M=M_{1}$ and $J=J_{1}$. Similarly, $r_{+1}$, $r_{-1}$ and $\kappa_{+1}$ are the locations of the event horizon, Cauchy horizon and the surface gravity of the event horizon, respectively, for the would-be black hole with mass $M_{1}$ and angular momentum $J_{1}$. Additionally, the objects, $\kappa_{1}^{(1)}$, $\kappa_{2}^{(2)}$, $q^{(1)}_{(\pi/2),1}$ and $q^{(2)}_{(\pi/2),1}$ have the following expressions,
\begin{align}
\kappa_{1}^{(1)}&=\frac{Qr_{+1}+2M_{1}r_{+1}^{2}}{(r_{+1}^{2}+a_{1}^{2})^{2}}-\frac{4\kappa_{+1}r_{+1}^{2}}{(r_{+1}^{2}+a_{1}^{2})}~,
\\
\kappa_{1}^{(2)}&=\frac{M_{1}r_{+1}^{2}}{(r_{+1}^{2}+a_{1}^{2})^{2}}-\frac{4r_{+1}^{2}\left(Qr_{+1}+2M_{1}r_{+1}^{2}\right)}{(r_{+1}^{2}+a_{1}^{2})^{3}}+\frac{\kappa_{+1}r_{+1}^{2}\left(10r_{+1}^{2}-2a_{1}^{2}\right)}{(r_{+1}^{2}+a_{1}^{2})^{2}}~,
\\
q^{(1)}_{(\pi/2),1}&=\frac{4r_{+1}^{2}\left(r_{+1}^{2}+a_{1}^{2}\right)-r_{+1}\left(r_{+1}-r_{-1}\right)a_{1}^{2}}{\left(r_{+1}^{2}+a_{1}^{2}\right)^{2}}~,
\\
q^{(2)}_{(\pi/2),1}&=\frac{2r_{+1}^{2}\left(r_{+1}^{2}+a_{1}^{2}\right)+4r_{+1}^{4}-a_{1}^{2}r_{+1}^{2}}{\left(r_{+1}^{2}+a_{1}^{2}\right)^{2}}~.
\end{align}
These expressions can be substituted in the rate of change of area of the surface of the exotic compact object at various orders in $\epsilon$ and hence $(dA_{\rm ECO1}/dt)^{(n)}$ can be derived as a function of the black hole hairs $M_{1}$, $J_{1}$ and $Q$, respectively, for $n=0,1,2$. Note that, just as the above expressions, namely \ref{area_change_eco0} to \ref{area_change_eco2} provide the rate of change of BECO1, due to the tidal field of BECO2. Similarly, by interchanging $1\leftrightarrow 2$, we can derive the rate of chnage of the area of the surface of BECO2, given the tidal field of BECO1. Finally, the rate of change of angular momentum can be expressed in terms of the rate of change of area in the case of a stationary companion, following \ref{rate_ang_mom}, as,
\begin{align}
\frac{1}{M_{1}}\frac{dJ_{1}}{dt}&=\frac{\left(\frac{dA_{\rm ECO1}}{dt}\right)^{(0)}}{\left(\frac{\partial A_{\rm ECO1}^{(0)}}{\partial a_{1}}\right)}+\epsilon \left[\frac{\left(\frac{dA_{\rm ECO1}}{dt}\right)^{(1)}}{\left(\frac{\partial A_{\rm ECO1}^{(0)}}{\partial a_{1}}\right)}-\frac{\left(\frac{\partial A_{\rm ECO1}^{(1)}}{\partial a_{1}}\right)}{\left(\frac{\partial A_{\rm ECO1}^{(0)}}{\partial a_{1}}\right)}\right]
\nonumber
\\
&+\epsilon^{2}\left[\frac{\left(\frac{dA_{\rm ECO1}}{dt}\right)^{(2)}}{\left(\frac{\partial A_{\rm ECO1}^{(0)}}{\partial a_{1}}\right)}-\frac{\left(\frac{\partial A_{\rm ECO1}^{(2)}}{\partial a_{1}}\right)}{\left(\frac{\partial A_{\rm ECO1}^{(0)}}{\partial a_{1}}\right)} 
+\frac{\left(\frac{\partial A_{\rm ECO1}^{(1)}}{\partial a_{1}}\right)^{2}}{\left(\frac{\partial A_{\rm ECO1}^{(0)}}{\partial a_{1}}\right)^{2}}
-\frac{\left(\frac{dA_{\rm ECO1}}{dt}\right)^{(1)}}{\left(\frac{\partial A_{\rm ECO1}^{(0)}}{\partial a_{1}}\right)}\frac{\left(\frac{\partial A_{\rm ECO1}^{(1)}}{\partial a_{1}}\right)}{\left(\frac{\partial A_{\rm ECO1}^{(0)}}{\partial a_{1}}\right)} \right]~.
\end{align}
In the above expression all the relevant quantities have already been derived earlier, e.g., the expressions for the rate of change of area at various orders of $\epsilon$ has been given in \ref{area_change_eco0} to \ref{area_change_eco2}. Similarly, expressions for $A^{(0)}_{\rm ECO1}$ is identical to that of a black hole given in \ref{area_bh}, while expressions for $A^{(1)}_{\rm ECO1}$ and $A^{(2)}_{\rm ECO1}$ have been given in \ref{first_order_area_eco} and \ref{second_order_area_eco}, respectively. Thus, we have presented the rate of change of area and the rate of change of angular momentum for an ECO in the braneworld scenario due to its stationary companion as a functions of the mass, angular momentum and tidal charge, as an expansion in the smallness parameter $\epsilon$. 
\subsection{First order effects for stationary exotic compact objects on the brane}

Let us briefly review the first order effects, i.e., phenomenon happening at $\mathcal{O}(\epsilon)$ due to tidal forces of the companion, for an exotic compact object on the brane. As in \cite{Datta:2020rvo}, in the present context as well, the rate of change of area to $\mathcal{O}(\epsilon)$ has an overall factor of $|\mathcal{T}|^{2}$ along with terms depending on the mass and the dimensionless combinations $\chi=(a/M)$ and $q=(Q/M^{2})$, respectively. Therefore, the rate of change of area of the exotic compact object behaves as, $\mathcal{A}^{(0)}+\epsilon|\mathcal{T}|^{2}\mathcal{A}^{(1)}$, where $\mathcal{A}^{(0)}$ and $\mathcal{A}^{(1)}$ are dependent on the black hole hairs alone. Except for different dependences on the dimensionless spin parameter $\chi$ and dimensionless tidal charge parameter $q$, both $\mathcal{A}^{(0)}$ and $\mathcal{A}^{(1)}$ have similar magnitudes (again, consistent with \cite{Datta:2020rvo}). Therefore, upto first order in $\epsilon$, the reflectivity (or, equivalently, the transmitivity) appears combined with $\epsilon$ in the manner $\epsilon |\mathcal{T}|^{2}$ and hence there is a degeneracy between the two. As we will see, second order effects will provide further new avenues to look for the signatures of exotic compact objects. 

\subsection{Second order effects for stationary exotic compact objects on the brane}

In this section, we will discuss certain aspects of the second order effects, i.e., we will highlight certain properties of the coefficient of the $\mathcal{O}(\epsilon^{2})$ term in the expansion of the rate of change of area of the surface of an ECO. As in the previous section, here also we will content ourselves with the stationary case, whose generalization for non-stationary system will be presented in a future work. Nevertheless, even in the case of stationary companion, there are some interesting features worth highlighting. First of all, from \ref{area_change_eco0} it follows that, 
\begin{equation}
\dot{A}_{\rm ECO} = |\mathcal{T}|^{2}\dot{A}_{\rm BH} + \mathcal{O}(\epsilon)~.
\end{equation}
Here, `dot' denotes derivative of the respective quantity with respect to the Boyer-Lindquist time coordinate $t$. Since the rate of change of area and mass are related by overall constant factors, it follows that an identical relation will hold true among the rate of change of mass of the ECO and the black hole as well, which is consistent with the findings of \cite{Datta:2020rvo}. In an identical footing, to first order in $\epsilon$, as it follows from \ref{area_change_eco1}, there is an overall factor of $|\mathcal{T}|^{2}$, again consistent with \cite{Datta:2020rvo}. However, when expanded till $\mathcal{O}(\epsilon^2)$ certain interesting features comes into existence. In particular, we find that the rate of change of area to $\mathcal{O}(\epsilon^{2})$ (presented in \ref{area_change_eco2}) is not just proportional to $|\mathcal{T}|^2$. Rather, the rate of change of area depends explicitly on the combination $(\mathcal{R}\mathcal{T}^{*}x^{-2\gamma_{m}}+\mathcal{R}^{*}\mathcal{T}x^{2\gamma_{m}})\equiv \Gamma_{m}$, where $\gamma_{m}$ and $x$ have been defined in \ref{redefinition}. Noting the expression of $x$ in terms of the tortoise coordinate $r_{*}$ and using the result that the surface of the ECO is located at $r=r_{\rm s}=r_{+}(1+\epsilon)$, upto the leading order in $\epsilon$ it follows that,
\begin{align}
x=\epsilon \left(\frac{r_{+}}{r_{+}-r_{-}}\right)e^{\eta r_{+}\epsilon}\left(\frac{r_{+}}{r_{-}}\epsilon+\frac{r_{+}-r_{-}}{r_{-}}\right)^{-\delta}
=\Bar{\eta}\epsilon+\mathcal{O}(\epsilon^{2})~,\,\,\,\,\,\,\Bar{\eta}\equiv \frac{r_{+} \left(r_{-}\right)^{\delta}}{\left(r_{+}-r_{-}\right)^{\delta+1}}~.
\end{align}
Here, $\eta \equiv (r_{+}-r_{-})/(r_{+}^{2}+a^{2})$ and $\delta \equiv (r_{-}^{2}+a^{2})/(r_{+}^{2}+a^{2})$. Note that, in the above expression for the tortoise coordinate $x$, terms of $\mathcal{O}(\epsilon^{2})$ will not be relevant for our purpose. Since in the expression for the rate of change of area, $\Gamma_{m}$ appears multiplied with $\epsilon^{2}$ and hence we can restrict ourselves by keeping terms upto leading order in $\epsilon$. Further, rewriting the complex quantities, namely the reflectivity $\mathcal{R}$ and the transmissivity $\mathcal{T}$ as,
\begin{equation}
\mathcal{R}=|\mathcal{R}|e^{i\phi_{\mathcal{R}}}, \,\,\,\,\mathcal{T} = |\mathcal{T}|e^{i\phi_{\mathcal{T}}},
\end{equation}
it follows that, 
\begin{align}
\Gamma_{m}&=|\mathcal{R}||\mathcal{T}|e^{i(\phi_{\mathcal{R}}-\phi_{\mathcal{T}})}\exp[-2\gamma_{m}\ln\left(\Bar{\eta}\epsilon\right)]
+|\mathcal{R}||\mathcal{T}|e^{-i(\phi_{\mathcal{R}}-\phi_{\mathcal{T}})}\exp[+2\gamma_{m}\ln\left(\Bar{\eta}\epsilon\right)]
\nonumber
\\
&= 2|\mathcal{R}||\mathcal{T}|\cos\left(2i\gamma_m\ln(\Bar{\eta} \epsilon)+\phi_{\mathcal{R}}-\phi_{\mathcal{T}}\right)~.
\end{align}
This shows the remarkable fact that the rate of change of area of an ECO due to the tidal field from a stationary companion depends on the $|\ln \epsilon|$ term, where $\epsilon$ is the fractional difference between the location of the surface of an ECO and the would be black hole horizon. It is worth pointing out that a similar behaviour can also be seen in the context of tidal deformability, where it was argued that the tidal deformability of an ECO $\sim \frac{1}{|\ln \epsilon|}$\cite{Cardoso:2017cfl}. In the context of tidal heating also we find this $\ln\epsilon$ behaviour, but its contribution is bounded by $\pm 1$ due to the cosine dependence. The expression also shows that not only the magnitude of $\mathcal{T}$ and $\mathcal{R}$, but also their respective phases can affect the result as long as $\phi_{\mathcal{R}}\neq \phi_{\mathcal{T}}$.

\begin{figure}[ht]
\includegraphics[scale=0.41]{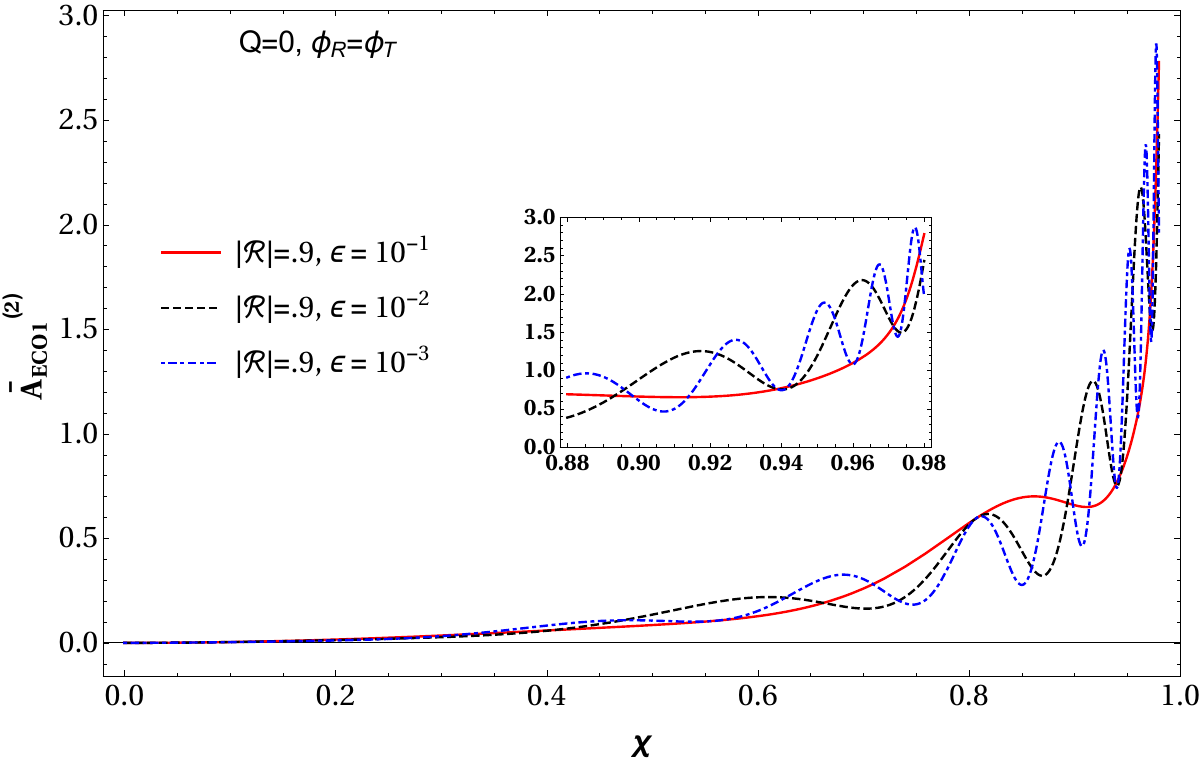}
\hskip .5 cm
\includegraphics[scale=0.41]{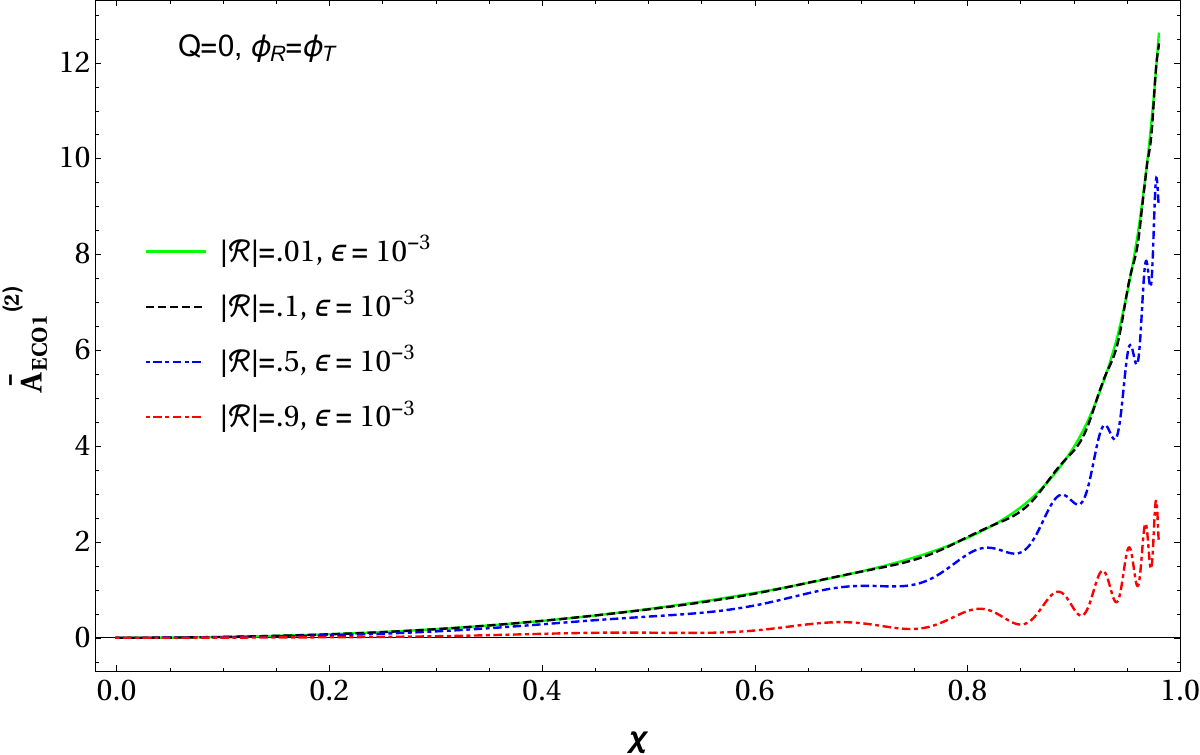}
\caption{The dimensionless quantity $\Bar{A}^{(2)}_{\rm ECO1} =(b^6/M_2^2)\dot{A}^{(2)}_{\rm ECO}\{\kappa_{+1}^{3}/2\left(r_{+1}^{2}+a_{1}^{2}\right)\}$ has been plotted against the spin parameter $\chi$ for zero tidal charge and for different choices of the reflectivity $|\mathcal{R}|$ and $\epsilon$. We have assumed $\phi_{\mathcal{R}}=\phi_{\mathcal{T}}$ in the above in order to illustrate all the dependence on $\epsilon$. See text for more discussions.}
\label{lnepsilon}
\end{figure}
   
To illustrate the effect of the $\ln \epsilon$ term in the phenomenon of tidal heating, we have plotted the dimensionless quantity $\Bar{A}^{(2)}_{\rm ECO1} =(b^6/M_2^2)\dot{A}^{(2)}_{\rm ECO}\{\kappa_{+1}^{3}/2\left(r_{+1}^{2}+a_{1}^{2}\right)\}$ against the spin parameter $\chi$, with zero tidal charge and for different choices of the reflectivity $|\mathcal{R}|$ and $\epsilon$ in \ref{lnepsilon}. Both the plots clearly demonstrates that the rate of change of area to the second order in $\epsilon$ has a sinusoidal behaviour. The plot on the left hand side shows that as the value of $\epsilon$ decreases the frequency of the variation of the rate of change of area with the spin parameter $\chi$ increases. However, as the plot on the right hand side suggests, with the increase of reflectivity keeping $\epsilon$ fixed, the frequency of variation does not change, but the amplitude increases.  

Finally, let us comment on the observability of $\epsilon$ and how it is intertwined with the presence of the tidal charge $Q$. First of all, the leading order expression for the rate of change of area is just the black hole result, multiplied by the transmission coefficient, which will be different from the case of vacuum four-dimensional general relativity due to the presence of non-zero tidal charge and can be considered as an observable for the tidal charge. The first order result, on the other hand, depends on the tidal charge $Q$ and linearly on the parameter $\epsilon$, but is structurally identical to the one presented in \cite{Datta:2020rvo} and harbours no surprise. In the second order, the dependence on $\epsilon$ is also interesting by itself. As elaborated above, the second order rate of change of area depends explicitly on $\epsilon^{2}\cos(\ln \epsilon)$, along with the tidal charge $Q$, such that the rate of change of area depicts significant variability with the dimensionless spin parameter $\chi$ (see \ref{lnepsilon}). In particular, if we consider variation of the rate of change of area of a compact object with its dimensionless spin parameter, it is the presence of the oscillatory behaviour, due to the $\cos(\ln \epsilon)$ term, which provides a distinct signature for the ECOs through the existence of a non-zero $\epsilon$. Since this behaviour does not arise from the presence of the tidal charge, it can be considered as an exclusive effect of the existence of ECO from the study of tidal heating for compact binary systems. Despite such intriguing behaviour for the rate of change of area, the effects discussed above are still suppressed by the $\epsilon^{2}$ term and hence is possibly not detectable in the near future. Additionally, the effect due to the $\cos(\ln \epsilon)$ term starts to dominate at larger values of the dimensionless spin parameter $\chi$, where superradiant instabilities are also pronounced. In this context, the tidal charge parameter comes to the rescue, as \cite{Dey:2020pth} has explicitly demonstrated, presence of the tidal charge significantly reduces the superradiant instability. Hence even larger values of the spin parameter can be supported by the rotating braneworld ECOs. Further, presence of the tidal charge can actually enhance this oscillatory behaviour, thus making possible detection of an ECO from tidal heating more promising. 
\section{Discussion and concluding remarks}\label{conclusion}

The tidal effects of a compact object on its companion in a binary system provide plethora of information about the gravitational interaction and also about the nature of the compact object. The tidal field will deform the compact object, thereby changing the area of the surface of the compact object. If the compact object is a black hole, then the rate of change of area is directly related to the rate of change of the hairs of the black hole. For example, in the case of a Kerr black hole, the rate of change of area is related to the rate of change of mass and angular momentum with appropriate coefficients. Thus given the rate of change of area and angular momentum, the rate of change of mass can be determined. This is achieved by first working out the rate of change of area in the case of stationary companion, using the Newman-Penrose formalism and then appropriately generalizing the same to the non-stationary situation. As long as the two compact objects in the binary system are far apart, in comparison to the characteristic size of the objects themselves, the above rate of change of the black hole hairs can be integrated in order to obtain the total change in the respective quantities as well. These have significant influence on the in-spiral phase of the gravitational wave signal.

Following this general outline, we have computed the effect of the tidal field of a companion on a compact object in a binary system in the presence of an extra spatial dimension. The black hole is living on the four-dimensional hypersurface, i.e., on the brane and inherits a \emph{negative} tidal charge parameter $Q$, in addition to the mass and the angular momentum, from the extra dimension. The tidal charge slips in the computations of the rate of change of various geometrical entities, thereby modifying the respective expressions in comparison to general relativity. It turns out that, depending on the fact whether the orbital angular momentum and spin of the braneworld black hole are aligned or, anti-aligned, the rate of change of mass and angular momentum are positive or, negative of one another. Additionally, the presence of the tidal charge $Q$ makes the maximum spin of the object greater than unity, for which the deviation from the general relativistic prediction is the most. For anti-aligned binary system, the mass rate is always positive and it increases as the relative velocity between the two objects increase. While, the rate of change of angular momentum is always negative and it decreases as the relative velocity between the two objects increase. On the other hand, for aligned spin the rate of change of mass and angular momentum can be positive or, negative, they can increase or, decrease depending on the mass ratio. For equal mass binary, both the rates are negative and they decreases for large spin and tidal charge, but for extreme mass ratio binary, for large relative velocity both the rates are positive and they increase irrespective of the spin and tidal charge. 

The total change in the mass and angular momentum are larger for equal mass binary, than for the extreme mass ratio binary system, by a factor of $\sim 10^{3}$. More importantly, the total change in the dimensionless angular momentum $(\Delta J/M^{2})$ can be as large as $75\%$ for large value of $Q$ ($(Q/M^{2})\sim 0.5$), when compared to the general relativistic scenario, in the case of equal mass ratio in-spiral. On the other hand, the total change in the mass can be as large as $175\%$, when the $(Q/M^{2})=0.5$ case is compared to general relativity in the extreme mass ratio binary in-spiral. Thus the deviation from general relativity is significant as far as total change in the mass and angular momentum is concerned and can have observable consequences in future gravitational wave detectors.

For non-equatorial motion of braneworld black holes in a binary system, the averaged rate of change of mass and angular momentum also shows interesting behaviour. For extreme mass ratio in-spiral, the averaged angular momentum rate is always negative, while the averaged mass rate is always positive. However, for moderate mass ratio binary system, both the average mass and angular momentum rate can be positive or negative depending on the choice of the tidal charge and spin parameter. In this case, the average rate of change of mass in the presence of tidal charge differs from the general relativistic prediction by $\sim 350\%$, while the rate of change of angular momentum differs from the general relativistic prediction by $\sim 300\%$, as the spin parameter takes near extremal values. Thus even for motion off the equatorial plane, presence of the tidal charge leads to significant departure from the general relativistic prediction, which can be used to impose significant constraints on the tidal charge from future gravitational wave observations. Further, it turns out that one can indeed derive a stringent constraint on the tidal charge parameter, using the close resemblance of the rate of change of mass in the presence of extra dimension with that of ECOs. Using the constraint on the reflectivity of an ECO, following \cite{Datta:2019epe}, we have derived an order-of-magnitude estimate for the constraint on the dimensionless tidal charge parameter $q$, which reads $q \lesssim 10^{-5}$. This is obtained by comparing the mismatch between two gravitational wave waveforms, one with a reflective membrane and the other within the premise of general relativity, and the associated SNR. The parameter space is that of EMRIs and is most suited for detection through LISA.

The braneworld black holes, on the other hand, must be modified by the semi-classical corrections due to the CFT living on the brane, thanks to the AdS/CFT correspondence. Additionally, as emphasized earlier, it is more appropriate to consider the black hole horizon as an apparent horizon, rather than an event horizon, since the extension of the brane-localized black hole to the full bulk spacetime is not known. These motivates one to study the braneworld solutions as exotic compact objects. This brings in certain modifications to the tidal heating scenario. For example, the rate of change of geometrical quantities will involve corrections over and above the black hole result and will depend on the difference, denoted by the dimensionless parameter $\epsilon$, in the location of the surface of the ECO with that of the would-be black hole horizon. In particular, we have presented all the  analytical expressions of these corrections in the relevant section. Intriguingly, the zeroth order rate of change of geometrical entities are related to their black hole counterparts by a factor of $|\mathcal{T}|^{2}$, where $\mathcal{T}$ is the transmissivity of the surface of the ECO. To  first order as well the rate of change of geometrical quantities depend solely on the transmissivity, through the term $|\mathcal{T}|^{2}$. While, as first pointed out in this work, the reflectivity of the surface $\mathcal{R}$ appears at the second order in $\epsilon$ and has a characteristic form $\sim \cos(\ln \epsilon)$. This provides a characteristic variation of the rate of change of area with the spin parameter, which holds information about $\epsilon$ and hence about the exotic compact object. Thus it provides a new window to look for exotic compact objects from the phenomenon of tidal heating. 

The estimation of the constraint on the tidal charge parameter $q$, presented in this work, is through an indirect avenue. It would be interesting to derive the same from a more direct analysis, where the modifications to the gravitational wave waveform due to the tidal charge parameter $q$ will be directly incorporated.
Additionally, in connection with the ECO, we have discussed the case of stationary companion alone. Possible generalization to the non-stationary scenario and the total change in various geometrical quantities for equatorial motion must also be explored. Moreover, implications for motion on the non-equatorial plane needs to be discussed at length. These issues we leave for the future.

\section*{Acknowledgements}

The authors are indebted to Sukanta Bose for useful comments and suggestions for the betterment of the article. Research of S.C. is funded by the INSPIRE Faculty fellowship from the DST, Government of India (Reg. No. DST/INSPIRE/04/2018/000893) and by the Start-Up Research Grant from SERB, DST, Government of India (Reg. No. SRG/2020/000409). S.D. would like to thank the University Grants Commission (UGC), India, for financial support as a senior research fellow. We would like to thank the essential workers who put their health at risk during the COVID-19 pandemic, whose contributions helped us complete this work.
\appendix
\labelformat{section}{Appendix #1} 
\labelformat{subsection}{Appendix #1}
\begin{appendices}
\section{Deriving the contribution from the angular average of the Weyl scalar}\label{AppC}

In this appendix, we provide a derivation of the quantity, $\sum_{m\neq 0}|C_{m}|^{2}$, used in the main text and necessary to obtain the rate of change of area of a braneworld black hole in a binary, due to its companion. For this purpose, we start from \ref{mag_arb_cons} and realizing that the expression is even in "m", we immediately obtain,
\begin{flalign}
\sum^{2}_{m\neq 0}|C_{m}|^{2} & = 2 |C_{1}|^{2} + 2 |C_{2}|^{2} 
\nonumber
\\
& = \dfrac{64\pi^{2} M_{2}^{2}}{3\times 25 b^{6}} \dfrac{\chi_{1}^{2}}{4\left(1-\chi_{1}^{2}+\frac{Q}{M_{1}^{2}}\right)} 
\dfrac{\left(4-3\chi_{1}^{2}+\frac{4Q}{M_{1}^{2}}\right)}{4\left(1-\chi_{1}^{2}+\frac{Q}{M_{1}^{2}}\right)} 
\dfrac{\left(4+\frac{4Q}{M_{1}^{2}}\right)}{4\left(1-\chi_{1}^{2}+\frac{Q}{M_{1}^{2}}\right)} \times \dfrac{1}{4} \left(\dfrac{15}{2\pi}\right) \sin^{2}\theta_{0} \cos^{2} \theta_{0} 
\nonumber
\\
&\hskip 1 cm + \dfrac{64\pi^{2}M_{2}^{2}}{3\times 25 b^{6}} \dfrac{\chi _{1}^{2}}{\left(1-\chi_{1}^{2}+\frac{Q}{M_{1}^{2}}\right)} 
\dfrac{\left(4+\frac{4Q}{M_{1}^{2}}\right)}{4\left(1-\chi_{1}^{2}+\frac{Q}{M_{1}^{2}}\right)} 
\dfrac{\left(4+12\chi_{1}^{2}+\frac{4Q}{M_{1}^{2}}\right)}{4\left(1-\chi_{1}^{2}+\frac{Q}{M_{1}^{2}}\right)} \times \dfrac{1}{16} \left(\dfrac{15}{2\pi} \right) \sin^{4} \theta_{0} 
\nonumber
\\
&= \dfrac{64\pi^{2}M_{2}^{2}}{3\times 25 b^{6}} \left(\dfrac{15}{2\pi}\right) \dfrac{\chi_{1}^{2}}{\left(1-\chi_{1}^{2}+\frac{Q}{M_{1}^{2}}\right)^{3}} \bigg[\dfrac{\left(4-3\chi_{1}^{2}+\frac{4Q}{M_{1}^{2}}\right)\left(1+\frac{Q}{M_{1}^{2}}\right)}{64} \cos^{2} \theta_{0}  
\nonumber
\\
&\hspace{5.2cm}+ \dfrac{\left(1+\frac{Q}{M_{1}^{2}}\right)\left(1+3\chi_{1}^{2}+\frac{Q}{M_{1}^{2}}\right)}{16}\sin^{2}\theta_{0}\bigg] \sin^{2}\theta_{0}  
\nonumber
\\
&=\dfrac{2\pi M_{2}^{2}}{5b^{6}} \dfrac{\chi_{1}^{2} \sin^{2}\theta_{0}}{\left(1-\chi_{1}^{2}+\frac{Q}{M_{1}^{2}}\right)^{3}} 
\bigg[\left(1+\frac{Q}{M_{1}^{2}}\right)\left(1+\frac{Q}{M_{1}^{2}}+3\chi_{1}^{2} \right)\sin^{2}\theta_{0} 
\nonumber
\\
&\hspace{3cm}+ \dfrac{1}{4}\left(4+\frac{4Q}{M_{1}^{2}}-3\chi_{1}^{2}\right)\left(1+\frac{Q}{M_{1}^{2}}\right)(1-\sin^{2}\theta_{0}) \bigg] 
\nonumber
\\
&=\dfrac{2\pi M_{2}^{2}}{5b^{6}} \dfrac{\chi_{1}^{2} \sin^{2}\theta_{0}}{\left(1-\chi_{1}^{2}+\frac{Q}{M_{1}^{2}}\right)^{3}} 
\bigg[1-\dfrac{3}{4}\chi_{1}^{2}+\frac{Q}{M_{1}^{2}} +\frac{Q}{M_{1}^{2}} \left(1- \dfrac{3}{4}\chi_{1}^{2}+\frac{Q}{M_{1}^{2}}\right) 
\nonumber
\\
&\hskip 1 cm +\bigg\{1+\frac{Q}{M_{1}^{2}}+\frac{Q}{M_{1}^{2}}+\left(\frac{Q}{M_{1}^{2}}\right)^{2}+3\chi_{1}^{2}+3\chi_{1}^{2}\frac{Q}{M_{1}^{2}}  -1 
\nonumber
\\
&\hskip 2 cm -\frac{Q}{M_{1}^{2}} +\dfrac{3}{4}\chi_{1}^{2}-\frac{Q}{M_{1}^{2}}-\left(\frac{Q}{M_{1}^{2}}\right)^{2}+\dfrac{3}{4}\chi_{1}^{2}+\frac{Q}{M_{1}^{2}}\bigg\} \sin^{2}\theta_{0}\bigg] 
\nonumber
\\
&= \dfrac{2\pi M_{2}^{2}}{5b^{6}} \dfrac{\chi_{1}^{2}\sin^{2}\theta_{0}}{\left(1-\chi_{1}^{2}+\frac{Q}{M_{1}^{2}}\right)^{3}} 
\bigg[1- \dfrac{3}{4}\chi_{1}^{2}+\frac{Q}{M_{1}^{2}}\left\{2-\dfrac{3}{4}\chi_{1}^{2}+\frac{Q}{M_{1}^{2}}\right\}
+\dfrac{15}{4}\chi_{1}^{2}\left\{1+\frac{Q}{M_{1}^{2}}\right\}\sin^{2}\theta_{0}\bigg]
\end{flalign}
This is the expression we have used to derive \ref{ang_avg_final} in the main text. 
\section{Rate of change of area for braneworld black hole}\label{AppD}

In this appendix, we will derive the rate of change of area, which will be used in the main text. For this purpose, we need to compute the expression for the surface gravity $\kappa_{+}$ associated with the event horizon of the braneworld black hole, which takes the following form, 
\begin{align}
\kappa_{+}=\frac{r_{+}-r_{-}}{2(r_{+}^{2}+a^{2})}=\frac{\sqrt{M^{2}-a^{2}+Q}}{2Mr_{+}+Q}=\frac{1}{2M}\frac{\sqrt{1-\chi^{2}+\frac{Q}{M^{2}}}}{\left[1+\sqrt{1-\chi^{2}+\frac{Q}{M^{2}}}+\frac{Q}{2M^{2}}\right]}~,
\end{align}
where, $\chi=(a/M)$. Another useful result in this context is the connection between the shear $\sigma$ in the Hartle-Hawking frame and the tidal part of the Weyl scalar $\Psi_{0}$, which is given by,
\begin{align}
|\sigma_{\rm HH}|^{2}=\frac{1}{\kappa_{+}^{2}}|\Psi_{0}^{\rm HH}|^{2}~.
\end{align}
Therefore, using the above expression for $|\sigma_{\rm HH}|^{2}$ in \ref{stat_ini_area} and using \ref{ang_avg_final}, we obtain the rate of change of area to yield,
\begin{align}
\frac{dA_{\rm BH1}}{dt}&=\frac{2(r_{+1}^{2}+a_{1}^{2})}{\kappa_{+1}}\int |\sigma_{\rm HH}|^{2}d\Omega
=\frac{2(r_{+1}^{2}+a_{1}^{2})}{\kappa_{+1}^{3}}\int |\Psi_{0}^{\rm HH}|^{2}d\Omega
\nonumber
\\
&=\frac{2(r_{+1}^{2}+a_{1}^{2})}{\kappa_{+1}^{3}}\dfrac{2\pi M_{2}^{2}}{5b^{6}}\dfrac{\left(1-\chi_{1}^{2}+\frac{Q}{M_{1}^{2}}\right)\chi_{1}^{2}\sin^{2}\theta_{0}}{\left[\left(1+\sqrt{1-\chi_{1}^{2}+\frac{Q}{M_{1}^{2}}}\right)+\frac{Q}{2M_{1}^{2}}\right]^{4}}\Bigg[A+B\sin^{2}\theta_{0}\Bigg]
\nonumber
\\
&=2(2M_{1}r_{+1}+Q)\left[2M_{1}\frac{\left[1+\sqrt{1-\chi_{1}^{2}+\frac{Q}{M_{1}^{2}}}+\frac{Q}{2M_{1}^{2}}\right]}{\sqrt{1-\chi_{1}^{2}+\frac{Q}{M_{1}^{2}}}}\right]^{3}
\nonumber
\\
&\hskip 2 cm \times \dfrac{2\pi M_{2}^{2}}{5b^{6}}\dfrac{\left(1-\chi_{1}^{2}+\frac{Q}{M_{1}^{2}}\right)\chi_{1}^{2}\sin^{2}\theta_{0}}{\left[\left(1+\sqrt{1-\chi_{1}^{2}+\frac{Q}{M_{1}^{2}}}\right)+\frac{Q}{2M_{1}^{2}}\right]^{4}}\Bigg[A+B\sin^{2}\theta_{0}\Bigg]
\nonumber
\\
&=4M_{1}^{2}\left[1+\sqrt{1-\chi_{1}^{2}+\frac{Q}{M_{1}^{2}}}+\frac{Q}{2M_{1}^{2}}\right]\left[8M_{1}^{3}\frac{\left[1+\sqrt{1-\chi_{1}^{2}+\frac{Q}{M_{1}^{2}}}+\frac{Q}{2M_{1}^{2}}\right]^{3}}{\left(1-\chi_{1}^{2}+\frac{Q}{M_{1}^{2}}\right)^{3/2}}\right]
\nonumber
\\
&\hskip 2 cm \times \dfrac{2\pi M_{2}^{2}}{5b^{6}}\dfrac{\left(1-\chi_{1}^{2}+\frac{Q}{M_{1}^{2}}\right)\chi_{1}^{2}\sin^{2}\theta_{0}}{\left[\left(1+\sqrt{1-\chi_{1}^{2}+\frac{Q}{M_{1}^{2}}}\right)+\frac{Q}{2M_{1}^{2}}\right]^{4}}\Bigg[A+B\sin^{2}\theta_{0}\Bigg]
\nonumber
\\
&=32M_{1}^{5}\left(\dfrac{2\pi M_{2}^{2}}{5b^{6}}\right)\frac{1}{\sqrt{1-\chi_{1}^{2}+\frac{Q}{M_{1}^{2}}}}\chi_{1}^{2}\sin^{2}\theta_{0}\Bigg[A+B\sin^{2}\theta_{0}\Bigg]~,
\end{align}
which have been used in the main text.
\section{Expansion of the area of an exotic compact object on the brane}\label{AppE}

In this section, we will expand the area of the surface of the ECO in powers of the difference between the horizon location of the would-be black hole and the surface of the ECO. For the braneworld black hole, we have the following relations: $r_{+}^{2}+a^{2}=2Mr_{+}+Q$ and $a^{2}= r_{+}(2M-r_{+})+Q$. Using which, the coefficient of the linear power of $\epsilon$ in the expansion of the area of the surface of the ECO, yields,
\begin{flalign}
A_{1}&=\dfrac{4\pi r_{+}}{3(r_{+}^{2}+a^{2})} \left[6 r_{+}(r_{+}^{2}+a^{2}) -a^{2}(r_{+}-r_{-}) \right] 
\nonumber
\\
&=\dfrac{4\pi r_{+}}{3(2Mr_{+}+Q)}\left[6r_{+}(2Mr_{+}+Q)-2a^{2}(r_{+}-M) \right]
\nonumber
\\
&=\dfrac{4\pi}{3M\left(1+\dfrac{Q}{2Mr_{+}}\right)}\left[3r_{+}(2Mr_{+}+Q) - \{r_{+}(2M-r_{+})+Q\}(r_{+}-M)\right]~.
\end{flalign}
Similarly, the coefficient of the quadratic power of $\epsilon$ in the expansion of the area of the surface of the ECO, provides the following expression,
\begin{flalign}
A_{2}&=\dfrac{4\pi r_{+}^{2}}{15(r_{+}^{2}+a^{2})^{3}} \Bigg[ 15(r_{+}^{2}+a^{2})^{3}-5a^{2} (r_{+}^{2}+a^{2})^{2}+20 a^{2}r_{+} (r_{+}-M)(r_{+}^{2}+a^{2})-4a^{4}(r_{+}-M)^{2}\Bigg] 
\nonumber
\\
&=\dfrac{4\pi r_{+}^{2}}{15 (2Mr_{+}+Q)^{3}}\Bigg[15\left(8M^{3}r_{+}^{3}+ 12M^{2}r_{+}^{2}Q+6Mr_{+}Q^{2}+Q^{3}\right)-5 a^{2} \left( 4M^{2}r_{+}^{2} +4Mr_{+}Q+Q^{2}\right) 
\nonumber
\\
&\hskip 1 cm +20 r_{+}(r_{+}-M)\left\{2Mr_{+}^{2}(2M-r_{+})+2Mr_{+}Q+r_{+}(2M-r_{+})Q+Q^{2} \right\} 
\nonumber
\\
&\hskip 1 cm -4(r_{+}-M)^{2} \left\{r_{+}^{2}(r_{+}-2M)^{2}-2r_{+}(r_{+}-2M)Q+Q^{2} \right\}\Bigg] 
\nonumber
\\
&=\dfrac{4\pi r_{+}^{2}}{15(2Mr_{+})^{3}\left(1+\dfrac{Q}{2Mr_{+}}\right)^{3}}\Bigg[120 M^{3}r_{+}^{3}-20(2Mr_{+}-r_{+}^{2}+Q)M^{2}r_{+}^{2}+40Mr_{+}^{3}(r_{+}-M)(2M-r_{+}) 
\nonumber
\\
&\hskip 1 cm  -4r_{+}^{2}(r_{+}-M)^{2}(r_{+}-2M)^{2} 
\nonumber
\\
&\hskip 1 cm -Q\bigg\{-180M^{2}r_{+}^{2} +20Mr_{+}(2Mr_{+}-r_{+}^{2}+Q)-20r_{+}(r_{+}-M)(4Mr_{+}-r_{+}^{2}) -8r_{+}(r_{+}-2M)(r_{+}-M)^{2}\bigg\} 
\nonumber
\\
&\hskip 1 cm +Q^{2}\bigg\{90Mr_{+}-5(2Mr_{+}-r_{+}^{2}+Q)+20r_{+}(r_{+}-M)-4(r_{+}-M)^{2} \bigg\}+15Q^{3} \Bigg] 
\nonumber
\\
&=\dfrac{2\pi r_{+}}{15M^{3}\left(1+\dfrac{Q}{2Mr_{+}}\right)^{3}\times 4r_{+}^{2}}\Bigg[  -4r_{+}^{6}-16Mr_{+}^{5}+88M^{2}r_{+}^{4}+48 M^{3}r_{+}^{3}- 16M^{4}r_{+}^{2} 
\nonumber
\\
&\hskip 1 cm -Q\bigg\{ 12r_{+}^{4}-88Mr_{+}^{3}-80M^{2}r_{+}^{2}+16M^{3}r_{+}\bigg\}
+Q^{2} \bigg\{21r_{+}^{2}+48 Mr_{+}-4M^{2}\bigg\}-Q^{3} \left\{-10\right\} \Bigg]~.
\end{flalign} 
These are the expressions, which we have used in the main text.

\section{Rate of change of area of an exotic compact object on the brane}\label{AppG}

In this appendix, we will derive the expression for the rate of change of area of braneworld ECO, which will be used in the main text. Since the surface of the ECO is supposed to be very close to the would-be-horizon of the black hole, we can take the rate of change of area to be,
\begin{align} 
\frac{dA_{\rm ECO1}}{dt}&=\frac{2}{\kappa_{\rm s}}\int |\sigma_{\rm HH}|^{2}\sqrt{q}d^{2}x
\nonumber
\\
&=\frac{2}{\kappa_{\rm s}}\int \sqrt{(r_{\rm s}^{2}+a^{2})^{2}-a^{2}\Delta_{\rm s}\sin^{2}\theta}~|\sigma_{\rm HH}|^{2}d\Omega~,
\end{align}
where, we have used the expression for the induced metric on the surface of the ECO from \ref{tran_det} and $\kappa_{s}$ is related to one of the Newman-Penrose spin coefficient in the Hartle-Hawking tetrad on the surface of the ECO. Expanding the induced metric on the surface of the ECO in power series in $\epsilon$, we obtain,
\begin{align}\label{expansion_2metric}
(r_{\rm s}^{2}+a^{2})^{2}-a^{2}\Delta_{\rm s}\sin^{2}\theta
&=\left(r_{+}^{2}+a^{2}+2\epsilon r_{+}^{2}+\epsilon^{2}r_{+}^{2}\right)^{2}-\epsilon a^{2}r_{+}\left(r_{+}-r_{-}+\epsilon r_{+}\right)\sin^{2}\theta
\nonumber
\\
&=\left(r_{+}^{2}+a^{2}\right)^{2}+2\left(r_{+}^{2}+a^{2}\right)\left(2\epsilon r_{+}^{2}+\epsilon^{2}r_{+}^{2}\right)+\left(2\epsilon r_{+}^{2}+\epsilon^{2}r_{+}^{2}\right)^{2}
\nonumber
\\
&\hskip 2 cm -\epsilon r_{+}\left(r_{+}-r_{-}\right)a^{2}\sin^{2}\theta-\epsilon^{2}a^{2}r_{+}^{2}\sin^{2}\theta
\nonumber
\\
&=\left(r_{+}^{2}+a^{2}\right)^{2}+\epsilon \left[4r_{+}^{2}\left(r_{+}^{2}+a^{2}\right)-r_{+}\left(r_{+}-r_{-}\right)a^{2}\sin^{2}\theta\right]
\nonumber
\\
&\hskip 2 cm +\epsilon^{2}\left[2r_{+}^{2}\left(r_{+}^{2}+a^{2}\right)+4r_{+}^{4}-a^{2}r_{+}^{2}\sin^{2}\theta\right]~.
\end{align}
Thus, we obtain,
\begin{align}
\sqrt{(r_{\rm s}^{2}+a^{2})^{2}-a^{2}\Delta_{\rm s}\sin^{2}\theta}
&=\left(r_{+}^{2}+a^{2}\right)\Bigg\{1+\epsilon \frac{\left[4r_{+}^{2}\left(r_{+}^{2}+a^{2}\right)-r_{+}\left(r_{+}-r_{-}\right)a^{2}\sin^{2}\theta\right]}{\left(r_{+}^{2}+a^{2}\right)^{2}}
\nonumber
\\
&\hskip 3 cm +\epsilon^{2}\frac{\left[2r_{+}^{2}\left(r_{+}^{2}+a^{2}\right)+4r_{+}^{4}-a^{2}r_{+}^{2}\sin^{2}\theta\right]}{\left(r_{+}^{2}+a^{2}\right)^{2}} \Bigg\}^{\frac{1}{2}}
\nonumber
\\
&\equiv \left(r_{+}^{2}+a^{2}\right)\left[1+\epsilon q^{(1)}+\epsilon^{2} q^{(2)}\right]^{\frac{1}{2}}
\nonumber
\\
&= \left(r_{+}^{2}+a^{2}\right)\left[1+\epsilon \left(\frac{q^{(1)}}{2}\right)+\epsilon^{2}\left(\frac{q^{(2)}}{2}-\frac{\left(q^{(1)}\right)^{2}}{8} \right) \right]~.
\end{align}
Finally, the quantity $\kappa_{\rm s}$ can be determined along the following lines: First of all, in the Weyl tetrad $(\ell^{\mu},k^{\mu},m^{\mu},\bar{m}^{\mu})$, used in the earlier computation, the relevant spin-coefficient is equal to $\kappa$, which identically vanishes. When transformed to the Hartle-Hawking tetrad, the vectors $m^{\mu}$ and $\bar{m}^{\mu}$ remains unchanged and the vector $\ell^{\mu}$ gets mapped to $\xi^{\mu}_{\rm H}=[\Delta/2(r^{2}+a^{2})]\ell^{\mu}$, while $k^{\mu}$ is scaled by the inverse quantity. Since the vectors $m^{\mu}$ and $\bar{m}^{\mu}$ are still parallel transported along the null generator, it follows that even for the Hartle-Hawking tetrad $\kappa$ provides a measure of the respective spin coefficient. Thus we obtain, on any $r=\textrm{constant}$ surface, the following expression,
\begin{align}
\kappa&=k^{\rm H}_{\mu}\left(\xi^{\alpha}_{\rm H}\nabla_{\alpha}\xi^{\mu}_{\rm H}\right)
=\frac{2(r^{2}+a^{2})}{\Delta}\left[\frac{\Delta}{2(r^{2}+a^{2})}\ell^{\mu}\nabla_{\mu}\left(\frac{\Delta}{2(r^{2}+a^{2})}\right)\right]
\nonumber
\\
&=\partial_{r}\left(\frac{\Delta}{2(r^{2}+a^{2})}\right)=\frac{\Delta'}{2(r^{2}+a^{2})}-\frac{2r\Delta}{2(r^{2}+a^{2})^{2}}
=\frac{\Delta'(r^{2}+a^{2})-2r\Delta}{2(r^{2}+a^{2})^{2}}
\nonumber
\\
&=\frac{\left(r-M\right)(r^{2}+a^{2})-r\left(r^{2}+a^{2}-2Mr-Q\right)}{(r^{2}+a^{2})^{2}}=\frac{M(r^{2}-a^{2})+Qr}{(r^{2}+a^{2})^{2}}
\end{align}
Evaluation of the above quantity on the surface of the ECO, yields the following expression for $\kappa_{\rm s}$,
\begin{align}
\kappa_{\rm s}=\frac{M(r_{\rm s}^{2}-a^{2})+Qr_{\rm s}}{(r_{\rm s}^{2}+a^{2})^{2}}~.
\end{align}
Substituting, $r_{\rm s}=r_{+}(1+\epsilon)$ and then expanding in a power series in $\epsilon$, we obtain,
\begin{align}\label{expansion_kappa}
\kappa_{\rm s}&=\frac{M(r_{+}^{2}-a^{2}+2\epsilon r_{+}^{2}+\epsilon^{2}r_{+}^{2})+Qr_{+}(1+\epsilon)}{(r_{+}^{2}+a^{2}+2\epsilon r_{+}^{2}+\epsilon^{2}r_{+}^{2})^{2}}
\nonumber
\\
&=\frac{M(r_{+}^{2}-a^{2})+Qr_{+}+\epsilon \left(Qr_{+}+2Mr_{+}^{2}\right)+\epsilon^{2}Mr_{+}^{2}}{(r_{+}^{2}+a^{2})^{2}}\left[1+\frac{2\epsilon r_{+}^{2}+\epsilon^{2}r_{+}^{2}}{(r_{+}^{2}+a^{2})}\right]^{-2}
\nonumber
\\
&=\left(\kappa_{+}+\epsilon \frac{Qr_{+}+2Mr_{+}^{2}}{(r_{+}^{2}+a^{2})^{2}}+\epsilon^{2}\frac{Mr_{+}^{2}}{(r_{+}^{2}+a^{2})^{2}}\right)\left[1-\frac{4\epsilon r_{+}^{2}+2\epsilon^{2}r_{+}^{2}}{(r_{+}^{2}+a^{2})}+\epsilon^{2}\frac{12r_{+}^{4}}{(r_{+}^{2}+a^{2})^{2}}\right]
\nonumber
\\
&=\kappa_{+}+\epsilon \left[\frac{Qr_{+}+2Mr_{+}^{2}}{(r_{+}^{2}+a^{2})^{2}}-\frac{4\kappa_{+}r_{+}^{2}}{(r_{+}^{2}+a^{2})}\right]
+\epsilon^{2}\left[\frac{Mr_{+}^{2}}{(r_{+}^{2}+a^{2})^{2}}-\frac{4r_{+}^{2}\left(Qr_{+}+2Mr_{+}^{2}\right)}{(r_{+}^{2}+a^{2})^{3}}
+\frac{\kappa_{+}r_{+}^{2}\left(10r_{+}^{2}-2a^{2}\right)}{(r_{+}^{2}+a^{2})^{2}}\right]
\nonumber
\\
&\equiv \kappa_{+}+\epsilon \kappa^{(1)}+\epsilon^{2}\kappa^{(2)}~.
\end{align}
This expansion for the surface gravity $\kappa_{\rm s}$ on the surface of the ECO will be important in what follows. Another important result in this context is the connection between the shear $\sigma$ in the Hartle-Hawking tetrad and the tidal part of Weyl scalar $\Psi_{0}$, which is given by,
\begin{align}
|\sigma_{\rm HH}|^{2}=\frac{1}{\kappa_{\rm s}^{2}}|\Psi_{0}^{\rm HH}|^{2}~.
\end{align}
Therefore, using the above expression for $|\sigma_{\rm HH}|^{2}$ in \ref{stat_ini_area} and using \ref{ang_avg_final}, we finally obtain the rate of change of area to yield,
\begin{align}
\frac{dA_{\rm ECO1}}{dt}&=\frac{2}{\kappa_{\rm s}^{3}}\int \sqrt{(r_{\rm s}^{2}+a^{2})^{2}-a^{2}\Delta_{\rm s}\sin^{2}\theta}~|\Psi^{\rm HH}_{0}|^{2}d\Omega
\nonumber
\\
&=\frac{2\left(r_{+}^{2}+a^{2}\right)}{\kappa_{+}^{3}}\left(1+\epsilon \frac{\kappa^{(1)}}{\kappa_{+}}+\epsilon^{2}\frac{\kappa^{(2)}}{\kappa_{+}}\right)^{-3}
\int d\Omega \left[1+\epsilon \left(\frac{q^{(1)}}{2}\right)+\epsilon^{2}\left(\frac{q^{(2)}}{2}-\frac{\left(q^{(1)}\right)^{2}}{8} \right) \right]~|\Psi^{\rm HH}_{0}|^{2}~,
\end{align}
where, $q^{(1)}$ and $q^{(2)}$ have been defined in \ref{expansion_2metric} and $\kappa^{(1)}$ and $\kappa^{(2)}$ have been defined in \ref{expansion_kappa}. 

However, there is one problematic aspect associated with the above expression, namely, the angular average of $|\Psi^{\rm HH}_{0}|^{2}$ in the presence of the terms $q^{(1)}$ and $q^{(2)}$, both of which are dependent on the angular coordinate $\theta$. This is because, the angular dependence of the Weyl scalar is through the spin-weighted spherical harmonics and its angular average over functions of the angular coordinate $\theta$ is not known. Thus for our purpose, it will suffice to replace the functions $q^{(1)}$ and $q^{(2)}$ with their respective values on the equatorial plane, since we are interested in the rate of change of area and angular momentum for motion on the equatorial plane. Thus, we obtain,
\begin{align}
\frac{dA_{\rm ECO1}}{dt}&=\frac{2\left(r_{+}^{2}+a^{2}\right)}{\kappa_{+}^{3}}\left(1+\epsilon \frac{\kappa^{(1)}}{\kappa_{+}}+\epsilon^{2}\frac{\kappa^{(2)}}{\kappa_{+}}\right)^{-3}\left[1+\epsilon \left(\frac{q^{(1)}_{(\pi/2)}}{2}\right)+\epsilon^{2}\left(\frac{q^{(2)}_{(\pi/2)}}{2}-\frac{\left(q^{(1)}_{(\pi/2)}\right)^{2}}{8} \right) \right]
\nonumber
\\
&\hskip 6 cm \times \int d\Omega ~|\Psi^{\rm HH}_{0}|^{2}
\end{align}
Using \ref{expansion_eco_weyl}, we finally obtain,
\begin{align}
\frac{dA_{\rm ECO1}}{dt}&=\frac{2\left(r_{+}^{2}+a^{2}\right)}{\kappa_{+}^{3}}\left(1+\epsilon \frac{\kappa^{(1)}}{\kappa_{+}}+\epsilon^{2}\frac{\kappa^{(2)}}{\kappa_{+}}\right)^{-3}\left[1+\epsilon \left(\frac{q^{(1)}_{(\pi/2)}}{2}\right)+\epsilon^{2}\left(\frac{q^{(2)}_{(\pi/2)}}{2}-\frac{\left(q^{(1)}_{(\pi/2)}\right)^{2}}{8} \right) \right]
\nonumber
\\
&\times \Bigg\{\dfrac{\left(1-\chi^{2}+\frac{Q}{M^{2}}\right)^{4}}{\left[\left(1+\sqrt{1-\chi^{2}+\frac{Q}{M^{2}}}\right)+\frac{Q}{2M^{2}}\right]^{4}}\big|\mathcal{T}\big|^{2}\sum_{m=-2}^{2}\vert C_{m}^{\rm BH}\vert^{2}
\nonumber
\\
&+\epsilon \left(\frac{r_{+}(r_{+}-r_{-})^{7}}{16\left(r_{+}^{2}+a^{2}\right)^{4}}\right)\sum_{m=-2}^{2}\big|\mathcal{T}\big|^{2}\big|C_{m}^{\rm BH}\big|^{2}\frac{4\left(3-4\gamma_{m}^{2}\right)}{4\gamma_{m}^{2}-1}
\nonumber
\\
&+\epsilon^{2}\left(\frac{r_{+}^{2}(r_{+}-r_{-})^{6}}{16\left(r_{+}^{2}+a^{2}\right)^{4}}\right)\sum_{m=-2}^{2}\big|C_{m}^{\rm BH}\big|^{2}\Bigg[10\big|\mathcal{T}\big|^{2}+\mathcal{R}\mathcal{T}^{*}x^{-2\gamma_{m}}+\mathcal{R}^{*}\mathcal{T}x^{2\gamma_{m}}\Bigg]\Bigg\}
\end{align}
These are the expressions used in the main text, to determine the rate of change of area of the ECO in the braneworld scenario. 

\section{Basics of waveform mismatch}\label{mismatch}

To evaluate the strength of an effect to be measurable in a gravitational wave (GW) detector, with a noise power spectral density $S_n(f)$, it is needed to compute the overlap $\mathcal{O}$ between the two waveforms $h_1(t)$ and $h_2(t)$:
\begin{equation}\label{overlap}
\mathcal{O}(h_1|h_2) = \frac{\left\langle h_1|h_2\right\rangle}{\sqrt{\left\langle h_1|h_1\right\rangle \left\langle h_2|h_2\right\rangle}}\,,
\end{equation}
the inner product $\left\langle h_1|h_2\right\rangle$ is defined as follows:
\begin{equation}
\left\langle h_1|h_2\right\rangle = 4\Re\,\int_{0}^{\infty} \frac{\tilde{h}_1 \tilde{h}^*_2}{S_n(f)} df\,.
\end{equation}
Here, $\Re f$ depicts the real part of the function f and the quantities with tilde on the top are the Fourier transforms and the star represents complex conjugation. As the waveforms are defined up to an arbitrary time and phase shift, it is important to maximize the overlap, see \ref{overlap}, over them. This has been done by computing the following quantity \cite{Allen:2005fk}: 
\begin{equation}\label{overlap2}
\mathcal{O}(h_1|h_2) = \frac{4}{\sqrt{\left\langle h_1|h_1\right\rangle \left\langle h_2|h_2\right\rangle}}\max_{t_0} \left|\mathcal{F}^{-1}\left[\frac{\tilde{h}_1 \tilde{h}^*_2}{S_n(f)}\right](t_0)\right|\,,
\end{equation}
where, $\mathcal{F}^{-1}[g(f)](t) =\int_{-\infty}^{+\infty} g(f) e^{-2\pi i f t}df$ is the inverse Fourier transform. The definition implies that $\mathcal{O}=1$ indicates a perfect agreement between the two waveforms. Following which, the mismatch $(\mathfrak{M})$ is defined as follows:
\begin{equation}
\mathfrak{M}\equiv 1-{\mathcal{O}}
\end{equation}
This definition of mismatch has been used in the main text. 

\end{appendices}
\bibliography{References}

\bibliographystyle{./utphys1}
\end{document}